\documentclass[superscriptaddress,prb]{revtex4-2}
\usepackage[T1]{fontenc}
\usepackage[utf8]{inputenc}
\usepackage{booktabs}
\usepackage{amsmath}
\usepackage{amssymb}
\usepackage{graphicx}

\makeatletter

\providecommand{\tabularnewline}{\\}

\makeatother

\begin{document}
\title{Electron Thermalization and Relaxation in Laser-Heated Nickel by Few-Femtosecond
Core-Level Transient Absorption Spectroscopy}
\author{Hung-Tzu Chang}
\thanks{H.-T. C. and A. G. contributed equally to this work.}
\affiliation{Department of Chemistry, University of California, Berkeley, CA 94720,
USA}
\author{Alexander Guggenmos}
\thanks{H.-T. C. and A. G. contributed equally to this work.}
\altaffiliation{Current address: UltraFast Innovations GmbH, Am Coulombwall 1, 85748 Garching, Germany}

\affiliation{Department of Chemistry, University of California, Berkeley, CA 94720,
USA}
\author{Scott K. Cushing}
\altaffiliation{Current address: Division of Chemistry and Chemical Engineering, California Institute of Technology, Pasadena, CA 91125, USA}

\affiliation{Department of Chemistry, University of California, Berkeley, CA 94720,
USA}
\affiliation{Chemical Sciences Division, Lawrence Berkeley National Laboratory,
Berkeley, CA 94720, USA}
\author{Yang Cui}
\altaffiliation{Current address: Fakultät für Chemie, Technische Universität München, Lichtenbergstr. 4, 85748 Garching, Germany}

\affiliation{Max-Planck Institut für Quantenoptik, Hans-Kopfermann-Str. 1, 85748
Garching, Germany}
\affiliation{Fakultät für Physik, Ludwig-Maximilians-Universität München, Am Coulombwall
1, 85748 Garching, Germany}
\author{Naseem Ud Din}
\affiliation{Department of Physics, University of Central Florida, Orlando, FL
32816, USA}
\author{Shree Ram Acharya}
\affiliation{Department of Physics, University of Central Florida, Orlando, FL
32816, USA}
\author{Ilana J. Porter}
\affiliation{Department of Chemistry, University of California, Berkeley, CA 94720,
USA}
\affiliation{Chemical Sciences Division, Lawrence Berkeley National Laboratory,
Berkeley, CA 94720, USA}
\author{Ulf Kleineberg}
\affiliation{Max-Planck Institut für Quantenoptik, Hans-Kopfermann-Str. 1, 85748
Garching, Germany}
\affiliation{Fakultät für Physik, Ludwig-Maximilians-Universität München, Am Coulombwall
1, 85748 Garching, Germany}
\author{Volodymyr Turkowski}
\affiliation{Department of Physics, University of Central Florida, Orlando, FL
32816, USA}
\author{Talat S. Rahman}
\affiliation{Department of Physics, University of Central Florida, Orlando, FL
32816, USA}
\author{Daniel M. Neumark}
\email{dneumark@berkeley.edu}

\affiliation{Department of Chemistry, University of California, Berkeley, CA 94720,
USA}
\affiliation{Chemical Sciences Division, Lawrence Berkeley National Laboratory,
Berkeley, CA 94720, USA}
\author{Stephen R. Leone}
\email{srl@berkeley.edu}

\affiliation{Department of Chemistry, University of California, Berkeley, CA 94720,
USA}
\affiliation{Chemical Sciences Division, Lawrence Berkeley National Laboratory,
Berkeley, CA 94720, USA}
\affiliation{Department of Physics, University of California, Berkeley, CA 94720,
USA}
\date{\today}
\begin{abstract}
Direct measurements of photoexcited carrier dynamics in nickel are
made using few-femtosecond extreme ultraviolet (XUV) transient absorption
spectroscopy at the nickel M$_{2,3}$ edge. It is observed that the
core-level absorption lineshape of photoexcited nickel can be described
by a Gaussian broadening ($\sigma$) and a red shift ($\omega_{s}$)
of the ground state absorption spectrum. Theory predicts, and the
experimental results verify that after initial rapid carrier thermalization,
the electron temperature increase ($\Delta T$) is linearly proportional
to the Gaussian broadening factor $\sigma$, providing quantitative
real-time tracking of the relaxation of the electron temperature.
Measurements reveal an electron cooling time for 50 nm thick polycrystalline
nickel films of 640$\pm$80 fs. With hot thermalized carriers, the
spectral red shift exhibits a power-law relationship with the change
in electron temperature of $\omega_{s}\propto\Delta T^{1.5}$. Rapid
electron thermalization via carrier-carrier scattering accompanies
and follows the nominal 4 fs photoexcitation pulse until the carriers
reach a quasi-thermal equilibrium. Entwined with a <6 fs instrument
response function, carrier thermalization times ranging from 34 fs
to 13 fs are estimated from experimental data acquired at different
pump fluences and it is observed that the electron thermalization
time decreases with increasing pump fluence. The study provides an
initial example of measuring electron temperature and thermalization
in metals in real time with XUV light, and it lays a foundation for further investigation
of photoinduced phase transitions and carrier transport in metals
with core-level absorption spectroscopy. 
\end{abstract}
\maketitle

\section{Introduction}

Probing and harnessing the relaxation of hot carriers in metals and
semiconductors are vital to the development and design of photovoltaics
and photocatalysts \cite{rossEfficiencyHotcarrierSolar1982,wurfelSolarEnergyConversion1997,luqueElectronPhononEnergy2010a,jailaubekovHotChargetransferExcitons2013,kamideNonequilibriumTheoryConversion2018,zhangSurfacePlasmonDrivenHotElectron2018,zhangPlasmonDrivenCatalysisMolecules2019},
and to the understanding of mechanisms in various photoinduced phase
transitions \cite{beaurepaireUltrafastSpinDynamics1996,imadaMetalinsulatorTransitions1998,Kirilyuk2010,Johnson2017}.
After photoexcitation, carriers driven out of equilibrium quickly
form a thermalized hot carrier distribution within a few to tens of
femtoseconds through carrier-carrier scattering, before further cooling
takes place through carrier-phonon interactions at timescales ranging
from hundreds of femtoseconds to picoseconds \cite{shahUltrafastSpectroscopySemiconductors1996}.
Although the carrier cooling process typically involves complex interactions
between the electronic, phonon, and spin degrees of freedom, the dynamics
can be successfully described phenomenologically by a ``multi-temperature
model'' in a wide variety of systems \cite{anisimovElectronEmissionMetal1974,allenTheoryThermalRelaxation1987,singhTwotemperatureModelNonequilibrium2010,Johnson2017,rossEfficiencyHotcarrierSolar1982,shahUltrafastSpectroscopySemiconductors1996,luqueElectronPhononEnergy2010a,kamideNonequilibriumTheoryConversion2018}.
In such a model, the electronic, vibrational, and spin degrees of
freedom are regarded as individual heat reservoirs and the energy
transfer between the reservoirs is governed by a set of ``interaction
coefficients''. Each reservoir is presumed to be in quasi-thermal
equilibrium with a particular ``temperature'', based on the assumption
that the heat equilibration within each reservoir, e.g. due to electron-electron
scattering within electronic reservoir and anharmonic interactions
for phonon baths, is much faster than the inter-reservoir energy transfer
\cite{singhTwotemperatureModelNonequilibrium2010}. The usefulness
of multi-temperature models is widely evidenced in studies of energy
transfer in heterostructures \cite{vasileiadisUltrafastHeatFlow2018},
hot electron cooling in two-dimensional materials and superconductors \cite{kampfrathStronglyCoupledOptical2005,mansartTemperaturedependentElectronphononCoupling2013},
and photoinduced spin dynamics and phase transitions \cite{beaurepaireUltrafastSpinDynamics1996,rhieFemtosecondElectronSpin2003,kimelLaserinducedUltrafastSpin2004,stammFemtosecondModificationElectron2007,kachelTransientElectronicMagnetic2009,boeglinDistinguishingUltrafastDynamics2010,La-O-Vorakiat2012,raduTransientFerromagneticlikeState2011,staubPersistenceMagneticOrder2014,johnsonMagneticOrderDynamics2015,naseskaUltrafastDestructionRecovery2018,Johnson2017}.

Despite the success of the multi-temperature model in elucidating
a wide variety of photophysical phenomena, its applications are often
limited to systems with an already thermalized carrier distribution.
On the other hand, non-equilibrium hot carriers are known to facilitate
charge separation dynamics in organic heterojunctions \cite{jailaubekovHotChargetransferExcitons2013}
and play an important role in plasmon-induced photocatalysis \cite{zhangSurfacePlasmonDrivenHotElectron2018,zhangPlasmonDrivenCatalysisMolecules2019}.
Studies of ultrafast demagnetization and all-optical magnetic state
switching also demonstrate that the resulting spin dynamics can be
coherently driven by non-thermal photoexcited carriers \cite{kimelUltrafastNonthermalControl2005,bigotCoherentUltrafastMagnetism2009,Kirilyuk2010,gravesNanoscaleSpinReversal2013,eschenlohrUltrafastSpinTransport2013,Batignani2015,Bossini2016,siegristLightwaveDynamicControl2019}.
The importance of measuring the electron thermalization process is
further outlined in a recent optical pump-probe study on copper, which
shows that the electron thermalization timescale is strongly dependent
on the excitation fluence and at the low fluence limit, the thermalization
timescale can become comparable to the electron-phonon scattering
time \cite{obergfellTrackingTimeEvolution2020}. To measure the photoexcited
carrier distributions, time-resolved photoemission methods for valence
electrons probe energy and momentum resolved carrier distributions
in real time. However, the photoemission methods are restricted to
timescales greater than tens of femtoseconds, due to the relation
between the energy bandwidth and the duration of the pulses that eject
the photoelectrons, and thus have limited capacity in directly capturing
carrier dynamics below 20 fs while maintaining <0.2 eV energy resolution that is typically required to resolve carrier distributions in condensed matter. In addition, as alloy and multilayer
structures are intrinsic to the construction of photovoltaics and
many magnetic materials exhibiting photoinduced changes in magnetization
\cite{bigotCoherentUltrafastMagnetism2009,gravesNanoscaleSpinReversal2013,eschenlohrUltrafastSpinTransport2013,siegristLightwaveDynamicControl2019,el-ghazalyUltrafastMagnetizationSwitching2019,hofherrUltrafastOpticallyInduced2020},
insight into the properties and performance of these materials can
be obtained through understanding the carrier dynamics in each layer
or sub-domain in the system. Therefore, it is important to develop
a unified experimental approach that provides element specificity,
and thus domain or layer selectivity, can interrogate the sub-10 fs
dynamics of non-equilibrium carrier distributions, and is also capable
of presenting key parameters such as carrier temperature after thermalization
to facilitate the understanding of the interactions between the different
degrees of freedom in photoexcited materials.

Core-level transient absorption (TA) spectroscopy in the extreme ultraviolet
(XUV) has recently been developed and utilized to investigate carrier
dynamics in semiconductors \cite{Schultze2014,zurch2017direct,zurchUltrafastCarrierThermalization2017,schlaepferAttosecondOpticalfieldenhancedCarrier2018,linCarrierSpecificFemtosecondXUV2017a,verkampBottleneckFreeHotHole2019,Carneiro2017,Cushing2018,Porter2018,Cushing2019,cushingLayerresolvedUltrafastExtreme2020}.
Exploiting the element specificity of this method, Cushing et al.
investigated layer-specific carrier dynamics in a Si-TiO$_{2}$-Ni
trilayer structure \cite{cushingLayerresolvedUltrafastExtreme2020}.
In studies on germanium \cite{zurch2017direct}, lead iodide \cite{linCarrierSpecificFemtosecondXUV2017a},
and lead halide perovskites \cite{verkampBottleneckFreeHotHole2019},
the energy distribution of the carriers and their relaxation can be
directly extracted from XUV TA spectra. In addition, Volkov et al.
utilized XUV TA spectroscopy to explore effects due to the change
of electronic screening during photoexcitation of titanium \cite{volkovAttosecondScreeningDynamics2019a}.
However, despite numerous studies on electron dynamics in solids using
core-level TA spectroscopy, the methodology to extract the energy
distribution of photoexcited carriers or carrier temperature in metals
from core-level TA spectra is still lacking. For many semiconductors
with well-screened core holes, features of the core-level absorption
spectra can be mapped onto the conduction band (CB) density of states
(DOS) \cite{Rehr2003,zurch2017direct,linCarrierSpecificFemtosecondXUV2017a,verkampBottleneckFreeHotHole2019,attarSimultaneousObservationCarrierSpecific2020,Cushing2018},
and carrier dynamics can therefore be directly extracted from core-level
TA measurements. In metals, by contrast, many-body interactions of
electrons at the Fermi surface with the core hole potential strongly
renormalize the spectral lineshape of core-to-CB transitions \cite{mahanManyParticlePhysics2000},
resulting in strong resonances at the absorption edge, termed ``edge
singularities'' \cite{mahanExcitonsMetalsInfinite1967,rouletSingularitiesXRayAbsorption1969,nozieresSingularitiesXRayAbsorption1969,mahanCollectiveExcitationsXray1975,Ohtaka1990}.
As many-body interactions drastically reshape the core-level absorption
spectra beyond the CB DOS, it is thus highly challenging in metals
to unravel the carrier distributions and extract important parameters
such as carrier temperatures using core-level absorption spectroscopy. 

Here we employ nickel as a prototypical system and study the core-level
TA spectra at the nickel M$_{2,3}$ edge around 67 eV to develop a
framework to understand the core-level TA spectra of metals, extract
the electron temperatures, investigate the carrier cooling dynamics,
and explore electron thermalization. Nickel is a ferromagnetic material
exhibiting sub-picosecond demagnetization when irradiated with a femtosecond
laser pulse and has been extensively studied \cite{beaurepaireUltrafastSpinDynamics1996,hohlfeldNonequilibriumMagnetizationDynamics1997,conradUltrafastElectronMagnetization1999,regensburgerTimeresolvedMagnetizationinducedSecondharmonic2000,melnikovDemagnetizationFollowingOptical2002,rhieFemtosecondElectronSpin2003,Kampen2005,stammFemtosecondModificationElectron2007,bigotCoherentUltrafastMagnetism2009,youRevealingNatureUltrafast2018,Tengdin2018,siegristLightwaveDynamicControl2019,hofherrUltrafastOpticallyInduced2020}.
Time-resolved photoemission measurements indicate that the photoexcited
electrons in nickel thermalize on a sub-30 fs timescale \cite{Tengdin2018}
and electron-phonon relaxation times ranging from 200 fs to 1 ps have
been derived from optical transient reflectivity and time-resolved
second harmonic generation measurements \cite{hohlfeldNonequilibriumMagnetizationDynamics1997,conradUltrafastElectronMagnetization1999,regensburgerTimeresolvedMagnetizationinducedSecondharmonic2000,melnikovDemagnetizationFollowingOptical2002,rhieFemtosecondElectronSpin2003,Kampen2005}.
On the other hand, ultrafast electron diffraction experiments show
that a non-thermal phonon distribution persists over several picoseconds
after optical excitation \cite{maldonadoTrackingUltrafastNonequilibrium2020}.
The many studies on photoexcited carrier dynamics in nickel thus provide
suitable benchmarks for the methodology development here to reveal
electron dynamics in metals using XUV core-level spectroscopy. 

In this work, it is observed that the core-level absorption of laser-heated
nickel can be described by a red-shifted and Gaussian broadened static
absorption spectrum. In Sec. \ref{sec:power-dependence}, the results
of a set of power dependence measurements are shown and the resulting
Gaussian broadening exhibits a linear dependence with respect to the
electron temperature change. In addition, the fitted spectral shift
exhibits a power-law relationship with the electron temperature change.
A theory for the linear relation between the Gaussian broadening and
electron temperature change is derived in Sec. \ref{subsec:Temperature-Gaussian}
to complement the analyses, and Sec. \ref{sec:Discussion} presents
a conjecture based on analogy to the works on many-body interactions
in the core-excited state after optical excitation \cite{volkovAttosecondScreeningDynamics2019a}
to explain the nonlinear relationship between the spectral shift and
electron temperature. Section \ref{sec:Electron-Cooling-Dynamics}
displays the extraction of electron temperature from core-level TA
spectra of photoexcited nickel according to the linear relationship
between the spectral broadening and electron temperature rise, enabling
real-time tracking of the carrier cooling process, and an electron
cooling time of 640$\pm$80 fs is obtained. The measurement results
indicate that the contribution to spectral changes from phonon heating
is negligible and the cause of the spectral red shift in all measurements
within this work is purely electronic. In Sec. \ref{sec:Electron-Thermalization},
fluence dependence measurements reveal a decrease of XUV TA signal
rise time from 35 fs to 15 fs as the final electron temperature, viz.
the maximum electron temperature after thermalization, increases from
2100 K to 3100 K. The measured fluence-dependent electron thermalization
times are in good agreement with theoretical predictions \cite{Mueller2013}.
By comparing the <6 fs long instrument response function with the
<40 fs growth dynamics of the spectral features that become the profile
of thermalized carrier distributions in the fluence dependence measurements,
a range of electron thermalization times between 34 fs and 13 fs is
obtained.

\section{Results \label{sec:Results}}

\begin{figure}
\includegraphics[width=0.98\textwidth]{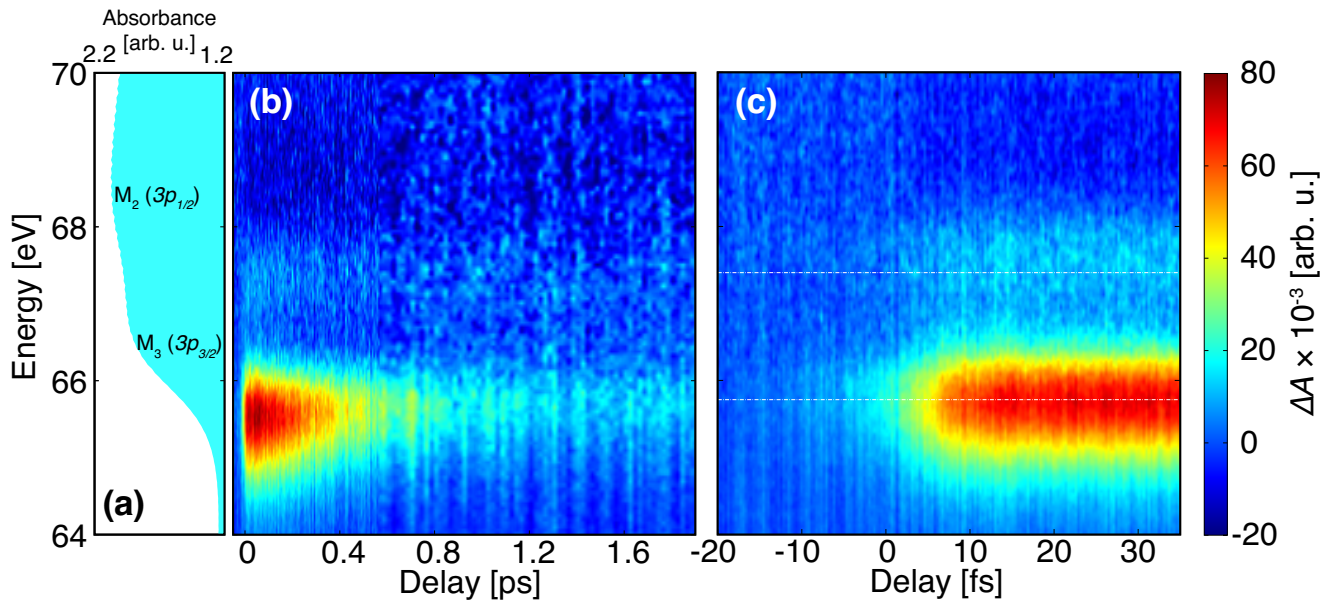}

\caption{(a) Static absorption spectrum of nickel M$_{2,3}$ edge and (b),
the measured XUV TA spectra of nickel between -50 fs and 1.9 ps time
delay. (c) displays the experimental XUV TA spectra between -20 and
35 fs time delay. The experimental pump fluence of the results in
(b) and (c) are 41 mJ/cm$^{2}$ and 33 mJ/cm$^{2}$, respectively.
\label{fig:2exp_results}}

\end{figure}

The core-level TA experiment on nickel was carried out with a table-top
XUV source based on high-harmonic generation. The experiment is depicted
in Fig. \ref{fig:experimental_setup} and details of the experimental
setup are described in Appendix \ref{sec:Experimental-Apparatus}.
Briefly, 50 nm thick polycrystalline nickel thin films deposited on
30 nm thick silicon nitride windows (Appendix \ref{sec:Sample-Preparation})
were excited at normal incidence by a $4.3\pm0.2$ fs long (Gaussian
FWHM), broadband optical pulse with a spectrum extending from 500
nm to 1000 nm and linear polarization. After optical excitation, the
sample was probed by a time-delayed broadband linearly polarized XUV
pulse, which is produced by high-harmonic generation in argon with
a $3.6\pm0.1$ fs long laser pulse centered at 730 nm. The polarization
of the XUV pulse is parallel to the optical pump and the XUV spectrum
spans 40 -- 73 eV (Fig. \ref{fig:experimental_setup}(b)). The XUV
spectrum covers the nickel M$_{2,3}$ edges located at 66.2 and 68.0
eV, consisting of excitations from nickel $3p$ levels to the conduction
band \cite{Dietz1980}. The static absorption spectrum of nickel
M$_{2,3}$ edges, shown in Fig. \ref{fig:2exp_results}(a), exhibits
a steep rising edge at approximately 66 eV due to absorption from
the Ni $3p_{3/2}$ core level (M$_{3}$) and another small absorption
feature at approximately 68.5 eV from the Ni $3p_{1/2}$ level excitation
(M$_{2}$). Above 69 eV, the absorbance slowly decreases with increasing
energy. 
\begin{figure}
\includegraphics[width=0.495\textwidth]{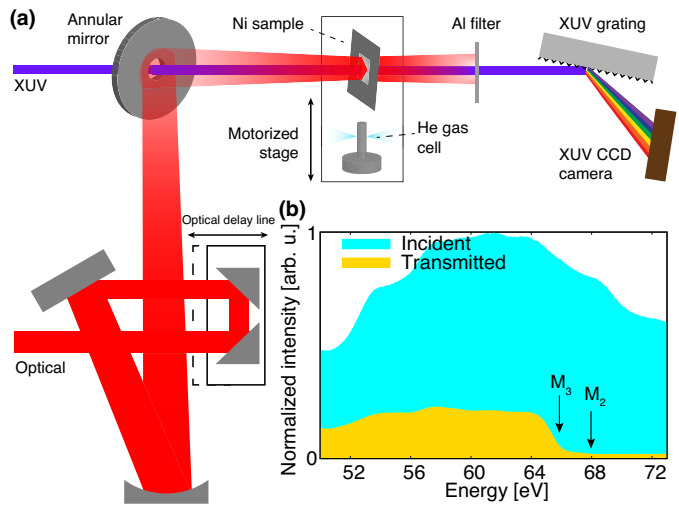}

\caption{(a) Experimental setup of the XUV TA experiment. (b) shows the spectra
of incident and transmitted XUV pulses through the 50 nm thick Ni
sample. }
\label{fig:experimental_setup}
\end{figure}

Dynamics following photoexcitation of nickel were probed by the change
of core-level absorbance $\Delta A$ at variable time delays between
the pump and probe pulses. A set of typical XUV TA spectra between
-50 fs and 1.9 ps time delay is displayed in Fig. \ref{fig:2exp_results}(b)
alongside the static absorption spectrum in Fig. \ref{fig:2exp_results}(a).
Two positive features (increased absorption) are observed at 65.7
and 67.4 eV, below the nickel M$_{3}$ and M$_{2}$ edge, respectively.
The two features decay within 1 ps, a duration conforming to the electron
cooling time in nickel due to electron-phonon interactions \cite{Kampen2005}.
To probe the electron thermalization dynamics, the XUV TA results
ranging from -20 fs to +35 fs time delay with 0.33 fs time steps are
plotted in Fig. \ref{fig:2exp_results}(c). The XUV TA spectra show
no significant changes between 15 fs and 35 fs time delay and no energetically
shifting spectral features are observed. In the following, we analyze
the results by first considering the interpretation of the core-level
TA spectra and extraction of electron temperature. Details of the
electron cooling and thermalization dynamics are discussed in Sec.
\ref{sec:Electron-Cooling-Dynamics} and \ref{sec:Electron-Thermalization},
respectively. 

To understand the core-level absorption spectra of optically excited
nickel, we focus on the XUV TA profile right after photoexcitation.
A core-level TA spectrum at 40 fs pump-probe delay is plotted in Fig.
\ref{fig:NiStatic_power_dependence}(a). Here, apart from the increase
in absorption (positive $\Delta A$) below the nickel M$_{3}$ and
M$_{2}$ edge (<67.4 eV), a shallow negative feature occurs above
the nickel M$_{2}$ edge (68 eV). Unlike the core-level TA spectra
at the nickel L$_{2,3}$ edge, where the magnitude of absorption changes
below and above the edge is highly symmetric \cite{stammFemtosecondModificationElectron2007,kachelTransientElectronicMagnetic2009},
in the M edge TA spectrum the positive features are much stronger
than the the negative feature and the integrated area of the TA profile
($\int\Delta A(\omega)d\omega$) is clearly nonzero. The asymmetry
of the TA profile indicates that it cannot be directly interpreted
by electronic occupation below and above the Fermi level in contrast
to the TA spectra at the nickel L edge \cite{carvaInfluenceLaserexcitedElectron2009}.
While the cause of the asymmetric TA profile is beyond the scope of
this work, it may stem from the asymmetric Fano broadening of the
nickel M$_{2,3}$ edge due to Coster-Kronig decay of the core hole
\cite{dietzLineShapeExcitation1974,davisInterpretation3pcoreexcitationSpectra1976,Dietz1980},
the many-body interactions between the electrons at the Fermi surface
and the core hole \cite{volkovAttosecondScreeningDynamics2019a},
and the splitting of core-levels with different angular and magnetic
quantum numbers \cite{valenciaQuadraticXRayMagnetoOptical2010}.

As Tengdin et al. showed that the electron thermalization time in
nickel is <30 fs \footnote{Tengdin et al. detected thermalized hot electron distribution with
time-resolved angle-resolved photoemission 24 fs after photoexcitation
by pulses centered at 780 nm with fluence of <6 mJ/cm$^{2}$ \cite{Tengdin2018}.
The reported pulse duration in Ref. \cite{Tengdin2018} is 28 fs.
As the rate of electron scattering increases with carrier temperature,
the carrier thermalization time is expected to be <30 fs long at fluences
used in this study (8 -- 62 mJ/cm$^{2}$).}, a hot, thermalized electron distribution is expected to be established
in the CB by 40 fs after photoexcitation, and as the electron-phonon
scattering time in nickel is on the order of a few hundred femtoseconds,
energy loss to the phonon bath can be ignored. Almbladh and Minnhagen
\cite{almbladhThermalBroadeningCore1978}, Ohtaka and Tanabe \cite{Ohtaka1983,Tanabe1984,ohtakaGoldenruleApproachSoftxrayabsorption1984},
and Ortner and coworkers \cite{adamjanXrayabsorptionProblemMetals1995,ortnerXrayabsorptionProblemMetals1996,Ortner1997}
have independently shown that the increase of electron temperature
in metals can impose a broadening to the core-level absorption edge.
In addition, the change of electronic screening in CB due to photoexcitation
can cause an energy shift of the core-to-CB transitions \cite{johanssonCorelevelBindingenergyShifts1980,stammFemtosecondModificationElectron2007,volkovAttosecondScreeningDynamics2019a}.
Thus motivated, we consider a model where the core-level absorption
of photoexcited nickel ($I(\omega)$) is simulated by a Gaussian broadening
$\sigma$ of the static absorption spectrum ($I_{0}(\omega)$) with
an overall energy shift $\omega_{s}$: 
\begin{align}
I(\omega,\omega_{s},\sigma) & =\int d\omega^{\prime}\:I_{0}(\omega-\omega_{s}-\omega^{\prime})f(\omega^{\prime},\sigma),\label{eq:fitting_model}\\
f(\omega,\sigma) & =\frac{1}{\sigma\sqrt{2\pi}}\exp(-\frac{\omega^{2}}{2\sigma^{2}}).\nonumber 
\end{align}
The fitting of the experimental TA spectrum (Fig. \ref{fig:NiStatic_power_dependence}(a),
blue line) with Eq. (\ref{eq:fitting_model}) (Fig. \ref{fig:NiStatic_power_dependence}(a),
red line) exhibits good agreement between the fitted and the experimental
result.

\subsection{Power Dependence Measurements \label{sec:power-dependence}}

\begin{figure}
\includegraphics[width=0.48\textwidth]{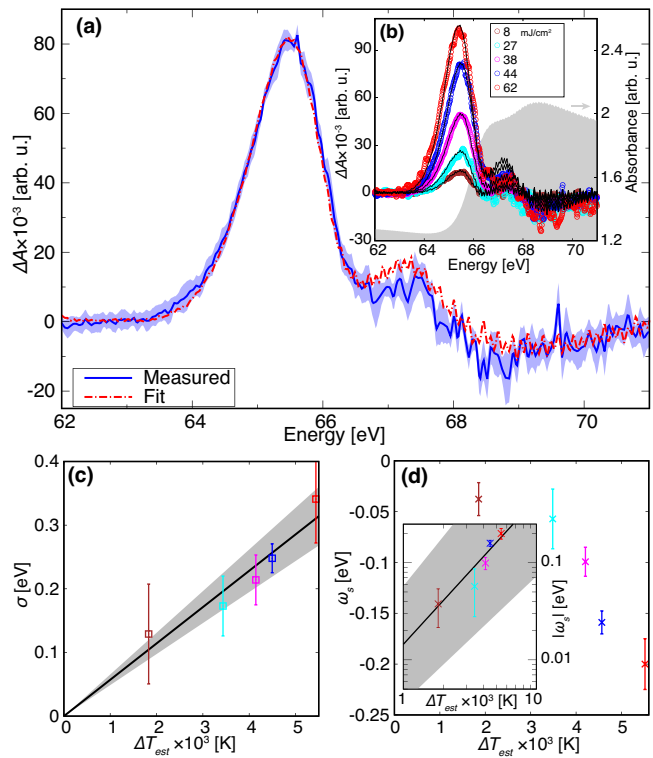}

\caption{(a) A typical XUV TA spectrum of nickel at 40 fs time delay (blue
line). $\Delta A$ denotes the change of XUV absorbance and the blue
shade in (a) shows the uncertainty of the TA spectrum. The red line
is the fitting result of Eq. (\ref{eq:fitting_model}) on the experimental
data. The Gaussian broadening and shift obtained from the fitting
are $0.25\pm0.02$ eV and $-0.16\pm0.01$ eV, respectively. (b) Absorbance
change $\Delta A$ (circles) after optical excitation at 40 fs time
delay with 5 different laser fluences. The static absorption spectrum
of nickel M$_{2,3}$ edge is displayed in gray as a reference. Results
of the fitting of measured data with Eq. (\ref{eq:fitting_model})
are shown in black lines and the obtained $\sigma$ and $\omega_{s}$
as a function of the simulated electron temperature rise $\Delta T_{est}$
are shown in (c) and (d), respectively. The results of linear fitting
of $\sigma$ versus $\Delta T_{est}$ is displayed as a black line
in (c) and the inset in (d) exhibits the fitting of $|\omega_{s}|$
versus $\Delta T_{est}$ with a power function (black line) in a log-log
plot. The uncertainties in the fitting are shown as gray areas in
(c) and (d), respectively.}
\label{fig:NiStatic_power_dependence}
\end{figure}

To quantify the effects of electron temperature on the core-level
absorption spectra and extract the physical properties related to
the broadening and shift, we performed power dependence measurements
at five different laser fluences (8 -- 62 mJ/cm$^{2}$). The core-level
TA spectra at 40 fs time delay and their fitting with Eq. (\ref{eq:fitting_model})
are displayed in Fig. \ref{fig:NiStatic_power_dependence}(b). The
magnitude of all TA features increases with increasing fluence and
the fitting results show that Eq. (\ref{eq:fitting_model}) can accurately
describe the measured spectral changes (Fig. \ref{fig:NiStatic_power_dependence}(b),
black lines). The spectral broadening and shift obtained from the
fitting with the corresponding electron temperature change, calculated
from the pump fluence and electron heat capacity of nickel (Appendix
\ref{sec:Estimating-Electron-Temperature}), are plotted in Figs.
\ref{fig:NiStatic_power_dependence}(c) and (d), respectively. Here
it is observed that the spectral broadening $\sigma$ and the corresponding
calculated electron temperature change $\Delta T_{est}$ can be fitted
by a formula $\sigma=a\Delta T_{est}$ (Fig. \ref{fig:NiStatic_power_dependence}(c),
black line), with $a=(5.7\pm0.8)\times10^{-5}$ eV/K. The negative
spectral shift $\omega_{s}$ indicates the absorption edge red-shifts
with increasing electron temperature (Fig. \ref{fig:NiStatic_power_dependence}(d))
and it exhibits a power-law relationship with electron temperature
change ($\omega_{s}\propto\Delta T_{est}^{\alpha}$), with exponent
$\alpha=1.5\pm0.2$ (Fig. \ref{fig:NiStatic_power_dependence}(d),
inset). We defer discussion on the cause of the spectral shift to
Sec. \ref{subsec:spectral-shift} and first consider a theoretical
explanation of the observed relation between the electron temperature
change and Gaussian broadening.

\subsection{Relation between Gaussian Broadening and Electron Temperature Changes
\label{subsec:Temperature-Gaussian}}

The many-body theory of core-level absorption in metals at nonzero
temperatures, pioneered by Almbladh and Minnhagen \cite{almbladhThermalBroadeningCore1978},
Ohtaka and Tanabe \cite{Ohtaka1983,Tanabe1984,ohtakaGoldenruleApproachSoftxrayabsorption1984},
and Ortner and coworkers \cite{adamjanXrayabsorptionProblemMetals1995,ortnerXrayabsorptionProblemMetals1996,Ortner1997},
shows that the core-level absorption spectra $I^{\prime}(\omega)$
can be approximated by a simple formula (in atomic units): 
\begin{align}
I^{\prime}(\omega) & \propto\ensuremath{\frac{{1}}{2}}\text{{Re}}\int_{-\infty}^{\infty}e^{i(\omega-\omega_{0})t}I^{\prime}(t)dt\nonumber \\
I^{\prime}(t) & =\left(\frac{{\pi T}}{i\sinh(\pi Tt)}\right)^{1-\zeta}.\label{eq:O-T-model}
\end{align}
Here, $\omega_{0}$ represents the difference between the Fermi energy
and the excited core-level, $T$ is the electron temperature, and
$\zeta<1$ is a coefficient related to the phase shift of the scattered
electrons from the core hole potential. Although the expression only
includes the effect of electron temperature, assumes a slowly varying
density of states near the Fermi level, and ignores several factors
that can lead to spectral distortion and broadening such as Auger
decay of core holes, phonon heating \cite{olsonThermomodulatedExafsSoft1980,olsonTemperatureDependenceM21980},
and sample crystallinity and inhomogeneity, it provides a clear mathematical
representation of temperature effects on core-level absorption. Note
that while the formula is derived from many-body theory, it is still
valid for core-level absorption in metals when core-hole mediated
electron scattering at the Fermi surface is negligible.

To derive an expression relating spectral changes and electron temperature
rise $\Delta T$, we consider an \textit{ansatz} relating the absorption
lineshape $I(\omega,T_{0}+\Delta T)$ with electron temperature change
$\Delta T$ with respect to the absorption spectrum of the sample
at temperature $T_{0}$ before excitation: 
\begin{equation}
I(\omega,T_{0}+\Delta T)=\int d\omega^{\prime}\:I(\omega-\omega^{\prime},T_{0})f(\omega^{\prime},\Delta T),\label{eq:spectral_convolution}
\end{equation}
where the absorption lineshape at electron temperature $T_{0}$ is
convoluted by a function $f(\omega,\Delta T)$. The choice of convolution
in the expression is motivated from experimental observation (Fig.
\ref{fig:NiStatic_power_dependence} and Eq. (\ref{eq:fitting_model}))
and theoretical results \cite{ohtakaGoldenruleApproachSoftxrayabsorption1984}.
The combination of Eqs. (\ref{eq:O-T-model}) and (\ref{eq:spectral_convolution})
indicate that $f(\omega,\Delta T)$ can be found by approximating
the component of free-induction decay $I^{\prime}(t)$ in Eq. (\ref{eq:O-T-model})
by
\begin{equation} 
\frac{\pi(T_{0}+\Delta T)}{\sinh(\pi(T_{0}+\Delta T)t)}\approx\frac{\pi T_{0}}{\sinh(\pi T_{0}t)}\times e^{g(t,\Delta T)},\label{eq:sinh_expand}
\end{equation}
with $f(\omega,\Delta T)=\frac{1}{2\pi}\int_{-\infty}^{\infty}e^{i\omega t}e^{(1-\zeta)g(t,\Delta T)}dt.$
Using second order expansion of $\Delta T$ on both sides of Eq. (\ref{eq:sinh_expand}),
it is shown that the first order term in the expansion of $g(t,\Delta T)$
with respect to time $t$ is zero and the coefficient of the second
order term is proportional to $(\Delta T)^{2}$ when $T$ is small
compared to $\Delta T$ (Appendix \ref{app:cumulant_expansion}).
In other words, the term $e^{(1-\zeta)g(t,\Delta T)}$ can be approximated
by the expression 
\[
e^{(1-\zeta)g(t,\Delta T)}\propto\exp\left(-\frac{(\sigma t)^{2}}{2}\right),
\]
with $\sigma=a\Delta T$, where $a$ is a proportionality constant.
This suggests that the core-level absorption spectrum after laser
heating $I(\omega,T_{0}+\Delta T)$ can be described by the convolution
of the spectrum before heating $I(\omega,T_{0})$ with a Gaussian
function $f(\omega,\sigma)=\frac{1}{\sigma\sqrt{2\pi}}\exp\left(-\frac{\omega^{2}}{2\sigma^{2}}\right)$
and the broadening factor $\sigma$ is directly proportional to the
temperature rise $\Delta T$.

Comparing Eq. (\ref{eq:sinh_expand}) with Eq. (\ref{eq:fitting_model}),
it is observed that the reference spectrum at temperature $T_{0}$
in Eq. (\ref{eq:sinh_expand}) is represented by the static spectrum
in Eq. (\ref{eq:fitting_model}). The two expressions merely differ
by the spectral shift $\omega_{s}$, which is not present in the derivation
above because the overall energy shift of the core-excited state is
not included in either the many-body theory of core-level absorption
\cite{ohtakaGoldenruleApproachSoftxrayabsorption1984,Ortner1997}
or our proposed ansatz (Eq. (\ref{eq:spectral_convolution})). The
mathematical derivation thus justifies the fitting of TA profiles
with a Gaussian broadened static spectrum and suggests that this approach
can be extended to other metallic systems. In addition, the robustness
of the method (Eq. (\ref{eq:fitting_model})) is ensured by limiting
the fitting parameters to only the shift and broadening. This is because
the major factors that contribute to spectral distortions and broadenings
are implicitly included in the formalism. The reference spectrum $I(\omega,T_{0})$
in Eq. (\ref{eq:spectral_convolution}) and the static absorption
spectrum in Eq. (\ref{eq:fitting_model}) automatically incorporate
the spectral contribution from the intrinsic core hole lifetime as
well as sample geometry and crystallinity, which remain unchanged
throughout the measurement. The effects of interaction between the
core hole and laser-heated CB electrons are included in Eq. (\ref{eq:O-T-model}),
the starting point of the derivation of $f(\omega,\Delta T)$. Note,
however, the mathematical derivation only considers the spectral changes
due to variation in electron temperature; other contributions from
processes that would follow photoexcitation, such as phonon dynamics,
are not included. In addition, the derivation of $f(\omega,\Delta T)$
involves truncation in a series expansion and is not analytically
exact. Therefore, the application of this method (Eq. (\ref{eq:fitting_model}))
and the extraction of electron temperature from the broadening always
require verification that the relation between the electron temperature
and the broadening is linear and dynamics in the electronic domain
are the dominant contributor to the TA lineshape. The relationship
between projected electron temperature and the fitted spectral broadening
in power dependence measurements, and the correspondence between the
fitted spectral shift and broadening, can both serve as checkpoints
to examine the adequacy of this analysis approach.

\subsection{Electron Cooling Dynamics \label{sec:Electron-Cooling-Dynamics}}

\begin{figure}
\includegraphics[width=0.48\textwidth]{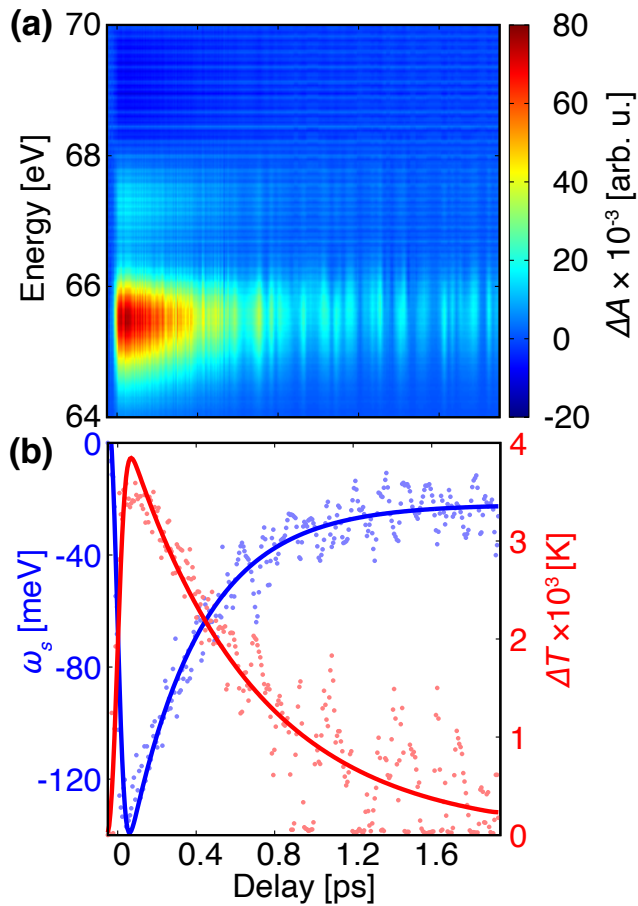}
\caption{(a) Results of fitting the TA spectra in Fig. \ref{fig:2exp_results}(b)
with Eq. (\ref{eq:fitting_model}) and the fitting parameters $\omega_{s}$
and $\Delta T$ at different time delays are shown as dots in (b).
The fitting of the changes of $\omega_{s}$ and $\Delta T$ as a function
of time delay with single exponential decay convoluted with a Gaussian
instrument response function are shown as the blue and red line, respectively.
\label{fig:long-time-dynamics}}
\end{figure}

Equipped with the formalism to understand the spectral change in core-level
excitations in nickel with hot thermalized electrons at a short 40
fs pump-probe delay (Eq. (\ref{eq:fitting_model})), we consider here
the electron cooling dynamics at >40 fs timescales (Fig. \ref{fig:2exp_results}(b)).
To analyze the measured dynamics, we first apply Eq. (\ref{eq:fitting_model})
to fit the TA spectrum at each time delay. The fitted spectra are
shown in Fig. \ref{fig:long-time-dynamics}(a), displaying good agreement
with the experimental results (Fig. \ref{fig:2exp_results}(b)). Note
that phonon heating is expected to occur from electron-phonon scattering
processes during the cooling of the electron bath \cite{Kampen2005}
and the adequacy of relating electron temperature with spectral broadening
requires verification. It has been shown that the rise of non-thermal
phonon occupation persists over timescales above 2 ps \cite{maldonadoTrackingUltrafastNonequilibrium2020}
and the electronic reservoir cools down below 1 ps \cite{Kampen2005}.
If phonon dynamics contribute significantly to the XUV TA spectra,
it is expected that at >1 ps timescales the contribution from phonons
would dominate and the magnitude of the XUV TA signal would rise with
increasing time delay. Contrary to the expectations, however, the
experimental TA signal diminishes to zero with increasing time delay
and its magnitude at 1.9 ps delay is barely above noise level (Fig.
\ref{fig:2exp_results}(b)). This suggests that the contribution from
heated phonons to the observed core-level TA signal is negligible
compared to the dynamics in the electronic domain \footnote{Note that the results only indicate that the TA signal is not sensitive
to the particular phonon heating dynamics in the experiment. It does
not imply that core-level absorption spectroscopy is insensitive to
phonon dynamics overall and depending on the system measured, lattice
dynamics can contribute to core-level TA signals (e.g. Ref. \cite{rothenbachMicroscopicNonequilibriumEnergy2019}).}. As such, the obtained spectral broadening $\sigma$ can still be
connected with $\Delta T$ in the analysis of XUV TA spectra at hundreds
of femtoseconds to picoseconds. 

Figure \ref{fig:long-time-dynamics}(b) shows the fitted spectral
shift (blue dots) and electron temperature ($\Delta T=\sigma/a$,
red dots) derived from the fitted spectral broadening at different
time delays. Both the spectral shift and electron temperature as a
function of time delay can be fitted by a single exponential decay
convoluted with the instrument response function (Fig. \ref{fig:long-time-dynamics}(b),
lines) and the time constants for electron temperature decay $\tau_{\sigma}=640\pm80$
fs and spectral shift $\tau_{s}=380\pm30$ fs are obtained, respectively.
Note that the shift $\omega_{s}$ and broadening $\sigma$ are the
only variables in the fitting procedure (Eq. (\ref{eq:fitting_model})),
and no additional parameters are introduced. The $640\pm80$ fs cooling
time is consistent with the reported values from optical pump-probe
measurements \cite{hohlfeldNonequilibriumMagnetizationDynamics1997,conradUltrafastElectronMagnetization1999,melnikovDemagnetizationFollowingOptical2002}.
The discrepancy between $\tau_{\sigma}$ and $\tau_{s}$ also agrees
with the observed relationship between the electron temperature and
the spectral shift in the power dependence measurements (Fig. \ref{fig:NiStatic_power_dependence}(d)).
As the spectral shift is related to the electron temperature change
by a power law ($\omega_{s}\propto\Delta T^{\alpha}$), it is expected
that the decay dynamics of spectral shift to follow the relation $\omega_{s}(t)\propto\Delta T(t)^{\alpha}\propto e^{-\alpha t/\tau_{\sigma}},$with
$\alpha=\tau_{\sigma}/\tau_{s}$. Here, the obtained $\tau_{\sigma}/\tau_{s}$
is approximately 1.7, agreeing with the $\alpha=1.5\pm0.2$ obtained
from the power dependence measurements (Fig. \ref{fig:NiStatic_power_dependence}(d),
inset). It is thus observed that the relation between the spectral
shift and broadening is maintained between 40 fs and 1.9 ps time delay.
At 40 fs delay, the effect of electron-phonon interaction is negligible
because the timescale is well below the electron-phonon scattering
time \cite{Kampen2005}, whereas at picosecond timescales, non-equilibrium
phonon dynamics are paramount \cite{Kampen2005,Waldecker2016,maldonadoTrackingUltrafastNonequilibrium2020}.
The consistency of the behavior between the spectral shift and broadening
from 40 fs to picosecond time delays further corroborate that the
fitted spectral broadening and shift are related to physical properties
in the electronic domain because the relation between the two persists
regardless of the generation of phonons out of thermal equilibrium.

\subsection{Electron Thermalization Dynamics \label{sec:Electron-Thermalization}}

\begin{figure}
\includegraphics[width=0.49\textwidth]{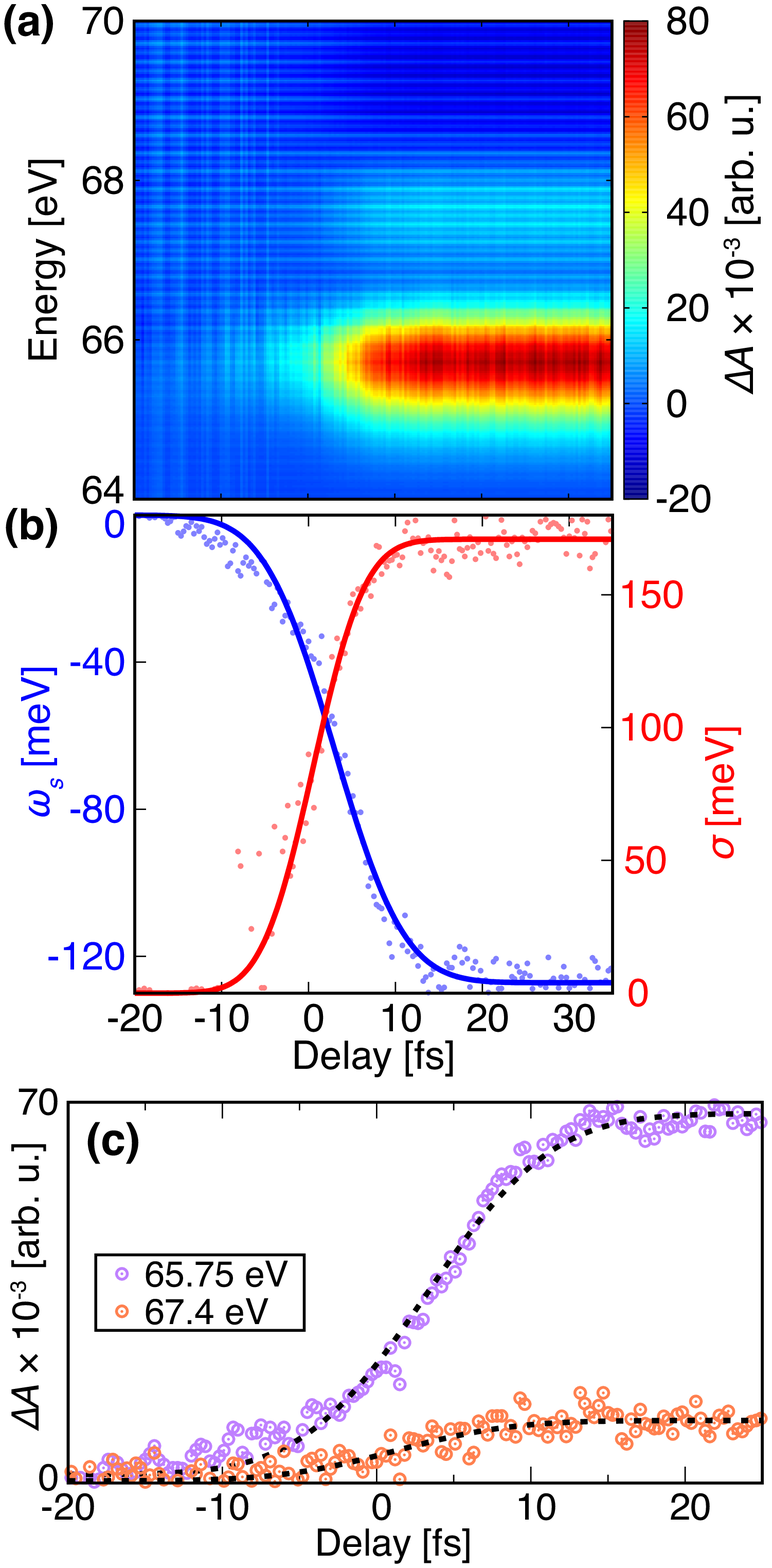}
\caption{(a) Fitting results of Fig. \ref{fig:2exp_results}(c) with Eq. (\ref{eq:fitting_model}).
The fitted edge shift ($\omega_{s}$) and broadening ($\sigma$) as
a function of time delay are plotted in (b) as dots and the fitting
of $\omega_{s}(t)$ and $\sigma(t)$ with Eq. (\ref{eq:error_function_fit})
are depicted as blue and red lines, respectively. (c) shows the lineouts
of $\Delta A$, shown in circled dots, at 65.75 and 67.4 eV (Fig.
\ref{fig:2exp_results}(c), white dashed lines) and their fitting
results (dashed black lines) with Eq. (\ref{eq:error_function_fit}).\label{fig:short-time-dynamics}}
\end{figure}

In this section, we focus on the few-femtosecond dynamics of carrier
photoexcitation and thermalization. The XUV TA spectra between -20
and 35 fs pump-probe delay with optical pump fluence at 33 mJ/cm$^{2}$
are plotted in Fig. \ref{fig:2exp_results}(c), which exhibits two
positive features at 65.75 eV and 67.4 eV that increase in magnitude
within -10 fs to +15 fs time delay and reach a plateau at >15 fs. The
lack of changes in the TA signal between 15 fs and 35 fs suggests
that the electrons thermalize within 15 fs \footnote{Note that, however, because the core-level absorption spectrum at
the Ni M$_{2,3}$ edge cannot be directly mapped on to the CB DOS,
it is impossible to directly quantify the deviation of the carrier
distribution from a hot Fermi-Dirac function with the core-level TA
spectra.}. The absence of energetically shifting spectral features within -10
fs to +15 fs time delay, which could be used to directly signify the
scattering and decay of the initial non-equilibrium photoexcited electrons
to form a hot thermalized distribution, initially implies that the
electrons thermalize within the pulse duration of the pump. However,
the analysis of the lineouts of TA features at 65.75 eV and 67.4 eV
(Fig. \ref{fig:short-time-dynamics}(c) (dots)) shows that the duration
of the growth of the TA features is significantly longer than the
optical pump pulse. By fitting the lineouts with a modified Gaussian
error function (Eq. (\ref{eq:error_function_fit})), TA signal rise
times $\tau_{rise}=15.1\pm0.4$ fs and $14\pm2$ fs for energies 65.75 eV
and 67.4 eV are obtained, respectively (Fig. \ref{fig:short-time-dynamics}(c)
(dashed lines)). The cross-correlation time between the optical and
XUV pulses, whose upper limit is set by the cross-correlation between
the pump pulse and the driving field for high-harmonic generation,
is less than $\sqrt{4.3^{2}+3.6^{2}}\approx5.6$ fs. The 15 fs rise
time of the features at 65.75 eV and 67.4 eV (Fig. \ref{fig:short-time-dynamics}(c)),
which is much longer than the <5.6 fs cross-correlation time between
the pump and the probe, suggests that the electron thermalization
time is longer than the pump pulse duration and the lengthened rise
time is connected to non-thermal electron relaxation. We present experimental
evidence to support this hypothesis, and a conjecture on the cause
of the absence of spectral signatures of non-equilibrium electron
distribution is discussed in Sec. \ref{subsec:Non-Equilibrium-Electron}.

To verify whether the electrons directly thermalize during photoexcitation,
the XUV TA spectra (Fig. \ref{fig:2exp_results}(c)) are fitted with
Eq. (\ref{eq:fitting_model}), and the quality of the fitting and
the relation between the fitted broadening $\sigma(t)$ and energy
shift $\omega_{s}(t)$ are examined. Figure \ref{fig:short-time-dynamics}(a)
shows the fitting results and the fitted broadening and energy shift
as a function of time delay are plotted in Fig. \ref{fig:short-time-dynamics}(b).
The good agreement between the fitting results (Fig. \ref{fig:short-time-dynamics}(a))
and experimental data (Fig. \ref{fig:2exp_results}(c)) initially
suggests that the electrons thermalize within the duration of the
optical pulse. However, if the electrons already thermalize within
the timescale of photoexcitation, the resulting spectral broadening
$\sigma$ and shift $\omega_{s}$ should follow the relation $\omega_{s}(t)\propto\sigma(t)^{1.5}$
as shown in Sec. \ref{sec:Electron-Cooling-Dynamics}. In Fig. \ref{fig:short-time-dynamics}(b),
the changes in the broadening and the shift as a function of time
delay are fitted to a modified Gaussian error function (Eq. (\ref{eq:error_function_fit}))
and the obtained error function rise times for the broadening and
the shift are $\tilde{\tau}_{\sigma}=11.0\pm0.4$ fs and $\tilde{\tau}_{s}=15.0\pm0.3$
fs, respectively. Given $\omega_{s}\propto\sigma^{1.5}$ for a thermalized
electron distribution, the increase of magnitude in the spectral shift
$\omega_{s}(t)$ should be steeper than the broadening $\sigma(t)$
($\tilde{\tau}_{\sigma}>\tilde{\tau}_{s}$), which is opposite to
the fitting results. This indicates that the electron thermalization
time in nickel is either comparable or longer than the optical pulse
(4.3 fs), and during the increase of magnitudes in the broadening
and spectral red shift, the broadening $\sigma$ cannot be directly
related to an electron temperature.

In addition to the inconsistent behavior of the spectral shift and
broadening compared to the results obtained at $\geq40$ fs time delay,
we consider the fluence dependence of the TA signal rise time (Fig.
\ref{fig:short-time-dynamics}(c)). Fermi liquid theory indicates
that the electron collision rate is proportional to the square of
electron temperature \cite{pinesTheoryQuantumLiquids1989}. Using
Boltzmann collision integrals, Mueller and Rethfeld predict that the
electron thermalization time in nickel decreases by an order of magnitude
as the ``final'' electron temperature, viz. the electron temperature
after thermalization, increases from 2000 K to 8000 K \cite{Mueller2013}.
The fluence dependence of electron thermalization has been experimentally
observed by Obergfell and Demsar in Cu by optical pump-probe measurements
\cite{obergfellTrackingTimeEvolution2020}. Here we examine the connection
between electron thermalization time and the rise time of the core-level
TA signal (Fig. \ref{fig:short-time-dynamics}(c)) by a set of power
dependence measurements at three different pump fluences. In Table
\ref{tab:thermalization-time-fluence}, we list the final electron
temperatures $T_{e}$ and the rise times $\tau_{rise}$ of TA signal
at 65.75 eV and a clear decrease in the rise time with respect to
increasing electron temperature is observed. As the final electron
temperature rises from approximately 2100 K to 3100 K, the measured
rise time decreases from 35 fs to 15 fs. The observed behavior of
the rise time as a function of final electron temperature is consistent
with theoretical predictions of the electron thermalization time and
we estimate this time $\tau_{th}$ by deconvolving the growth dynamics
with the pump-probe cross-correlation ($\tau_{th}\approx\sqrt{\tau_{rise}^{2}-5.6^{2}}$).
The electron thermalization times obtained by deconvolution are listed
in Table \ref{tab:thermalization-time-fluence}, showing that the
electron thermalization time decreases from 34 fs to 13 fs as the
electron temperature rises from 2100 K to 3100 K. The obtained thermalization
times are on the same order of the theoretically predicted thermalization
time in nickel \cite{Mueller2013}.

\begin{table}
\caption{Experimentally obtained TA signal rise time $\tau_{rise}$ and extracted
electron thermalization time $\tau_{th}$ versus electron temperature
$T_{e}$ from fluence dependence measurements. The asymmetry in the
uncertainty of $\tau_{rise}$ for entries with $T_{e}=$2129 K and
2552 K is due to the drift of time delay within the experiment which
cannot be compensated, leading to a stretch in the rise time. Further
explanation on the time delay drift is described in Appendix \ref{sec:Experimental-Apparatus}.
\label{tab:thermalization-time-fluence}}

\begin{tabular}{ccc}
\toprule 
$T_{e}$ {[}K{]} & $\tau_{rise}$ {[}fs{]} & $\tau_{th}$ {[}fs{]}\tabularnewline
\midrule
\midrule 
2129 & $35\substack{+10\\
-15
}
$ & 34\tabularnewline
\midrule 
2552 & $22\substack{+5\\
-10
}
$ & 21\tabularnewline
\midrule 
3060 & $15\pm3$ & 13\tabularnewline
\bottomrule
\end{tabular}
\end{table}

\section{Discussion \label{sec:Discussion}}

Despite the success of fitting the TA profile with spectral shift
and broadening (Eq. (\ref{eq:fitting_model})), the physical origin
of the spectral shift has yet to be clarified. In addition, while
the electron thermalization time is extracted from the rise of TA
signal (Fig. \ref{fig:short-time-dynamics}(c)), the absence of spectral
signatures of the non-equilibrium electron distribution during photoexcitation
has not been explained. Here we discuss the potential origin of the
spectral shift and the absence of spectral features of non-thermalized
electrons.

\subsection{Origin of Spectral Shift \label{subsec:spectral-shift}}

The spectral shift in the core-level absorption spectra can be interpreted
as an overall change in the energy of the core-excited state in laser-heated
nickel. This can be caused by electron-phonon interactions \cite{zurch2017direct},
and in particular, lattice displacement and heating due to optical
excitations \cite{attarSimultaneousObservationCarrierSpecific2020,rothenbachMicroscopicNonequilibriumEnergy2019}.
However, as the experimental results suggest that the direct influence
of phonon excitations on the TA profile is negligible, we restrict
the discussion within the electronic domain. Electronically, the spectral
red shift of core-level absorption after optical pump illumination
can originate from the lowering of the chemical potential in the CB
or stabilization of core-excited state due to many-body interactions
\cite{volkovAttosecondScreeningDynamics2019a}. The lowering of the
chemical potential as a possible cause can be eliminated as Lin et
al. showed that the chemical potential of the CB of nickel increases
rather than decreases with rising electron temperature \cite{linTemperatureDependencesElectronphonon2007,linElectronphononCouplingElectron2008,bevillonFreeelectronPropertiesMetals2014},
in contrast to the experimental observations here. Due to the complexity
in simulating the effect of many-body interactions in the core-excited
state of nickel with hot, thermalized CB electrons, here we provide
a subjective explanation for the nonlinear relationship between $\omega_{s}$
and $\Delta T$ based on related works by analogy and invite future
theoretical works to verify the validity of the conjecture. 

In attosecond TA studies of titanium, Volkov et al. showed that the
optical excitation of electrons increases the occupation of the localized
Ti $3d$ orbitals, which further causes a spectral blue shift due
to the increase of electronic repulsion in the core-excited state
\cite{volkovAttosecondScreeningDynamics2019a}. In nickel, the increase
of electron temperature causes the transfer of Ni $3d$ electrons
to the higher-lying $4s$ and $4p$ bands \cite{bevillonFreeelectronPropertiesMetals2014},
which would reduce the electronic repulsion in the localized $3d$
orbitals in contrast to the repulsion increase observed when heating
an early transition metal such as Ti. The cause of the opposing behavior
between Ni and Ti is theorized from the inversion in the orbital character
with respect to band energies. In titanium, the occupied $4s$ bands
are below the largely unoccupied $3d$ bands \cite{volkovAttosecondScreeningDynamics2019a},
whereas in nickel, the occupied bands are composed primarily of $3d$
orbitals and the unoccupied bands comprise an increased $4s$ and
$4p$ character \cite{kasapSpringerHandbookElectronic2017}. The
reduction in electronic repulsion stabilizing the core-excited state
of laser-heated nickel thus presents a plausible explanation to the
spectral red shift observed in the XUV TA spectra. As the electron
repulsion and reduction of population in the nickel $3d$ bands are
not linearly proportional to electron temperature, the nonlinearity
in the relation between $\omega_{s}$ and $\Delta T$ is also potentially
clarified.

\subsection{Non-Equilibrium Electron Relaxation \label{subsec:Non-Equilibrium-Electron}}

\begin{figure}
\includegraphics[width=0.45\textwidth]{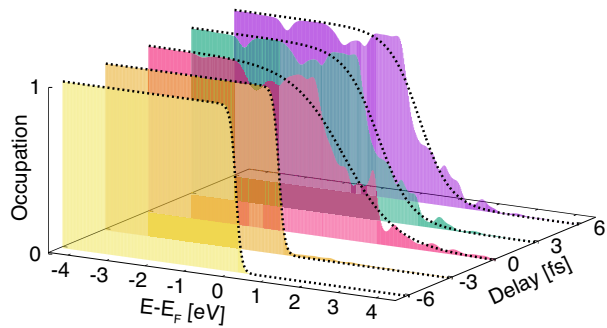}

\caption{Electronic occupation as a function of energy during photoexcitation.
The results of fitting each time slice to a Fermi-Dirac function are
shown in black dashed lines. \label{fig:Electronic-occupation}}

\end{figure}

While the fluence dependent rise time of the TA signal (Fig. \ref{fig:short-time-dynamics}(c)
and Table \ref{tab:thermalization-time-fluence}) indicate that the
electron thermalization time is longer than the pump pulse duration,
no energetically shifting spectral features are observed within the
electron thermalization timescale in the TA spectra (Fig. \ref{fig:2exp_results}(c))
to represent the thermalization of non-equilibrium carrier distribution.
To understand this phenomenon, we simulated the dynamics of photoexcitation
in nickel through a density matrix formalism based on the band structure
of nickel calculated by density functional theory (DFT) (Appendix
\ref{sec:Simulation-of-Photoexcitation}). Snapshots of electronic
occupation near the Fermi-level ($E_{F}$) at different time delays
with respect to the optical pulse are shown in Fig. \ref{fig:Electronic-occupation}.
In the simulated electron distributions, occupation around 1.5 eV
below the Fermi level decreases following photoexcitation. However,
the photoexcited electron distribution at and above the Fermi energy
still closely resembles a Fermi-Dirac function (Fig. \ref{fig:Electronic-occupation},
black dashed lines). This implies that because the initial photoexcited
electronic occupation does not significantly deviate from a hot thermalized
distribution, possible spectral signatures of a non-equilibrium CB
electron distribution will not be easily distinguished. Note that
in the density matrix formalism, the effects of scattering between
the photoexcited electrons are not included. Therefore, it is expected
that the true photoexcited carrier distribution will feature an even
smaller deviation from the thermalized electron distribution than
the simulated results (Fig. \ref{fig:Electronic-occupation}). This
provides a potential explanation to the absence of spectral features
of non-thermalized carriers.

\section{Conclusion}

In summary, it is observed that the transient absorption spectra of
optically excited nickel at the nickel M$_{2,3}$ edge can be simulated
with a spectral red shift and Gaussian broadening of the static spectrum.
For a hot thermalized electron distribution, the Gaussian broadening
is experimentally revealed and theoretically derived to be linearly
related to the change of electron temperature and can be used to track
the electron temperature. The increase of spectral red shift
with rising electron temperature can be plausibly explained by the
reduction of electron repulsion due to the repartitioning of localized
$3d$ electrons to $4s$ and $4p$ bands through laser heating. For
thermalized electrons, the red shift displays a power-law relationship
with the electron temperature change by a power $\alpha\approx1.5$.
While probing the sub-40 fs dynamics of optically excited nickel,
the relation between the spectral shift and electron temperature,
and thus the broadening for thermalized carriers, is utilized to determine
that the electrons do not thermalize instantaneously during the optical
excitation, even though the core-level absorption lineshape at sub-15
fs timescale closely resembles the spectra with thermalized electrons.
In the core-level transient absorption spectroscopy of nickel at Ni
M$_{2,3}$ edge, the electron thermalization process is represented
by a lengthened growth of spectral features for a thermalized electron
distribution, indicating that electron scattering and thermalization
accompany and follow the photoexcitation and finally create a hot
thermalized carrier distribution. A fluence-dependent electron thermalization
timescale ranging between 34 fs and 13 fs is extracted by deconvolving
the rise of the transient absorption signal from the instrument response.

The results in this work indicate that core-level absorption spectroscopy
can be utilized to extract the electron temperature of metallic samples
and to assess both the timescale of electron thermalization and the
validity of using a multi-temperature model. As such, this work brings
a unification of the observation of >30 fs dynamics of thermalized
electrons, and the few-femtosecond dynamics of non-equilibrium electron
relaxation. The former can be readily probed with time-resolved photoemission
methods but has been difficult to extract and interpret from core-level
absorption spectroscopy, while the latter can be interrogated with
few-femtosecond core-level spectroscopy but is inaccessible through
photoemission techniques. The methodology developed within this study
facilitates the understanding of core-level absorption spectra of
laser-heated metals with a simple and intuitive picture, and the approach
can be readily extended to treat other metallic systems or to investigate
photoinduced phase transitions in metallic films and multilayers.
\begin{acknowledgments}
The authors would like to thank Dr. Xun Shi, Phoebe Tengdin, Wenjing
You, and Dr. David Prendergast for fruitful discussions. Investigations
were supported by the Defense Advanced Research Projects Agency PULSE
Program Grant W31P4Q-13-1-0017 (concluded), the U.S. Air Force Office
of Scientific Research Nos. FA9550-19-1-0314, FA9550-20-1-0334, FA9550-15-0037
(concluded), and FA9550-14-1-0154 (concluded), the Army Research Office
No. WN911NF-14-1-0383, and the W.M. Keck Foundation award No. 046300-002.
H.-T. C. acknowledges support from Air Force Office of Scientific
Research (AFOSR) (FA9550-15-1-0037 and FA9550-19-1-0314) and W. M.
Keck Foundation (No. 046300); A. G. acknowledges support from German
Research Foundation (GU 1642/1-1); S. K. C. acknowledges support by
the Department of Energy, Office of Energy Efficiency and Renewable
Energy (EERE) Postdoctoral Research Award under the EERE Solar Energy
Technologies Office; I. J. P. is supported by U.S. Department of Energy,
Office of Science, Office of Basic Energy Sciences, Materials Sciences
and Engineering Division, under Contract No. DEAC02-05-CH11231, within
the Physical Chemistry of Inorganic Nanostructures Program (KC3103).
N. U. D., S. R. A., V. T. and T. S. R. acknowledge support from U.S.
Department of Energy (Grant No. DE-FG02-07ER46354). 
\end{acknowledgments}

\appendix

\section{Experimental Apparatus\label{sec:Experimental-Apparatus}}

\begin{figure}
\includegraphics[width=0.98\textwidth]{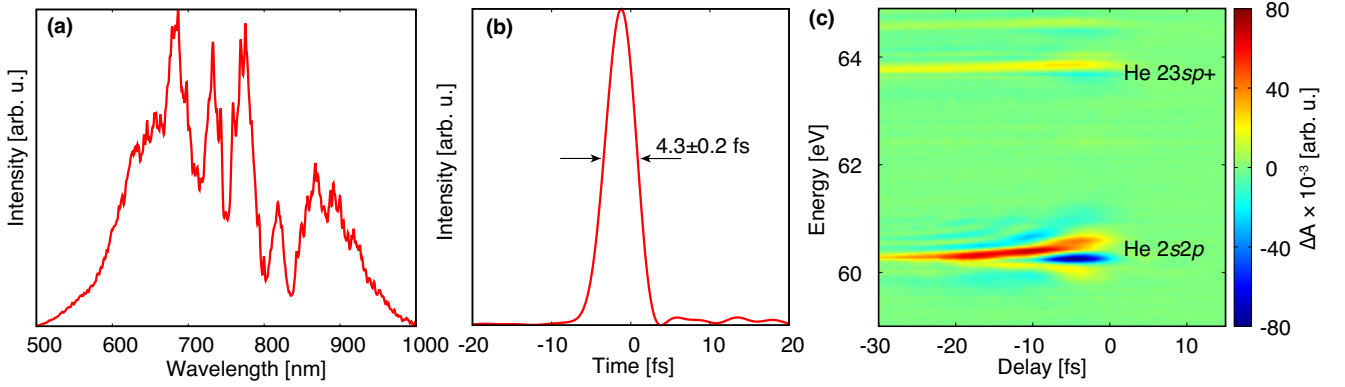}

\caption{(a) The typical spectrum and (b) temporal profile of the pump pulses
measured by dispersion scan \cite{Silva2014}. (c) displays the typical
transient absorption spectra of helium \textit{2snp} autoionization
states taken subsequently after each scan through all time delay points
on the nickel sample for time delay calibration.\label{fig:A1-exp-appratus}}
\end{figure}

The table-top XUV TA setup consists of a Ti:sapphire laser with 1.8
mJ output pulse energy, 30 fs pulse duration operating at 1 kHz repetition
rate. The laser pulses centered at 790 nm wavelength are then focused
into a 1 m long, Ne filled hollow-core fiber for supercontinuum generation,
resulting in pulses with a spectrum spanning between 500 and 1000
nm. A mechanical chopper is installed at the exit of the hollow-core
fiber to chop the beam repetition rate down to 100 Hz so as to prevent
sample damage from optical heating due to the poor heat conductivity
in nanometer thick thin films. In addition, the sample is raster scanned
during the measurement to prevent laser damage due to long time exposure
and systematic error due to sample inhomogeneity. Static absorption
measurements are taken before and after each transient absorption
experiment to assess whether sample damage occurs during the experiment.
The spectrally broadened pulses are subsequently compressed by a set
of broadband double-angle chriped mirrors and split by a 1:9 broadband
beam splitter into the pump and probe arm, respectively. The third-order
dispersion of the laser pulses is compensated by transmitting the
beam through a 2 mm thick ammonium dihydrogen phosphate crystal \cite{Timmers2017}.
The fine tuning of dispersion in the pump and probe arm is achieved
by a pair of glass wedges installed in each arm, and the pulse duration
of the pump and probe pulses measured by dispersion scan \cite{Silva2014}
are 4.3$\pm$0.2 fs and 3.6$\pm$0.1 fs long, respectively. The duration
of the pump pulse is slightly longer than the probe because of the
limited bandwidth of the broadband beamsplitter. The typical spectrum
and temporal profile of the pump pulse are shown in Fig. \ref{fig:A1-exp-appratus}(a)
and \ref{fig:A1-exp-appratus}(b), respectively. The intensity of
the pump beam is controlled by an iris and the beam is time-delayed,
focused, and recombined collinearly into the probe arm with an annular
mirror (Fig. \ref{fig:experimental_setup}(a)). For each individual
experiment, the beam profile of the pump is measured by a CMOS camera
directly at the sample position for the determination of the intensity
and fluence of the pump pulse. The beam in the probe arm is focused
into an Ar gas jet to generate broadband XUV pulses spectrally spanning
40 -- 73 eV (Fig. \ref{fig:experimental_setup}(b)). After filtering
the driving near-IR field with a 100 nm thick Al filter, the XUV beam
is refocused by a gold coated toroidal mirror onto the measurement
target. After passing through the measurement target, the XUV beam
transmits through another 100 nm thick Al filter that blocks the pump
light and the transmitted XUV is then dispersed by a flat-field grating
onto an XUV CCD camera with 16-bit bit depth that provides a dynamic
range of approximately 5 orders of magnitude. With the approximately
1 OD (optical density) edge jump at the nickel edge, the noise floor
of the transient measurement is approximately 5 mOD, which is mainly 
contributed by the fluctuations of the XUV light source. The spectral
resolution of the apparatus at the photon energies of the experiment
is approximately 60 meV. The measurement delay step sizes used in
the results shown in Fig. \ref{fig:experimental_setup}(c) are 1.3
fs (0.2 $\mu$m) between -66.7 fs (-10 $\mu$m) to +133.3 fs (+20
$\mu$m) delay, 3.3 fs (0.5 $\mu$m) between +133.3 fs (+20 $\mu$m)
and +560 fs (+84 $\mu$m) delay, and 20 fs (3 $\mu$m) between +560
fs (+84 $\mu$m) and +1.98 ps (+297 $\mu$m) delay. The measurement
delay step size used in the results shown in Fig. \ref{fig:experimental_setup}(d)
is 0.33 fs (0.05 $\mu$m) between -50 fs (-7.5 $\mu$m) and +50 fs
(+7.5 $\mu$m) delay. Here the positions of the optical delay stage
in micron are listed in parentheses. As the retroreflector folds the
optical beam once (Fig. \ref{fig:experimental_setup}(a)), 1 $\mu$m
change of the optical delay stage translates to approximately 6.6
fs delay.

\begin{figure}
\includegraphics[width=0.49\textwidth]{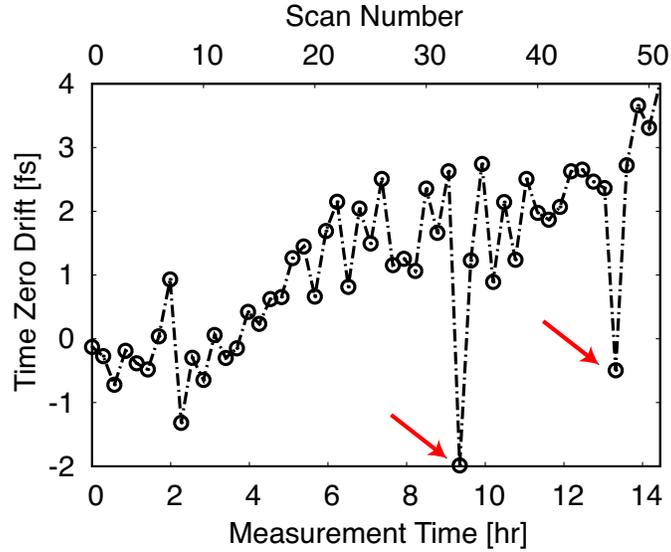}

\caption{A sample time zero drift trace taken in an XUV TA experiment on nickel.
The time zero is extracted from XUV TA measurements on He autoionization
lines. The red arrows indicate the scans to be discarded from data
analysis due to >3 fs drifts over a single scan. \label{fig:A-timezero}}

\end{figure}

To avoid the slow drift of time delays during the experiment, we ran
one transient absorption measurement on the $2snp$ autoionizing states
of helium after each scan through all time delay points on the nickel
sample \cite{Kaldun2016}. A typical transient absorption trace on
the He $2snp$ autoionizing states is shown in Fig. \ref{fig:A1-exp-appratus}(c).
The transient absorption signal of the He $2s2p$ state near time
zero was then fitted to an error function to determine the exact zero
time overlap between the XUV and optical pulses. With the calibrated
time zero of each scan, the changes of absorbance for each scan were
then interpolated onto a gridline and averaged together \cite{zurch2017direct}.
A sample time zero drift trace over the course of an XUV TA measurement
on nickel is shown in Fig. \ref{fig:A-timezero}. Note that as the
fluence dependence on the TA signal for the He autoionization lines
is highly nonlinear \cite{Kaldun2016}, reliable TA measurements
for time zero correction at fluences lower than 30 mJ/cm$^{2}$ have
not been obtained. This leads to smearing and stretching of dynamics
with respect to time delay for low fluence measurements. To estimate
the amount of time-axis smearing in the observed dynamics (e.g. the
rise of TA signal shown in Sec. \ref{sec:Electron-Thermalization})
by the time zero drift, XUV TA measurements of He autoionization lines
were conducted over the same amount of time as the acquisition time
for the XUV TA experiments on nickel and we focus on the overall time
delay drift over the entire measurement time. Specifically for the
measurements displayed in Table \ref{tab:thermalization-time-fluence},
an overall time zero drift of 5 fs is obtained. This leads to an increase
of negative uncertainty in the first two entries of $\tau_{rise}$
in Table \ref{tab:thermalization-time-fluence}.

\section{Sample Preparation\label{sec:Sample-Preparation}}

The nickel thin films used in this experiment were prepared by dual
ion-beam deposition of 50 nm thick nickel onto 30 nm thick silicon
nitride membranes with a free-standing window size of 3 mm $\times$
3 mm \cite{guggenmosAperiodicCrScMultilayer2013}, utilizing (neutralized)
600 eV krypton ions at a background pressure of $10^{-7}$ Pa. The
layer thickness was controlled via the deposition time where typical
sputter rates are below 0.1 nm per second and were calibrated using
surface profilometry as well as \textit{in situ} spectral ellipsometry.
The sputter time calculations are based on a numerical model \cite{guggenmosAperiodicCrScMultilayer2013},
to compensate both inter-diffusion losses and systematic deposition
variations due to, for example, shutter response times. The substrate
holder spun during deposition with a spinning frequency of 40 rpm
and an $R/\theta$ shaper was used for shaping the particle flux laterally
for a high lateral homogeneity film thickness growth. A film of 4
nm thick boron carbide was then deposited above the nickel thin film
to prevent oxidation through exposure to ambient air.

\section{Estimating Electron Temperature Rise Due to Optical Absorption \label{sec:Estimating-Electron-Temperature}}

The electron temperature rise after optical excitation of nickel at
40 fs time delay is calculated from the spectrum of the optical pulse
(Fig. \ref{fig:experimental_setup}(d)), the wavelength-dependent
thin film absorbance, and the electronic heat capacity of nickel.
As the electron-phonon scattering time in nickel is a few hundred
femtoseconds \cite{Kampen2005}, the electronic cooling due to phonons
is neglected in the following calculation. The energy absorbed by
the nickel film at a specific wavelength $\lambda$ can be described
as $\mathcal{{E}}_{abs}(\lambda)=\frac{{hc}}{\lambda}N_{ph}(\lambda)f_{abs}(\lambda)$.
Here $N_{ph}(\lambda)$ is the number of incident photons as a function
of wavelength, which can be directly derived from the pump pulse spectrum.
$f_{abs}(\lambda)$ is the fraction of light absorbed in the nickel
film and is calculated using the transfer matrix method \cite{burkhardAccountingInterferenceScattering2010}.
The value of wavelength-dependent complex refractive indices of nickel
and silicon nitride are taken from Refs. \cite{johnsonOpticalConstantsTransition1974}
and \cite{vogtDevelopmentPhysicalModels2015}, respectively. Denoting
the electron heat capacity by $C_{e}(T)$, the estimated maximum electron
temperature $T_{est}$ is related to the total energy absorbed by
the nickel film $\mathcal{E}$ as
\[
\mathcal{{E}}=\int_{T_{0}=300K}^{T_{est}}C_{e}(T^{\prime})dT^{\prime},
\]
where the temperature-dependent electron heat capacity of nickel is
taken from Ref. \cite{linElectronphononCouplingElectron2008}. The
electron temperature rise $\Delta T_{est}$ is $T_{est}-T_{0}$.

\section{Expansion of $I^{\prime}(t)$\label{app:cumulant_expansion}}

The expression of the core-level absorption lineshape $I^{\prime}(\omega)$
in Eq. (\ref{eq:O-T-model}) is the Fourier transform of its free-induction
decay
\begin{equation}
I^{\prime}(t)=\left(\frac{{\pi T}}{i\sinh(\pi Tt)}\right)^{1-\zeta}=(-i\pi S(t,T))^{1-\zeta}.\label{eq:FID-OT}
\end{equation}
To find $e^{g(t,\Delta T)}$ (cf. Eq. (\ref{eq:sinh_expand})), we
first expand $S(t,T)$ with respect to temperature:
\begin{align*}
S(t,T+\Delta T) & =S(t,T)+\partial_{T}S(t,T)\cdot\Delta T+\frac{1}{2}\partial_{T}^{2}S(t,T)\cdot(\Delta T)^{2}+\ldots.
\end{align*}
Letting $z=\pi t$, the derivatives of $S(t,T)$ are expressed 
\begin{align*}
\partial_{T}S(t,T) & =\frac{1}{\sinh(zT)}\left(1-zT\coth(zT)\right)\\
\partial_{T}^{2}S(t,T) & =\frac{T}{\sinh(zT)}\left(2\left(z^{2}\coth^{2}(zT)-\frac{z\coth(zT)}{T}\right)-z^{2}\right).
\end{align*}
Expressing the $m^{th}$ Taylor expansion term of $S(t,T+\Delta T)$
as $\partial_{T}^{m}S(t,T)(\Delta T)^{m}/m!=s_{m}S(t,T)$, we can
rewrite $S(t,T+\Delta T)$ as 
\[
S(t,T+\Delta T)=S(t,T)\exp\left(\ln\left(1+\sum_{m=1}^{\infty}s_{m}\right)\right),
\]
and by comparing the terms between the Taylor expansion of $S(t,T+\Delta T)$
and $\{s_{m}\}$, the second order expansion of $S(t,T+\Delta T)$
can be expressed as \cite{laxFranckCondonPrinciple1952,kuboGeneralizedCumulantExpansion1962}
\begin{align}
S(t,T+\Delta T) & \approx S(t,T)\exp\left(s_{1}+s_{2}-\frac{s_{1}^{2}}{2}\right).\label{eq:cumulant-expansion}
\end{align}
Here we concentrate on the behavior of $S(t,T)$ when $t$ is near
zero because the denominator $\sinh(zt)$ in the integrand $I^{\prime}(t)$
grows exponentially with $t$. This ``short time'' approximation
can be further justified by comparing the temperature induced broadening
and the natural linewidth of the core hole. While the temperature
induced broadening is typically on the scale of tens to hundreds of
meV (cf. Fig. \ref{fig:NiStatic_power_dependence}(c)), the core hole
lifetime broadening is typically larger by one order of magnitude
\cite{Dietz1980,grootCoreLevelSpectroscopy2008}. This indicates
that the ``true'' free induction decay of the core-level transitions
is far faster than the decay of $I^{\prime}(t)$.  Under such condition,
we can approximate the term $\coth(zT)$ using the asymptotic relation
$\coth x\approx1/x+x/3-x^{3}/45+\ldots$ \cite{abramowitzHandbookMathematicalFunctions1970}.
To first order with $\coth(zT)\approx1/(zT),$ we obtain 
\begin{align*}
s_{1} & =\left(\frac{1}{T}-z\coth(zT)\right)\Delta T\approx0\\
s_{2} & =\left(z^{2}\coth^{2}(zT)-\frac{z\coth(zT)}{T}-\frac{z^{2}}{2}\right)(\Delta T)^{2}\\
 & \approx-\frac{z^{2}}{2}(\Delta T)^2.
\end{align*}
Inserting the expansion terms back into Eq. (\ref{eq:cumulant-expansion}),
a preliminary expression for $S(t,T+\Delta T)$ is derived:
\begin{align*}
S(t,T+\Delta T) & \approx S(t,T)\cdot\exp\left(-\frac{(\pi\Delta Tt)^{2}}{2}\right),
\end{align*}
and therefore
\[
e^{g(t,\Delta T)}=\exp\left(-\frac{(\pi\Delta Tt)^{2}}{2}\right).
\]

To estimate the error of the approximated expression, we include the
second and the third term of the asymptotic expansion ($\coth(zT)\approx1/(zT)+zT/3-(zT)^3/45$)
and obtain
\begin{align*}
s_{1} & =\left(\frac{1}{T}-z\coth(zT)\right)\Delta T\approx\left(-\frac{z^{2}T}{3} +\frac{z^4 T^3}{45}\right) \Delta T \\
s_{2} & =\left(z^{2}\coth^{2}(zT)-\frac{z\coth(zT)}{T}-\frac{z^{2}}{2}\right)(\Delta T)^{2}\\
 & \approx\left(-\frac{z^{2}}{2}+\frac{z^{2}}{3}+\frac{4z^{4}T^{2}}{45}\right)(\Delta T)^{2}.
\end{align*}
The expression $S(t,T+\Delta T)$ becomes 
\begin{align*}
S(t,T+\Delta T) & \approx S(t,T)\cdot\exp\left(-z^2\left( \frac{\Delta T^2+2T\Delta T}{6}\right)\right)\times\\
 & \exp\left(z^4\left( \frac{T^3\Delta T}{45}+\frac{T^2(\Delta T)^2}{30} \right)\right).
\end{align*}
Note that in the equations above, we only include the terms up to $z^4$ because in the region where $z=\pi t$ is near zero, the higher order terms can be neglected. The same reasoning also applies
to leaving out the higher order terms in the asymptotic expansion
of $\coth(zT)$ as higher order terms will lead to expressions of
$z^{6}$ and above in $S(t,T+\Delta T)$. To explore the adequacy of the ``short time'' approximation, we compare the magnitude of the $z^2$ term to the $z^4$ term. Given the typical core hole lifetime broadening of $\sim$1 eV \cite{grootCoreLevelSpectroscopy2008}, room temperature $T\approx 0.03$ eV, and an overestimated $\Delta T=10000$ K $\approx 0.9$ eV, the magnitude of the second order term is
$$\left( \frac{\pi\hbar}{1\: \text{eV}}\right)^2 \frac{0.9^2+2\cdot 0.9\cdot 0.03}{6\hbar^2}(\text{eV})^2\approx 1.42,$$
and the magnitude of the $z^4$ term is
$$\left( \frac{\pi\hbar}{1\: \text{eV}}\right)^4 \left(\frac{0.9\cdot 0.03^3}{45}+\frac{0.03^2\cdot0.9^2}{30}\right ) (\frac{\text{eV}}{\hbar})^4\approx 0.002.$$
The magnitude of the fourth order term is three orders of magnitude smaller than the second order term, indicating that the $z^4$ term can also be neglected, yielding
\begin{align} \label{eq:err_2ndorder-expansion}
S(t,T+\Delta T) & \approx S(t,T)\cdot\exp\left(-\frac{(\Delta T^{2}+2T\Delta T)(\pi t)^{2}}{6}\right).
\end{align}
Here we observe that the expression remains a Gaussian function with
respect to $t$ and there is no first order term with respect to $t$
in the exponent. The Fourier transformed broadening factor in the
spectral domain is proportional to $\sqrt{\Delta T^{2}+2T\Delta T}$.
However, as the electron temperature change $\Delta T$ is at least
6 times larger than the temperature for the reference spectrum $T=T_{0}=300$
K in the measurements (Fig. \ref{fig:NiStatic_power_dependence}(c)),
the term $2T\Delta T$ is small compared to $\Delta T^{2}$. This
explains the linear relationship between the obtained spectral broadening
and electron temperature. As a corollary on the comparison of terms with different orders of $z$, note that the terms in the expansion of $\{s_m\}$ always have the form $z^nT^k(\Delta T)^l$, with $n,k,l\geq 0$ and $k+l=n$. This indicates that the ``short time'' approximation, or the truncation at the second order holds as long as the electron temperature change ($k_B\Delta T$) and electron temperature ($k_B T$) are smaller than the core hole broadening. 

Finally, we connect Eq.~(\ref{eq:err_2ndorder-expansion}) with Eq.~(\ref{eq:sinh_expand}) and observe
\begin{align*}
	\frac{\pi (T_0 + \Delta T)}{\sinh(\pi (T_0+\Delta T)t)}\approx S(t,T_0)\times\exp\left( -\frac{\Delta T^{2}}{6}(\pi t)^{2}\right)\approx \frac{\pi T_0}{\sinh(\pi T_0 t)}\times e^{g(t,\Delta T)},
\end{align*}
indicating $g(t,\Delta T)\approx \frac{(\pi \Delta T t)^2}{6}$. In the equation above, the term $2T\Delta T$ from Eq.~(\ref{eq:err_2ndorder-expansion}) is ignored as $(\Delta T)^2\gg 2T\Delta T$. The broadening function $f(\omega,\Delta T)=\frac{1}{2\pi}\int_{-\infty}^{\infty}e^{i\omega t}e^{(1-\zeta)g(t,\Delta T)}dt$ is thus
$$f(\omega,\Delta T)\approx \frac{1}{\sigma \sqrt{2\pi}}e^{-\frac{\omega^2}{2\sigma^2}},$$
with
$$\sigma=\pi\Delta T\sqrt{\frac{1-\zeta}{3}}.$$
Rewriting the equation in SI units, the broadening function reads
$$f(\hbar\omega,\Delta T)\approx \frac{1}{\sigma \sqrt{2\pi}}e^{-\frac{(\hbar\omega)^2}{2\sigma^2}},$$
and
$$\sigma=\pi k_B \Delta T\sqrt{\frac{1-\zeta}{3}}=a\Delta T.$$
Here $a=\pi k_B\sqrt{(1-\zeta)/3}$ and $\zeta<1$ is a constant phase factor related to the electron scattering from the core hole potential (Eq.~(\ref{eq:O-T-model})).

\section{Fitting with Modified Gaussian Error Function \label{sec:Fitting-with-Modified}}

To fit the sub-20 femtosecond dynamics of $\sigma(t)$, $\omega_{s}(t)$,
and the TA lineouts at 65.75 and 67.4 eV, a modified Gaussian error
function 
\begin{align}
\mathcal{{F}}(t,\tau) & =\frac{c_{1}}{2}\left(1+\frac{2}{\sqrt{\pi}}\int_{t_{0}}^{t}e^{-\left(\frac{2\sqrt{\ln2}(t^{\prime}-t_{0})}{\tilde{\tau}}\right)^{2}}dt^{\prime}\right)+c_{0}\label{eq:error_function_fit}
\end{align}
is utilized, where $t_{0}$ marks the timing of the dynamics relative
to zero time delay and $\tilde{\tau}$ the duration of the growth;
$c_{0}$ and $c_{1}$ are fitting coefficients for offset and amplitude
of the TA signal. The coefficient $2\sqrt{\ln2}$ enables direct comparison
between $\tilde{\tau}$ and the cross-correlation time between the
pump and probe pulses. If the electronic response is instantaneous
with respect to the excitation pulse, $\tilde{\tau}$ will be equal
to the cross-correlation time.

\section{Simulation of Photoexcitation Dynamics \label{sec:Simulation-of-Photoexcitation}}

The dynamics of photoexcitation are simulated by the density matrix
formalism based on the nickel band structure calculated by DFT, where
the density matrix $\rho_{\mathbf{{x}\mathbf{x^{\prime}}}}(t)=\langle c_{\mathbf{x}}^{\dagger}(t)c_{\mathbf{x^{\prime}}}(t)\rangle$.
$c$ is the annihilation operator and $\mathbf{x}=\{{\mathbf{k},m,s}$\}
denotes the combination of momentum $\mathbf{k}$, band index $m$,
and spin index $s$. The Hamiltonian for the Liouville equation $\dot{\rho}=-i[H,\rho]/\hbar$
is
\[
H=\sum_{\mathbf{x}}\mathcal{E}_{\mathbf{x}}c_{\mathbf{x}}^{\dagger}c_{\mathbf{x}}+\sum_{\mathbf{x},\mathbf{x^{\prime}}}V_{\mathbf{x},\mathbf{x^{\prime}}}(t)c_{\mathbf{x}}^{\dagger}c_{\mathbf{x}^{\prime}},
\]
where $\mathcal{E}$ denotes the band energy and 
\[
V_{\mathbf{x},\mathbf{x^{\prime}}}(t)=-\mathbf{d_{x,x^{\prime}}}\cdot\mathbf{E}(t).
\]
In the equation above, $\mathbf{d}$ is the dipole operator and the
magnitude of the electric field $\mathbf{E}(t)$ is $E_{0}\exp\left(-(2\sqrt{\ln2}t/\tau_{pulse})^{2}\right)$.
$E_{0}=2.5$ V/nm is derived from the peak intensity of the pulse
and $\tau_{pulse}=4.3$ fs. To obtain band energies $\mathcal{E}_{\mathbf{x}}$
and dipole operator $\mathbf{d_{x,x^{\prime}}}$, DFT calculations
were performed using the Quantum ESPRESSO package with Perdew-Burke-Ernzerhof
(PBE) exchange correlation functional and ultrasoft, scalar relativistic
pseudopotential \cite{perdewGeneralizedGradientApproximation1996,giannozziQUANTUMESPRESSOModular2009,giannozziAdvancedCapabilitiesMaterials2017}.
The simulation was conducted on a $15\times15\times15$ $k$-point
meshgrid using the Monkhorst-Pack scheme \cite{monkhorstSpecialPointsBrillouinzone1976},
and converged with cutoff energy at 60 Ryd. The occupation number
as a function of energy and time delay $O(\mathcal{E},t)$ is calculated
by summing the mapping of the diagonal terms of the density matrix
onto an energy grid and subsequently dividing by the density of states:
\[
O(\mathcal{E},t)=\sum_{\mathbf{x}}w_{\mathbf{x}}\rho_{\mathbf{xx}}(t)M(\mathcal{E},\mathcal{E}_{\mathbf{x}})/\sum_{\mathbf{x}}w_{\mathbf{x}}M(\mathcal{E},\mathcal{E}_{\mathbf{x}}).
\]
We use a Gaussian mapping function $M(\mathcal{E},\mathcal{E}_{\mathbf{x}})=\exp(-((\mathcal{E}-\mathcal{E}_{\mathbf{x}})/\delta\mathcal{E})^{2})$
with width $\delta\mathcal{E}=0.1$ eV. $w_{\mathbf{x}}$ is the weighting
coefficient within the Monkhorst-Pack scheme at point $\mathbf{k}$. 

\bibliographystyle{apsrev4-2}

\begin{thebibliography}{107}%
	\makeatletter
	\providecommand \@ifxundefined [1]{%
		\@ifx{#1\undefined}
	}%
	\providecommand \@ifnum [1]{%
		\ifnum #1\expandafter \@firstoftwo
		\else \expandafter \@secondoftwo
		\fi
	}%
	\providecommand \@ifx [1]{%
		\ifx #1\expandafter \@firstoftwo
		\else \expandafter \@secondoftwo
		\fi
	}%
	\providecommand \natexlab [1]{#1}%
	\providecommand \enquote  [1]{``#1''}%
	\providecommand \bibnamefont  [1]{#1}%
	\providecommand \bibfnamefont [1]{#1}%
	\providecommand \citenamefont [1]{#1}%
	\providecommand \href@noop [0]{\@secondoftwo}%
	\providecommand \href [0]{\begingroup \@sanitize@url \@href}%
	\providecommand \@href[1]{\@@startlink{#1}\@@href}%
	\providecommand \@@href[1]{\endgroup#1\@@endlink}%
	\providecommand \@sanitize@url [0]{\catcode `\\12\catcode `\$12\catcode
		`\&12\catcode `\#12\catcode `\^12\catcode `\_12\catcode `\%12\relax}%
	\providecommand \@@startlink[1]{}%
	\providecommand \@@endlink[0]{}%
	\providecommand \url  [0]{\begingroup\@sanitize@url \@url }%
	\providecommand \@url [1]{\endgroup\@href {#1}{\urlprefix }}%
	\providecommand \urlprefix  [0]{URL }%
	\providecommand \Eprint [0]{\href }%
	\providecommand \doibase [0]{https://doi.org/}%
	\providecommand \selectlanguage [0]{\@gobble}%
	\providecommand \bibinfo  [0]{\@secondoftwo}%
	\providecommand \bibfield  [0]{\@secondoftwo}%
	\providecommand \translation [1]{[#1]}%
	\providecommand \BibitemOpen [0]{}%
	\providecommand \bibitemStop [0]{}%
	\providecommand \bibitemNoStop [0]{.\EOS\space}%
	\providecommand \EOS [0]{\spacefactor3000\relax}%
	\providecommand \BibitemShut  [1]{\csname bibitem#1\endcsname}%
	\let\auto@bib@innerbib\@empty
	\bibitem [{\citenamefont {Ross}\ and\ \citenamefont
		{Nozik}(1982)}]{rossEfficiencyHotcarrierSolar1982}%
	\BibitemOpen
	\bibfield  {author} {\bibinfo {author} {\bibfnamefont {R.~T.}\ \bibnamefont
			{Ross}}\ and\ \bibinfo {author} {\bibfnamefont {A.~J.}\ \bibnamefont
			{Nozik}},\ }\href {https://doi.org/10.1063/1.331124} {\bibfield  {journal}
		{\bibinfo  {journal} {J. Appl. Phys.}\ }\textbf {\bibinfo {volume} {53}},\
		\bibinfo {pages} {3813} (\bibinfo {year} {1982})}\BibitemShut {NoStop}%
	\bibitem [{\citenamefont {W{\"u}rfel}(1997)}]{wurfelSolarEnergyConversion1997}%
	\BibitemOpen
	\bibfield  {author} {\bibinfo {author} {\bibfnamefont {P.}~\bibnamefont
			{W{\"u}rfel}},\ }\href {https://doi.org/10.1016/S0927-0248(96)00092-X}
	{\bibfield  {journal} {\bibinfo  {journal} {Solar Energy Materials and Solar
				Cells}\ }\textbf {\bibinfo {volume} {46}},\ \bibinfo {pages} {43} (\bibinfo
		{year} {1997})}\BibitemShut {NoStop}%
	\bibitem [{\citenamefont {Luque}\ and\ \citenamefont
		{Mart{\'i}}(2010)}]{luqueElectronPhononEnergy2010a}%
	\BibitemOpen
	\bibfield  {author} {\bibinfo {author} {\bibfnamefont {A.}~\bibnamefont
			{Luque}}\ and\ \bibinfo {author} {\bibfnamefont {A.}~\bibnamefont
			{Mart{\'i}}},\ }\href {https://doi.org/10.1016/j.solmat.2009.10.001}
	{\bibfield  {journal} {\bibinfo  {journal} {Solar Energy Materials and Solar
				Cells}\ }\textbf {\bibinfo {volume} {94}},\ \bibinfo {pages} {287} (\bibinfo
		{year} {2010})}\BibitemShut {NoStop}%
	\bibitem [{\citenamefont {Jailaubekov}\ \emph {et~al.}(2013)\citenamefont
		{Jailaubekov}, \citenamefont {Willard}, \citenamefont {Tritsch},
		\citenamefont {Chan}, \citenamefont {Sai}, \citenamefont {Gearba},
		\citenamefont {Kaake}, \citenamefont {Williams}, \citenamefont {Leung},
		\citenamefont {Rossky},\ and\ \citenamefont
		{Zhu}}]{jailaubekovHotChargetransferExcitons2013}%
	\BibitemOpen
	\bibfield  {author} {\bibinfo {author} {\bibfnamefont {A.~E.}\ \bibnamefont
			{Jailaubekov}}, \bibinfo {author} {\bibfnamefont {A.~P.}\ \bibnamefont
			{Willard}}, \bibinfo {author} {\bibfnamefont {J.~R.}\ \bibnamefont
			{Tritsch}}, \bibinfo {author} {\bibfnamefont {W.-L.}\ \bibnamefont {Chan}},
		\bibinfo {author} {\bibfnamefont {N.}~\bibnamefont {Sai}}, \bibinfo {author}
		{\bibfnamefont {R.}~\bibnamefont {Gearba}}, \bibinfo {author} {\bibfnamefont
			{L.~G.}\ \bibnamefont {Kaake}}, \bibinfo {author} {\bibfnamefont {K.~J.}\
			\bibnamefont {Williams}}, \bibinfo {author} {\bibfnamefont {K.}~\bibnamefont
			{Leung}}, \bibinfo {author} {\bibfnamefont {P.~J.}\ \bibnamefont {Rossky}},\
		and\ \bibinfo {author} {\bibfnamefont {X.-Y.}\ \bibnamefont {Zhu}},\ }\href
	{https://doi.org/10.1038/nmat3500} {\bibfield  {journal} {\bibinfo  {journal}
			{Nat. Mater.}\ }\textbf {\bibinfo {volume} {12}},\ \bibinfo {pages} {66}
		(\bibinfo {year} {2013})}\BibitemShut {NoStop}%
	\bibitem [{\citenamefont {Kamide}\ \emph {et~al.}(2018)\citenamefont {Kamide},
		\citenamefont {Mochizuki}, \citenamefont {Akiyama},\ and\ \citenamefont
		{Takato}}]{kamideNonequilibriumTheoryConversion2018}%
	\BibitemOpen
	\bibfield  {author} {\bibinfo {author} {\bibfnamefont {K.}~\bibnamefont
			{Kamide}}, \bibinfo {author} {\bibfnamefont {T.}~\bibnamefont {Mochizuki}},
		\bibinfo {author} {\bibfnamefont {H.}~\bibnamefont {Akiyama}},\ and\ \bibinfo
		{author} {\bibfnamefont {H.}~\bibnamefont {Takato}},\ }\href
	{https://doi.org/10.1103/PhysRevApplied.10.044069} {\bibfield  {journal}
		{\bibinfo  {journal} {Phys. Rev. Applied}\ }\textbf {\bibinfo {volume}
			{10}},\ \bibinfo {pages} {044069} (\bibinfo {year} {2018})}\BibitemShut
	{NoStop}%
	\bibitem [{\citenamefont {Zhang}\ \emph {et~al.}(2018)\citenamefont {Zhang},
		\citenamefont {He}, \citenamefont {Guo}, \citenamefont {Hu}, \citenamefont
		{Huang}, \citenamefont {Mulcahy},\ and\ \citenamefont
		{Wei}}]{zhangSurfacePlasmonDrivenHotElectron2018}%
	\BibitemOpen
	\bibfield  {author} {\bibinfo {author} {\bibfnamefont {Y.}~\bibnamefont
			{Zhang}}, \bibinfo {author} {\bibfnamefont {S.}~\bibnamefont {He}}, \bibinfo
		{author} {\bibfnamefont {W.}~\bibnamefont {Guo}}, \bibinfo {author}
		{\bibfnamefont {Y.}~\bibnamefont {Hu}}, \bibinfo {author} {\bibfnamefont
			{J.}~\bibnamefont {Huang}}, \bibinfo {author} {\bibfnamefont {J.~R.}\
			\bibnamefont {Mulcahy}},\ and\ \bibinfo {author} {\bibfnamefont {W.~D.}\
			\bibnamefont {Wei}},\ }\href {https://doi.org/10.1021/acs.chemrev.7b00430}
	{\bibfield  {journal} {\bibinfo  {journal} {Chem. Rev.}\ }\textbf {\bibinfo
			{volume} {118}},\ \bibinfo {pages} {2927} (\bibinfo {year}
		{2018})}\BibitemShut {NoStop}%
	\bibitem [{\citenamefont {Zhang}\ \emph {et~al.}(2019)\citenamefont {Zhang},
		\citenamefont {Zhang}, \citenamefont {Zheng},\ and\ \citenamefont
		{Xu}}]{zhangPlasmonDrivenCatalysisMolecules2019}%
	\BibitemOpen
	\bibfield  {author} {\bibinfo {author} {\bibfnamefont {Z.}~\bibnamefont
			{Zhang}}, \bibinfo {author} {\bibfnamefont {C.}~\bibnamefont {Zhang}},
		\bibinfo {author} {\bibfnamefont {H.}~\bibnamefont {Zheng}},\ and\ \bibinfo
		{author} {\bibfnamefont {H.}~\bibnamefont {Xu}},\ }\href
	{https://doi.org/10.1021/acs.accounts.9b00224} {\bibfield  {journal}
		{\bibinfo  {journal} {Acc. Chem. Res.}\ }\textbf {\bibinfo {volume} {52}},\
		\bibinfo {pages} {2506} (\bibinfo {year} {2019})}\BibitemShut {NoStop}%
	\bibitem [{\citenamefont {Beaurepaire}\ \emph {et~al.}(1996)\citenamefont
		{Beaurepaire}, \citenamefont {Merle}, \citenamefont {Daunois},\ and\
		\citenamefont {Bigot}}]{beaurepaireUltrafastSpinDynamics1996}%
	\BibitemOpen
	\bibfield  {author} {\bibinfo {author} {\bibfnamefont {E.}~\bibnamefont
			{Beaurepaire}}, \bibinfo {author} {\bibfnamefont {J.-C.}\ \bibnamefont
			{Merle}}, \bibinfo {author} {\bibfnamefont {A.}~\bibnamefont {Daunois}},\
		and\ \bibinfo {author} {\bibfnamefont {J.-Y.}\ \bibnamefont {Bigot}},\ }\href
	{https://doi.org/10.1103/PhysRevLett.76.4250} {\bibfield  {journal} {\bibinfo
			{journal} {Phys. Rev. Lett.}\ }\textbf {\bibinfo {volume} {76}},\ \bibinfo
		{pages} {4250} (\bibinfo {year} {1996})},\ \Eprint
	{https://arxiv.org/abs/cond-mat/9709264} {arXiv:cond-mat/9709264}
	\BibitemShut {NoStop}%
	\bibitem [{\citenamefont {Imada}\ \emph {et~al.}(1998)\citenamefont {Imada},
		\citenamefont {Fujimori},\ and\ \citenamefont
		{Tokura}}]{imadaMetalinsulatorTransitions1998}%
	\BibitemOpen
	\bibfield  {author} {\bibinfo {author} {\bibfnamefont {M.}~\bibnamefont
			{Imada}}, \bibinfo {author} {\bibfnamefont {A.}~\bibnamefont {Fujimori}},\
		and\ \bibinfo {author} {\bibfnamefont {Y.}~\bibnamefont {Tokura}},\ }\href
	{https://doi.org/10.1103/RevModPhys.70.1039} {\bibfield  {journal} {\bibinfo
			{journal} {Rev. Mod. Phys.}\ }\textbf {\bibinfo {volume} {70}},\ \bibinfo
		{pages} {1039} (\bibinfo {year} {1998})}\BibitemShut {NoStop}%
	\bibitem [{\citenamefont {Kirilyuk}\ \emph {et~al.}(2010)\citenamefont
		{Kirilyuk}, \citenamefont {Kimel},\ and\ \citenamefont
		{Rasing}}]{Kirilyuk2010}%
	\BibitemOpen
	\bibfield  {author} {\bibinfo {author} {\bibfnamefont {A.}~\bibnamefont
			{Kirilyuk}}, \bibinfo {author} {\bibfnamefont {A.~V.}\ \bibnamefont
			{Kimel}},\ and\ \bibinfo {author} {\bibfnamefont {T.}~\bibnamefont
			{Rasing}},\ }\href {https://doi.org/10.1103/RevModPhys.82.2731} {\bibfield
		{journal} {\bibinfo  {journal} {Rev. Mod. Phys.}\ }\textbf {\bibinfo {volume}
			{82}},\ \bibinfo {pages} {2731} (\bibinfo {year} {2010})}\BibitemShut
	{NoStop}%
	\bibitem [{\citenamefont {Johnson}\ \emph {et~al.}(2017)\citenamefont
		{Johnson}, \citenamefont {Savoini}, \citenamefont {Beaud}, \citenamefont
		{Ingold}, \citenamefont {Staub}, \citenamefont {Carbone}, \citenamefont
		{Castiglioni}, \citenamefont {Hengsberger},\ and\ \citenamefont
		{Osterwalder}}]{Johnson2017}%
	\BibitemOpen
	\bibfield  {author} {\bibinfo {author} {\bibfnamefont {S.~L.}\ \bibnamefont
			{Johnson}}, \bibinfo {author} {\bibfnamefont {M.}~\bibnamefont {Savoini}},
		\bibinfo {author} {\bibfnamefont {P.}~\bibnamefont {Beaud}}, \bibinfo
		{author} {\bibfnamefont {G.}~\bibnamefont {Ingold}}, \bibinfo {author}
		{\bibfnamefont {U.}~\bibnamefont {Staub}}, \bibinfo {author} {\bibfnamefont
			{F.}~\bibnamefont {Carbone}}, \bibinfo {author} {\bibfnamefont
			{L.}~\bibnamefont {Castiglioni}}, \bibinfo {author} {\bibfnamefont
			{M.}~\bibnamefont {Hengsberger}},\ and\ \bibinfo {author} {\bibfnamefont
			{J.}~\bibnamefont {Osterwalder}},\ }\href {https://doi.org/10.1063/1.4996176}
	{\bibfield  {journal} {\bibinfo  {journal} {Struct. Dyn.}\ }\textbf {\bibinfo
			{volume} {4}},\ \bibinfo {pages} {061506} (\bibinfo {year}
		{2017})}\BibitemShut {NoStop}%
	\bibitem [{\citenamefont
		{Shah}(1996)}]{shahUltrafastSpectroscopySemiconductors1996}%
	\BibitemOpen
	\bibfield  {author} {\bibinfo {author} {\bibfnamefont {J.}~\bibnamefont
			{Shah}},\ }\href@noop {} {\emph {\bibinfo {title} {Ultrafast Spectroscopy of
				Semiconductors and Semiconductornanostructures}}},\ \bibinfo {series}
	{Springer Series in Solid-State Sciences}\ No.\ \bibinfo {number} {115}\
	(\bibinfo  {publisher} {{Springer}},\ \bibinfo {address} {{Berlin ; New
			York}},\ \bibinfo {year} {1996})\BibitemShut {NoStop}%
	\bibitem [{\citenamefont {Anisimov}\ \emph {et~al.}(1974)\citenamefont
		{Anisimov}, \citenamefont {Kapeliovich},\ and\ \citenamefont
		{Perelman}}]{anisimovElectronEmissionMetal1974}%
	\BibitemOpen
	\bibfield  {author} {\bibinfo {author} {\bibfnamefont {S.~I.}\ \bibnamefont
			{Anisimov}}, \bibinfo {author} {\bibfnamefont {B.~L.}\ \bibnamefont
			{Kapeliovich}},\ and\ \bibinfo {author} {\bibfnamefont {T.~L.}\ \bibnamefont
			{Perelman}},\ }\href@noop {} {\bibfield  {journal} {\bibinfo  {journal}
			{Soviet JETP}\ }\textbf {\bibinfo {volume} {39}},\ \bibinfo {pages} {375}
		(\bibinfo {year} {1974})}\BibitemShut {NoStop}%
	\bibitem [{\citenamefont {Allen}(1987)}]{allenTheoryThermalRelaxation1987}%
	\BibitemOpen
	\bibfield  {author} {\bibinfo {author} {\bibfnamefont {P.~B.}\ \bibnamefont
			{Allen}},\ }\href {https://doi.org/10.1103/PhysRevLett.59.1460} {\bibfield
		{journal} {\bibinfo  {journal} {Phys. Rev. Lett.}\ }\textbf {\bibinfo
			{volume} {59}},\ \bibinfo {pages} {1460} (\bibinfo {year}
		{1987})}\BibitemShut {NoStop}%
	\bibitem [{\citenamefont
		{Singh}(2010)}]{singhTwotemperatureModelNonequilibrium2010}%
	\BibitemOpen
	\bibfield  {author} {\bibinfo {author} {\bibfnamefont {N.}~\bibnamefont
			{Singh}},\ }\href {https://doi.org/10.1142/S0217979210055366} {\bibfield
		{journal} {\bibinfo  {journal} {Int. J. Mod. Phys. B}\ }\textbf {\bibinfo
			{volume} {24}},\ \bibinfo {pages} {1141} (\bibinfo {year}
		{2010})}\BibitemShut {NoStop}%
	\bibitem [{\citenamefont {Vasileiadis}\ \emph {et~al.}(2018)\citenamefont
		{Vasileiadis}, \citenamefont {Waldecker}, \citenamefont {Foster},
		\citenamefont {Da~Silva}, \citenamefont {Zahn}, \citenamefont {Bertoni},
		\citenamefont {Palmer},\ and\ \citenamefont
		{Ernstorfer}}]{vasileiadisUltrafastHeatFlow2018}%
	\BibitemOpen
	\bibfield  {author} {\bibinfo {author} {\bibfnamefont {T.}~\bibnamefont
			{Vasileiadis}}, \bibinfo {author} {\bibfnamefont {L.}~\bibnamefont
			{Waldecker}}, \bibinfo {author} {\bibfnamefont {D.}~\bibnamefont {Foster}},
		\bibinfo {author} {\bibfnamefont {A.}~\bibnamefont {Da~Silva}}, \bibinfo
		{author} {\bibfnamefont {D.}~\bibnamefont {Zahn}}, \bibinfo {author}
		{\bibfnamefont {R.}~\bibnamefont {Bertoni}}, \bibinfo {author} {\bibfnamefont
			{R.~E.}\ \bibnamefont {Palmer}},\ and\ \bibinfo {author} {\bibfnamefont
			{R.}~\bibnamefont {Ernstorfer}},\ }\href
	{https://doi.org/10.1021/acsnano.8b01423} {\bibfield  {journal} {\bibinfo
			{journal} {ACS Nano}\ }\textbf {\bibinfo {volume} {12}},\ \bibinfo {pages}
		{7710} (\bibinfo {year} {2018})}\BibitemShut {NoStop}%
	\bibitem [{\citenamefont {Kampfrath}\ \emph {et~al.}(2005)\citenamefont
		{Kampfrath}, \citenamefont {Perfetti}, \citenamefont {Schapper},
		\citenamefont {Frischkorn},\ and\ \citenamefont
		{Wolf}}]{kampfrathStronglyCoupledOptical2005}%
	\BibitemOpen
	\bibfield  {author} {\bibinfo {author} {\bibfnamefont {T.}~\bibnamefont
			{Kampfrath}}, \bibinfo {author} {\bibfnamefont {L.}~\bibnamefont {Perfetti}},
		\bibinfo {author} {\bibfnamefont {F.}~\bibnamefont {Schapper}}, \bibinfo
		{author} {\bibfnamefont {C.}~\bibnamefont {Frischkorn}},\ and\ \bibinfo
		{author} {\bibfnamefont {M.}~\bibnamefont {Wolf}},\ }\href
	{https://doi.org/10.1103/PhysRevLett.95.187403} {\bibfield  {journal}
		{\bibinfo  {journal} {Phys. Rev. Lett.}\ }\textbf {\bibinfo {volume} {95}},\
		\bibinfo {pages} {187403} (\bibinfo {year} {2005})}\BibitemShut {NoStop}%
	\bibitem [{\citenamefont {Mansart}\ \emph {et~al.}(2013)\citenamefont
		{Mansart}, \citenamefont {Cottet}, \citenamefont {Mancini}, \citenamefont
		{Jarlborg}, \citenamefont {Dugdale}, \citenamefont {Johnson}, \citenamefont
		{Mariager}, \citenamefont {Milne}, \citenamefont {Beaud}, \citenamefont
		{Gr{\"u}bel}, \citenamefont {Johnson}, \citenamefont {Kubacka}, \citenamefont
		{Ingold}, \citenamefont {Prsa}, \citenamefont {R{\o}nnow}, \citenamefont
		{Conder}, \citenamefont {Pomjakushina}, \citenamefont {Chergui},\ and\
		\citenamefont
		{Carbone}}]{mansartTemperaturedependentElectronphononCoupling2013}%
	\BibitemOpen
	\bibfield  {author} {\bibinfo {author} {\bibfnamefont {B.}~\bibnamefont
			{Mansart}}, \bibinfo {author} {\bibfnamefont {M.~J.~G.}\ \bibnamefont
			{Cottet}}, \bibinfo {author} {\bibfnamefont {G.~F.}\ \bibnamefont {Mancini}},
		\bibinfo {author} {\bibfnamefont {T.}~\bibnamefont {Jarlborg}}, \bibinfo
		{author} {\bibfnamefont {S.~B.}\ \bibnamefont {Dugdale}}, \bibinfo {author}
		{\bibfnamefont {S.~L.}\ \bibnamefont {Johnson}}, \bibinfo {author}
		{\bibfnamefont {S.~O.}\ \bibnamefont {Mariager}}, \bibinfo {author}
		{\bibfnamefont {C.~J.}\ \bibnamefont {Milne}}, \bibinfo {author}
		{\bibfnamefont {P.}~\bibnamefont {Beaud}}, \bibinfo {author} {\bibfnamefont
			{S.}~\bibnamefont {Gr{\"u}bel}}, \bibinfo {author} {\bibfnamefont {J.~A.}\
			\bibnamefont {Johnson}}, \bibinfo {author} {\bibfnamefont {T.}~\bibnamefont
			{Kubacka}}, \bibinfo {author} {\bibfnamefont {G.}~\bibnamefont {Ingold}},
		\bibinfo {author} {\bibfnamefont {K.}~\bibnamefont {Prsa}}, \bibinfo {author}
		{\bibfnamefont {H.~M.}\ \bibnamefont {R{\o}nnow}}, \bibinfo {author}
		{\bibfnamefont {K.}~\bibnamefont {Conder}}, \bibinfo {author} {\bibfnamefont
			{E.}~\bibnamefont {Pomjakushina}}, \bibinfo {author} {\bibfnamefont
			{M.}~\bibnamefont {Chergui}},\ and\ \bibinfo {author} {\bibfnamefont
			{F.}~\bibnamefont {Carbone}},\ }\href
	{https://doi.org/10.1103/PhysRevB.88.054507} {\bibfield  {journal} {\bibinfo
			{journal} {Phys. Rev. B}\ }\textbf {\bibinfo {volume} {88}},\ \bibinfo
		{pages} {054507} (\bibinfo {year} {2013})}\BibitemShut {NoStop}%
	\bibitem [{\citenamefont {Rhie}\ \emph {et~al.}(2003)\citenamefont {Rhie},
		\citenamefont {D{\"u}rr},\ and\ \citenamefont
		{Eberhardt}}]{rhieFemtosecondElectronSpin2003}%
	\BibitemOpen
	\bibfield  {author} {\bibinfo {author} {\bibfnamefont {H.-S.}\ \bibnamefont
			{Rhie}}, \bibinfo {author} {\bibfnamefont {H.~A.}\ \bibnamefont {D{\"u}rr}},\
		and\ \bibinfo {author} {\bibfnamefont {W.}~\bibnamefont {Eberhardt}},\ }\href
	{https://doi.org/10.1103/PhysRevLett.90.247201} {\bibfield  {journal}
		{\bibinfo  {journal} {Phys. Rev. Lett.}\ }\textbf {\bibinfo {volume} {90}},\
		\bibinfo {pages} {247201} (\bibinfo {year} {2003})}\BibitemShut {NoStop}%
	\bibitem [{\citenamefont {Kimel}\ \emph {et~al.}(2004)\citenamefont {Kimel},
		\citenamefont {Kirilyuk}, \citenamefont {Tsvetkov}, \citenamefont {Pisarev},\
		and\ \citenamefont {Rasing}}]{kimelLaserinducedUltrafastSpin2004}%
	\BibitemOpen
	\bibfield  {author} {\bibinfo {author} {\bibfnamefont {A.~V.}\ \bibnamefont
			{Kimel}}, \bibinfo {author} {\bibfnamefont {A.}~\bibnamefont {Kirilyuk}},
		\bibinfo {author} {\bibfnamefont {A.}~\bibnamefont {Tsvetkov}}, \bibinfo
		{author} {\bibfnamefont {R.~V.}\ \bibnamefont {Pisarev}},\ and\ \bibinfo
		{author} {\bibfnamefont {T.}~\bibnamefont {Rasing}},\ }\href
	{https://doi.org/10.1038/nature02659} {\bibfield  {journal} {\bibinfo
			{journal} {Nature}\ }\textbf {\bibinfo {volume} {429}},\ \bibinfo {pages}
		{850} (\bibinfo {year} {2004})}\BibitemShut {NoStop}%
	\bibitem [{\citenamefont {Stamm}\ \emph {et~al.}(2007)\citenamefont {Stamm},
		\citenamefont {Kachel}, \citenamefont {Pontius}, \citenamefont {Mitzner},
		\citenamefont {Quast}, \citenamefont {Holldack}, \citenamefont {Khan},
		\citenamefont {Lupulescu}, \citenamefont {Aziz}, \citenamefont {Wietstruk},
		\citenamefont {D{\"u}rr},\ and\ \citenamefont
		{Eberhardt}}]{stammFemtosecondModificationElectron2007}%
	\BibitemOpen
	\bibfield  {author} {\bibinfo {author} {\bibfnamefont {C.}~\bibnamefont
			{Stamm}}, \bibinfo {author} {\bibfnamefont {T.}~\bibnamefont {Kachel}},
		\bibinfo {author} {\bibfnamefont {N.}~\bibnamefont {Pontius}}, \bibinfo
		{author} {\bibfnamefont {R.}~\bibnamefont {Mitzner}}, \bibinfo {author}
		{\bibfnamefont {T.}~\bibnamefont {Quast}}, \bibinfo {author} {\bibfnamefont
			{K.}~\bibnamefont {Holldack}}, \bibinfo {author} {\bibfnamefont
			{S.}~\bibnamefont {Khan}}, \bibinfo {author} {\bibfnamefont {C.}~\bibnamefont
			{Lupulescu}}, \bibinfo {author} {\bibfnamefont {E.~F.}\ \bibnamefont {Aziz}},
		\bibinfo {author} {\bibfnamefont {M.}~\bibnamefont {Wietstruk}}, \bibinfo
		{author} {\bibfnamefont {H.~A.}\ \bibnamefont {D{\"u}rr}},\ and\ \bibinfo
		{author} {\bibfnamefont {W.}~\bibnamefont {Eberhardt}},\ }\href
	{https://doi.org/10.1038/nmat1985} {\bibfield  {journal} {\bibinfo  {journal}
			{Nat. Mater.}\ }\textbf {\bibinfo {volume} {6}},\ \bibinfo {pages} {740}
		(\bibinfo {year} {2007})}\BibitemShut {NoStop}%
	\bibitem [{\citenamefont {Kachel}\ \emph {et~al.}(2009)\citenamefont {Kachel},
		\citenamefont {Pontius}, \citenamefont {Stamm}, \citenamefont {Wietstruk},
		\citenamefont {Aziz}, \citenamefont {D{\"u}rr}, \citenamefont {Eberhardt},\
		and\ \citenamefont {{de Groot}}}]{kachelTransientElectronicMagnetic2009}%
	\BibitemOpen
	\bibfield  {author} {\bibinfo {author} {\bibfnamefont {T.}~\bibnamefont
			{Kachel}}, \bibinfo {author} {\bibfnamefont {N.}~\bibnamefont {Pontius}},
		\bibinfo {author} {\bibfnamefont {C.}~\bibnamefont {Stamm}}, \bibinfo
		{author} {\bibfnamefont {M.}~\bibnamefont {Wietstruk}}, \bibinfo {author}
		{\bibfnamefont {E.~F.}\ \bibnamefont {Aziz}}, \bibinfo {author}
		{\bibfnamefont {H.~A.}\ \bibnamefont {D{\"u}rr}}, \bibinfo {author}
		{\bibfnamefont {W.}~\bibnamefont {Eberhardt}},\ and\ \bibinfo {author}
		{\bibfnamefont {F.~M.~F.}\ \bibnamefont {{de Groot}}},\ }\href
	{https://doi.org/10.1103/PhysRevB.80.092404} {\bibfield  {journal} {\bibinfo
			{journal} {Phys. Rev. B}\ }\textbf {\bibinfo {volume} {80}},\ \bibinfo
		{pages} {092404} (\bibinfo {year} {2009})}\BibitemShut {NoStop}%
	\bibitem [{\citenamefont {Boeglin}\ \emph {et~al.}(2010)\citenamefont
		{Boeglin}, \citenamefont {Beaurepaire}, \citenamefont {Halt{\'e}},
		\citenamefont {{L{\'o}pez-Flores}}, \citenamefont {Stamm}, \citenamefont
		{Pontius}, \citenamefont {D{\"u}rr},\ and\ \citenamefont
		{Bigot}}]{boeglinDistinguishingUltrafastDynamics2010}%
	\BibitemOpen
	\bibfield  {author} {\bibinfo {author} {\bibfnamefont {C.}~\bibnamefont
			{Boeglin}}, \bibinfo {author} {\bibfnamefont {E.}~\bibnamefont
			{Beaurepaire}}, \bibinfo {author} {\bibfnamefont {V.}~\bibnamefont
			{Halt{\'e}}}, \bibinfo {author} {\bibfnamefont {V.}~\bibnamefont
			{{L{\'o}pez-Flores}}}, \bibinfo {author} {\bibfnamefont {C.}~\bibnamefont
			{Stamm}}, \bibinfo {author} {\bibfnamefont {N.}~\bibnamefont {Pontius}},
		\bibinfo {author} {\bibfnamefont {H.~A.}\ \bibnamefont {D{\"u}rr}},\ and\
		\bibinfo {author} {\bibfnamefont {J.-Y.}\ \bibnamefont {Bigot}},\ }\href
	{https://doi.org/10.1038/nature09070} {\bibfield  {journal} {\bibinfo
			{journal} {Nature}\ }\textbf {\bibinfo {volume} {465}},\ \bibinfo {pages}
		{458} (\bibinfo {year} {2010})}\BibitemShut {NoStop}%
	\bibitem [{\citenamefont {{La-O-Vorakiat}}\ \emph {et~al.}(2012)\citenamefont
		{{La-O-Vorakiat}}, \citenamefont {Turgut}, \citenamefont {Teale},
		\citenamefont {Kapteyn}, \citenamefont {Murnane}, \citenamefont {Mathias},
		\citenamefont {Aeschlimann}, \citenamefont {Schneider}, \citenamefont {Shaw},
		\citenamefont {Nembach},\ and\ \citenamefont {Silva}}]{La-O-Vorakiat2012}%
	\BibitemOpen
	\bibfield  {author} {\bibinfo {author} {\bibfnamefont {C.}~\bibnamefont
			{{La-O-Vorakiat}}}, \bibinfo {author} {\bibfnamefont {E.}~\bibnamefont
			{Turgut}}, \bibinfo {author} {\bibfnamefont {C.~A.}\ \bibnamefont {Teale}},
		\bibinfo {author} {\bibfnamefont {H.~C.}\ \bibnamefont {Kapteyn}}, \bibinfo
		{author} {\bibfnamefont {M.~M.}\ \bibnamefont {Murnane}}, \bibinfo {author}
		{\bibfnamefont {S.}~\bibnamefont {Mathias}}, \bibinfo {author} {\bibfnamefont
			{M.}~\bibnamefont {Aeschlimann}}, \bibinfo {author} {\bibfnamefont {C.~M.}\
			\bibnamefont {Schneider}}, \bibinfo {author} {\bibfnamefont {J.~M.}\
			\bibnamefont {Shaw}}, \bibinfo {author} {\bibfnamefont {H.~T.}\ \bibnamefont
			{Nembach}},\ and\ \bibinfo {author} {\bibfnamefont {T.~J.}\ \bibnamefont
			{Silva}},\ }\href {https://doi.org/10.1103/PhysRevX.2.011005} {\bibfield
		{journal} {\bibinfo  {journal} {Phys. Rev. X}\ }\textbf {\bibinfo {volume}
			{2}},\ \bibinfo {pages} {011005} (\bibinfo {year} {2012})}\BibitemShut
	{NoStop}%
	\bibitem [{\citenamefont {Radu}\ \emph {et~al.}(2011)\citenamefont {Radu},
		\citenamefont {Vahaplar}, \citenamefont {Stamm}, \citenamefont {Kachel},
		\citenamefont {Pontius}, \citenamefont {D{\"u}rr}, \citenamefont {Ostler},
		\citenamefont {Barker}, \citenamefont {Evans}, \citenamefont {Chantrell},
		\citenamefont {Tsukamoto}, \citenamefont {Itoh}, \citenamefont {Kirilyuk},
		\citenamefont {Rasing},\ and\ \citenamefont
		{Kimel}}]{raduTransientFerromagneticlikeState2011}%
	\BibitemOpen
	\bibfield  {author} {\bibinfo {author} {\bibfnamefont {I.}~\bibnamefont
			{Radu}}, \bibinfo {author} {\bibfnamefont {K.}~\bibnamefont {Vahaplar}},
		\bibinfo {author} {\bibfnamefont {C.}~\bibnamefont {Stamm}}, \bibinfo
		{author} {\bibfnamefont {T.}~\bibnamefont {Kachel}}, \bibinfo {author}
		{\bibfnamefont {N.}~\bibnamefont {Pontius}}, \bibinfo {author} {\bibfnamefont
			{H.~A.}\ \bibnamefont {D{\"u}rr}}, \bibinfo {author} {\bibfnamefont {T.~A.}\
			\bibnamefont {Ostler}}, \bibinfo {author} {\bibfnamefont {J.}~\bibnamefont
			{Barker}}, \bibinfo {author} {\bibfnamefont {R.~F.~L.}\ \bibnamefont
			{Evans}}, \bibinfo {author} {\bibfnamefont {R.~W.}\ \bibnamefont
			{Chantrell}}, \bibinfo {author} {\bibfnamefont {A.}~\bibnamefont
			{Tsukamoto}}, \bibinfo {author} {\bibfnamefont {A.}~\bibnamefont {Itoh}},
		\bibinfo {author} {\bibfnamefont {A.}~\bibnamefont {Kirilyuk}}, \bibinfo
		{author} {\bibfnamefont {T.}~\bibnamefont {Rasing}},\ and\ \bibinfo {author}
		{\bibfnamefont {A.~V.}\ \bibnamefont {Kimel}},\ }\href
	{https://doi.org/10.1038/nature09901} {\bibfield  {journal} {\bibinfo
			{journal} {Nature}\ }\textbf {\bibinfo {volume} {472}},\ \bibinfo {pages}
		{205} (\bibinfo {year} {2011})}\BibitemShut {NoStop}%
	\bibitem [{\citenamefont {Staub}\ \emph {et~al.}(2014)\citenamefont {Staub},
		\citenamefont {{de Souza}}, \citenamefont {Beaud}, \citenamefont
		{{M{\"o}hr-Vorobeva}}, \citenamefont {Ingold}, \citenamefont {Caviezel},
		\citenamefont {Scagnoli}, \citenamefont {Delley}, \citenamefont {Schlotter},
		\citenamefont {Turner}, \citenamefont {Krupin}, \citenamefont {Lee},
		\citenamefont {Chuang}, \citenamefont {Patthey}, \citenamefont {Moore},
		\citenamefont {Lu}, \citenamefont {Yi}, \citenamefont {Kirchmann},
		\citenamefont {Trigo}, \citenamefont {Denes}, \citenamefont {Doering},
		\citenamefont {Hussain}, \citenamefont {Shen}, \citenamefont {Prabhakaran},
		\citenamefont {Boothroyd},\ and\ \citenamefont
		{Johnson}}]{staubPersistenceMagneticOrder2014}%
	\BibitemOpen
	\bibfield  {author} {\bibinfo {author} {\bibfnamefont {U.}~\bibnamefont
			{Staub}}, \bibinfo {author} {\bibfnamefont {R.~A.}\ \bibnamefont {{de
					Souza}}}, \bibinfo {author} {\bibfnamefont {P.}~\bibnamefont {Beaud}},
		\bibinfo {author} {\bibfnamefont {E.}~\bibnamefont {{M{\"o}hr-Vorobeva}}},
		\bibinfo {author} {\bibfnamefont {G.}~\bibnamefont {Ingold}}, \bibinfo
		{author} {\bibfnamefont {A.}~\bibnamefont {Caviezel}}, \bibinfo {author}
		{\bibfnamefont {V.}~\bibnamefont {Scagnoli}}, \bibinfo {author}
		{\bibfnamefont {B.}~\bibnamefont {Delley}}, \bibinfo {author} {\bibfnamefont
			{W.~F.}\ \bibnamefont {Schlotter}}, \bibinfo {author} {\bibfnamefont {J.~J.}\
			\bibnamefont {Turner}}, \bibinfo {author} {\bibfnamefont {O.}~\bibnamefont
			{Krupin}}, \bibinfo {author} {\bibfnamefont {W.-S.}\ \bibnamefont {Lee}},
		\bibinfo {author} {\bibfnamefont {Y.-D.}\ \bibnamefont {Chuang}}, \bibinfo
		{author} {\bibfnamefont {L.}~\bibnamefont {Patthey}}, \bibinfo {author}
		{\bibfnamefont {R.~G.}\ \bibnamefont {Moore}}, \bibinfo {author}
		{\bibfnamefont {D.}~\bibnamefont {Lu}}, \bibinfo {author} {\bibfnamefont
			{M.}~\bibnamefont {Yi}}, \bibinfo {author} {\bibfnamefont {P.~S.}\
			\bibnamefont {Kirchmann}}, \bibinfo {author} {\bibfnamefont {M.}~\bibnamefont
			{Trigo}}, \bibinfo {author} {\bibfnamefont {P.}~\bibnamefont {Denes}},
		\bibinfo {author} {\bibfnamefont {D.}~\bibnamefont {Doering}}, \bibinfo
		{author} {\bibfnamefont {Z.}~\bibnamefont {Hussain}}, \bibinfo {author}
		{\bibfnamefont {Z.~X.}\ \bibnamefont {Shen}}, \bibinfo {author}
		{\bibfnamefont {D.}~\bibnamefont {Prabhakaran}}, \bibinfo {author}
		{\bibfnamefont {A.~T.}\ \bibnamefont {Boothroyd}},\ and\ \bibinfo {author}
		{\bibfnamefont {S.~L.}\ \bibnamefont {Johnson}},\ }\href
	{https://doi.org/10.1103/PhysRevB.89.220401} {\bibfield  {journal} {\bibinfo
			{journal} {Phys. Rev. B}\ }\textbf {\bibinfo {volume} {89}},\ \bibinfo
		{pages} {220401} (\bibinfo {year} {2014})}\BibitemShut {NoStop}%
	\bibitem [{\citenamefont {Johnson}\ \emph {et~al.}(2015)\citenamefont
		{Johnson}, \citenamefont {Kubacka}, \citenamefont {Hoffmann}, \citenamefont
		{Vicario}, \citenamefont {{de Jong}}, \citenamefont {Beaud}, \citenamefont
		{Gr{\"u}bel}, \citenamefont {Huang}, \citenamefont {Huber}, \citenamefont
		{Windsor}, \citenamefont {Bothschafter}, \citenamefont {Rettig},
		\citenamefont {Ramakrishnan}, \citenamefont {Alberca}, \citenamefont
		{Patthey}, \citenamefont {Chuang}, \citenamefont {Turner}, \citenamefont
		{Dakovski}, \citenamefont {Lee}, \citenamefont {Minitti}, \citenamefont
		{Schlotter}, \citenamefont {Moore}, \citenamefont {Hauri}, \citenamefont
		{Koohpayeh}, \citenamefont {Scagnoli}, \citenamefont {Ingold}, \citenamefont
		{Johnson},\ and\ \citenamefont {Staub}}]{johnsonMagneticOrderDynamics2015}%
	\BibitemOpen
	\bibfield  {author} {\bibinfo {author} {\bibfnamefont {J.~A.}\ \bibnamefont
			{Johnson}}, \bibinfo {author} {\bibfnamefont {T.}~\bibnamefont {Kubacka}},
		\bibinfo {author} {\bibfnamefont {M.~C.}\ \bibnamefont {Hoffmann}}, \bibinfo
		{author} {\bibfnamefont {C.}~\bibnamefont {Vicario}}, \bibinfo {author}
		{\bibfnamefont {S.}~\bibnamefont {{de Jong}}}, \bibinfo {author}
		{\bibfnamefont {P.}~\bibnamefont {Beaud}}, \bibinfo {author} {\bibfnamefont
			{S.}~\bibnamefont {Gr{\"u}bel}}, \bibinfo {author} {\bibfnamefont {S.-W.}\
			\bibnamefont {Huang}}, \bibinfo {author} {\bibfnamefont {L.}~\bibnamefont
			{Huber}}, \bibinfo {author} {\bibfnamefont {Y.~W.}\ \bibnamefont {Windsor}},
		\bibinfo {author} {\bibfnamefont {E.~M.}\ \bibnamefont {Bothschafter}},
		\bibinfo {author} {\bibfnamefont {L.}~\bibnamefont {Rettig}}, \bibinfo
		{author} {\bibfnamefont {M.}~\bibnamefont {Ramakrishnan}}, \bibinfo {author}
		{\bibfnamefont {A.}~\bibnamefont {Alberca}}, \bibinfo {author} {\bibfnamefont
			{L.}~\bibnamefont {Patthey}}, \bibinfo {author} {\bibfnamefont {Y.-D.}\
			\bibnamefont {Chuang}}, \bibinfo {author} {\bibfnamefont {J.~J.}\
			\bibnamefont {Turner}}, \bibinfo {author} {\bibfnamefont {G.~L.}\
			\bibnamefont {Dakovski}}, \bibinfo {author} {\bibfnamefont {W.-S.}\
			\bibnamefont {Lee}}, \bibinfo {author} {\bibfnamefont {M.~P.}\ \bibnamefont
			{Minitti}}, \bibinfo {author} {\bibfnamefont {W.}~\bibnamefont {Schlotter}},
		\bibinfo {author} {\bibfnamefont {R.~G.}\ \bibnamefont {Moore}}, \bibinfo
		{author} {\bibfnamefont {C.~P.}\ \bibnamefont {Hauri}}, \bibinfo {author}
		{\bibfnamefont {S.~M.}\ \bibnamefont {Koohpayeh}}, \bibinfo {author}
		{\bibfnamefont {V.}~\bibnamefont {Scagnoli}}, \bibinfo {author}
		{\bibfnamefont {G.}~\bibnamefont {Ingold}}, \bibinfo {author} {\bibfnamefont
			{S.~L.}\ \bibnamefont {Johnson}},\ and\ \bibinfo {author} {\bibfnamefont
			{U.}~\bibnamefont {Staub}},\ }\href
	{https://doi.org/10.1103/PhysRevB.92.184429} {\bibfield  {journal} {\bibinfo
			{journal} {Phys. Rev. B}\ }\textbf {\bibinfo {volume} {92}},\ \bibinfo
		{pages} {184429} (\bibinfo {year} {2015})}\BibitemShut {NoStop}%
	\bibitem [{\citenamefont {Naseska}\ \emph {et~al.}(2018)\citenamefont
		{Naseska}, \citenamefont {Pogrebna}, \citenamefont {Cao}, \citenamefont {Xu},
		\citenamefont {Mihailovic},\ and\ \citenamefont
		{Mertelj}}]{naseskaUltrafastDestructionRecovery2018}%
	\BibitemOpen
	\bibfield  {author} {\bibinfo {author} {\bibfnamefont {M.}~\bibnamefont
			{Naseska}}, \bibinfo {author} {\bibfnamefont {A.}~\bibnamefont {Pogrebna}},
		\bibinfo {author} {\bibfnamefont {G.}~\bibnamefont {Cao}}, \bibinfo {author}
		{\bibfnamefont {Z.~A.}\ \bibnamefont {Xu}}, \bibinfo {author} {\bibfnamefont
			{D.}~\bibnamefont {Mihailovic}},\ and\ \bibinfo {author} {\bibfnamefont
			{T.}~\bibnamefont {Mertelj}},\ }\href
	{https://doi.org/10.1103/PhysRevB.98.035148} {\bibfield  {journal} {\bibinfo
			{journal} {Phys. Rev. B}\ }\textbf {\bibinfo {volume} {98}},\ \bibinfo
		{pages} {035148} (\bibinfo {year} {2018})}\BibitemShut {NoStop}%
	\bibitem [{\citenamefont {Kimel}\ \emph {et~al.}(2005)\citenamefont {Kimel},
		\citenamefont {Kirilyuk}, \citenamefont {Usachev}, \citenamefont {Pisarev},
		\citenamefont {Balbashov},\ and\ \citenamefont
		{Rasing}}]{kimelUltrafastNonthermalControl2005}%
	\BibitemOpen
	\bibfield  {author} {\bibinfo {author} {\bibfnamefont {A.~V.}\ \bibnamefont
			{Kimel}}, \bibinfo {author} {\bibfnamefont {A.}~\bibnamefont {Kirilyuk}},
		\bibinfo {author} {\bibfnamefont {P.~A.}\ \bibnamefont {Usachev}}, \bibinfo
		{author} {\bibfnamefont {R.~V.}\ \bibnamefont {Pisarev}}, \bibinfo {author}
		{\bibfnamefont {A.~M.}\ \bibnamefont {Balbashov}},\ and\ \bibinfo {author}
		{\bibfnamefont {T.}~\bibnamefont {Rasing}},\ }\href
	{https://doi.org/10.1038/nature03564} {\bibfield  {journal} {\bibinfo
			{journal} {Nature}\ }\textbf {\bibinfo {volume} {435}},\ \bibinfo {pages}
		{655} (\bibinfo {year} {2005})}\BibitemShut {NoStop}%
	\bibitem [{\citenamefont {Bigot}\ \emph {et~al.}(2009)\citenamefont {Bigot},
		\citenamefont {Vomir},\ and\ \citenamefont
		{Beaurepaire}}]{bigotCoherentUltrafastMagnetism2009}%
	\BibitemOpen
	\bibfield  {author} {\bibinfo {author} {\bibfnamefont {J.-Y.}\ \bibnamefont
			{Bigot}}, \bibinfo {author} {\bibfnamefont {M.}~\bibnamefont {Vomir}},\ and\
		\bibinfo {author} {\bibfnamefont {E.}~\bibnamefont {Beaurepaire}},\ }\href
	{https://doi.org/10.1038/nphys1285} {\bibfield  {journal} {\bibinfo
			{journal} {Nat. Phys.}\ }\textbf {\bibinfo {volume} {5}},\ \bibinfo {pages}
		{515} (\bibinfo {year} {2009})}\BibitemShut {NoStop}%
	\bibitem [{\citenamefont {Graves}\ \emph {et~al.}(2013)\citenamefont {Graves},
		\citenamefont {Reid}, \citenamefont {Wang}, \citenamefont {Wu}, \citenamefont
		{de~Jong}, \citenamefont {Vahaplar}, \citenamefont {Radu}, \citenamefont
		{Bernstein}, \citenamefont {Messerschmidt}, \citenamefont {M{\"u}ller},
		\citenamefont {Coffee}, \citenamefont {Bionta}, \citenamefont {Epp},
		\citenamefont {Hartmann}, \citenamefont {Kimmel}, \citenamefont {Hauser},
		\citenamefont {Hartmann}, \citenamefont {Holl}, \citenamefont {Gorke},
		\citenamefont {Mentink}, \citenamefont {Tsukamoto}, \citenamefont {Fognini},
		\citenamefont {Turner}, \citenamefont {Schlotter}, \citenamefont {Rolles},
		\citenamefont {Soltau}, \citenamefont {Str{\"u}der}, \citenamefont
		{Acremann}, \citenamefont {Kimel}, \citenamefont {Kirilyuk}, \citenamefont
		{Rasing}, \citenamefont {St{\"o}hr}, \citenamefont {Scherz},\ and\
		\citenamefont {D{\"u}rr}}]{gravesNanoscaleSpinReversal2013}%
	\BibitemOpen
	\bibfield  {author} {\bibinfo {author} {\bibfnamefont {C.~E.}\ \bibnamefont
			{Graves}}, \bibinfo {author} {\bibfnamefont {A.~H.}\ \bibnamefont {Reid}},
		\bibinfo {author} {\bibfnamefont {T.}~\bibnamefont {Wang}}, \bibinfo {author}
		{\bibfnamefont {B.}~\bibnamefont {Wu}}, \bibinfo {author} {\bibfnamefont
			{S.}~\bibnamefont {de~Jong}}, \bibinfo {author} {\bibfnamefont
			{K.}~\bibnamefont {Vahaplar}}, \bibinfo {author} {\bibfnamefont
			{I.}~\bibnamefont {Radu}}, \bibinfo {author} {\bibfnamefont {D.~P.}\
			\bibnamefont {Bernstein}}, \bibinfo {author} {\bibfnamefont {M.}~\bibnamefont
			{Messerschmidt}}, \bibinfo {author} {\bibfnamefont {L.}~\bibnamefont
			{M{\"u}ller}}, \bibinfo {author} {\bibfnamefont {R.}~\bibnamefont {Coffee}},
		\bibinfo {author} {\bibfnamefont {M.}~\bibnamefont {Bionta}}, \bibinfo
		{author} {\bibfnamefont {S.~W.}\ \bibnamefont {Epp}}, \bibinfo {author}
		{\bibfnamefont {R.}~\bibnamefont {Hartmann}}, \bibinfo {author}
		{\bibfnamefont {N.}~\bibnamefont {Kimmel}}, \bibinfo {author} {\bibfnamefont
			{G.}~\bibnamefont {Hauser}}, \bibinfo {author} {\bibfnamefont
			{A.}~\bibnamefont {Hartmann}}, \bibinfo {author} {\bibfnamefont
			{P.}~\bibnamefont {Holl}}, \bibinfo {author} {\bibfnamefont {H.}~\bibnamefont
			{Gorke}}, \bibinfo {author} {\bibfnamefont {J.~H.}\ \bibnamefont {Mentink}},
		\bibinfo {author} {\bibfnamefont {A.}~\bibnamefont {Tsukamoto}}, \bibinfo
		{author} {\bibfnamefont {A.}~\bibnamefont {Fognini}}, \bibinfo {author}
		{\bibfnamefont {J.~J.}\ \bibnamefont {Turner}}, \bibinfo {author}
		{\bibfnamefont {W.~F.}\ \bibnamefont {Schlotter}}, \bibinfo {author}
		{\bibfnamefont {D.}~\bibnamefont {Rolles}}, \bibinfo {author} {\bibfnamefont
			{H.}~\bibnamefont {Soltau}}, \bibinfo {author} {\bibfnamefont
			{L.}~\bibnamefont {Str{\"u}der}}, \bibinfo {author} {\bibfnamefont
			{Y.}~\bibnamefont {Acremann}}, \bibinfo {author} {\bibfnamefont {A.~V.}\
			\bibnamefont {Kimel}}, \bibinfo {author} {\bibfnamefont {A.}~\bibnamefont
			{Kirilyuk}}, \bibinfo {author} {\bibfnamefont {T.}~\bibnamefont {Rasing}},
		\bibinfo {author} {\bibfnamefont {J.}~\bibnamefont {St{\"o}hr}}, \bibinfo
		{author} {\bibfnamefont {A.~O.}\ \bibnamefont {Scherz}},\ and\ \bibinfo
		{author} {\bibfnamefont {H.~A.}\ \bibnamefont {D{\"u}rr}},\ }\href
	{https://doi.org/10.1038/nmat3597} {\bibfield  {journal} {\bibinfo  {journal}
			{Nat. Mater.}\ }\textbf {\bibinfo {volume} {12}},\ \bibinfo {pages} {293}
		(\bibinfo {year} {2013})}\BibitemShut {NoStop}%
	\bibitem [{\citenamefont {Eschenlohr}\ \emph {et~al.}(2013)\citenamefont
		{Eschenlohr}, \citenamefont {Battiato}, \citenamefont {Maldonado},
		\citenamefont {Pontius}, \citenamefont {Kachel}, \citenamefont {Holldack},
		\citenamefont {Mitzner}, \citenamefont {F{\"o}hlisch}, \citenamefont
		{Oppeneer},\ and\ \citenamefont
		{Stamm}}]{eschenlohrUltrafastSpinTransport2013}%
	\BibitemOpen
	\bibfield  {author} {\bibinfo {author} {\bibfnamefont {A.}~\bibnamefont
			{Eschenlohr}}, \bibinfo {author} {\bibfnamefont {M.}~\bibnamefont
			{Battiato}}, \bibinfo {author} {\bibfnamefont {P.}~\bibnamefont {Maldonado}},
		\bibinfo {author} {\bibfnamefont {N.}~\bibnamefont {Pontius}}, \bibinfo
		{author} {\bibfnamefont {T.}~\bibnamefont {Kachel}}, \bibinfo {author}
		{\bibfnamefont {K.}~\bibnamefont {Holldack}}, \bibinfo {author}
		{\bibfnamefont {R.}~\bibnamefont {Mitzner}}, \bibinfo {author} {\bibfnamefont
			{A.}~\bibnamefont {F{\"o}hlisch}}, \bibinfo {author} {\bibfnamefont {P.~M.}\
			\bibnamefont {Oppeneer}},\ and\ \bibinfo {author} {\bibfnamefont
			{C.}~\bibnamefont {Stamm}},\ }\href {https://doi.org/10.1038/nmat3546}
	{\bibfield  {journal} {\bibinfo  {journal} {Nat. Mater.}\ }\textbf {\bibinfo
			{volume} {12}},\ \bibinfo {pages} {332} (\bibinfo {year} {2013})}\BibitemShut
	{NoStop}%
	\bibitem [{\citenamefont {Batignani}\ \emph {et~al.}(2015)\citenamefont
		{Batignani}, \citenamefont {Bossini}, \citenamefont {Di~Palo}, \citenamefont
		{Ferrante}, \citenamefont {Pontecorvo}, \citenamefont {Cerullo},
		\citenamefont {Kimel},\ and\ \citenamefont {Scopigno}}]{Batignani2015}%
	\BibitemOpen
	\bibfield  {author} {\bibinfo {author} {\bibfnamefont {G.}~\bibnamefont
			{Batignani}}, \bibinfo {author} {\bibfnamefont {D.}~\bibnamefont {Bossini}},
		\bibinfo {author} {\bibfnamefont {N.}~\bibnamefont {Di~Palo}}, \bibinfo
		{author} {\bibfnamefont {C.}~\bibnamefont {Ferrante}}, \bibinfo {author}
		{\bibfnamefont {E.}~\bibnamefont {Pontecorvo}}, \bibinfo {author}
		{\bibfnamefont {G.}~\bibnamefont {Cerullo}}, \bibinfo {author} {\bibfnamefont
			{A.}~\bibnamefont {Kimel}},\ and\ \bibinfo {author} {\bibfnamefont
			{T.}~\bibnamefont {Scopigno}},\ }\href
	{https://doi.org/10.1038/nphoton.2015.121} {\bibfield  {journal} {\bibinfo
			{journal} {Nat. Photonics}\ }\textbf {\bibinfo {volume} {9}},\ \bibinfo
		{pages} {506} (\bibinfo {year} {2015})}\BibitemShut {NoStop}%
	\bibitem [{\citenamefont {Bossini}\ \emph {et~al.}(2016)\citenamefont
		{Bossini}, \citenamefont {Dal~Conte}, \citenamefont {Hashimoto},
		\citenamefont {Secchi}, \citenamefont {Pisarev}, \citenamefont {Rasing},
		\citenamefont {Cerullo},\ and\ \citenamefont {Kimel}}]{Bossini2016}%
	\BibitemOpen
	\bibfield  {author} {\bibinfo {author} {\bibfnamefont {D.}~\bibnamefont
			{Bossini}}, \bibinfo {author} {\bibfnamefont {S.}~\bibnamefont {Dal~Conte}},
		\bibinfo {author} {\bibfnamefont {Y.}~\bibnamefont {Hashimoto}}, \bibinfo
		{author} {\bibfnamefont {A.}~\bibnamefont {Secchi}}, \bibinfo {author}
		{\bibfnamefont {R.~V.}\ \bibnamefont {Pisarev}}, \bibinfo {author}
		{\bibfnamefont {T.}~\bibnamefont {Rasing}}, \bibinfo {author} {\bibfnamefont
			{G.}~\bibnamefont {Cerullo}},\ and\ \bibinfo {author} {\bibfnamefont {A.~V.}\
			\bibnamefont {Kimel}},\ }\href {https://doi.org/10.1038/ncomms10645}
	{\bibfield  {journal} {\bibinfo  {journal} {Nat. Commun.}\ }\textbf {\bibinfo
			{volume} {7}},\ \bibinfo {pages} {10645} (\bibinfo {year}
		{2016})}\BibitemShut {NoStop}%
	\bibitem [{\citenamefont {Siegrist}\ \emph {et~al.}(2019)\citenamefont
		{Siegrist}, \citenamefont {Gessner}, \citenamefont {Ossiander}, \citenamefont
		{Denker}, \citenamefont {Chang}, \citenamefont {Schr{\"o}der}, \citenamefont
		{Guggenmos}, \citenamefont {Cui}, \citenamefont {Walowski}, \citenamefont
		{Martens}, \citenamefont {Dewhurst}, \citenamefont {Kleineberg},
		\citenamefont {M{\"u}nzenberg}, \citenamefont {Sharma},\ and\ \citenamefont
		{Schultze}}]{siegristLightwaveDynamicControl2019}%
	\BibitemOpen
	\bibfield  {author} {\bibinfo {author} {\bibfnamefont {F.}~\bibnamefont
			{Siegrist}}, \bibinfo {author} {\bibfnamefont {J.~A.}\ \bibnamefont
			{Gessner}}, \bibinfo {author} {\bibfnamefont {M.}~\bibnamefont {Ossiander}},
		\bibinfo {author} {\bibfnamefont {C.}~\bibnamefont {Denker}}, \bibinfo
		{author} {\bibfnamefont {Y.-P.}\ \bibnamefont {Chang}}, \bibinfo {author}
		{\bibfnamefont {M.~C.}\ \bibnamefont {Schr{\"o}der}}, \bibinfo {author}
		{\bibfnamefont {A.}~\bibnamefont {Guggenmos}}, \bibinfo {author}
		{\bibfnamefont {Y.}~\bibnamefont {Cui}}, \bibinfo {author} {\bibfnamefont
			{J.}~\bibnamefont {Walowski}}, \bibinfo {author} {\bibfnamefont
			{U.}~\bibnamefont {Martens}}, \bibinfo {author} {\bibfnamefont {J.~K.}\
			\bibnamefont {Dewhurst}}, \bibinfo {author} {\bibfnamefont {U.}~\bibnamefont
			{Kleineberg}}, \bibinfo {author} {\bibfnamefont {M.}~\bibnamefont
			{M{\"u}nzenberg}}, \bibinfo {author} {\bibfnamefont {S.}~\bibnamefont
			{Sharma}},\ and\ \bibinfo {author} {\bibfnamefont {M.}~\bibnamefont
			{Schultze}},\ }\href {https://doi.org/10.1038/s41586-019-1333-x} {\bibfield
		{journal} {\bibinfo  {journal} {Nature}\ }\textbf {\bibinfo {volume} {571}},\
		\bibinfo {pages} {240} (\bibinfo {year} {2019})},\ \Eprint
	{https://arxiv.org/abs/1812.07420} {arXiv:1812.07420} \BibitemShut {NoStop}%
	\bibitem [{\citenamefont {Obergfell}\ and\ \citenamefont
		{Demsar}(2020)}]{obergfellTrackingTimeEvolution2020}%
	\BibitemOpen
	\bibfield  {author} {\bibinfo {author} {\bibfnamefont {M.}~\bibnamefont
			{Obergfell}}\ and\ \bibinfo {author} {\bibfnamefont {J.}~\bibnamefont
			{Demsar}},\ }\href {https://doi.org/10.1103/PhysRevLett.124.037401}
	{\bibfield  {journal} {\bibinfo  {journal} {Phys. Rev. Lett.}\ }\textbf
		{\bibinfo {volume} {124}},\ \bibinfo {pages} {037401} (\bibinfo {year}
		{2020})}\BibitemShut {NoStop}%
	\bibitem [{\citenamefont {{El-Ghazaly}}\ \emph {et~al.}(2019)\citenamefont
		{{El-Ghazaly}}, \citenamefont {Tran}, \citenamefont {Ceballos}, \citenamefont
		{Lambert}, \citenamefont {Pattabi}, \citenamefont {Salahuddin}, \citenamefont
		{Hellman},\ and\ \citenamefont
		{Bokor}}]{el-ghazalyUltrafastMagnetizationSwitching2019}%
	\BibitemOpen
	\bibfield  {author} {\bibinfo {author} {\bibfnamefont {A.}~\bibnamefont
			{{El-Ghazaly}}}, \bibinfo {author} {\bibfnamefont {B.}~\bibnamefont {Tran}},
		\bibinfo {author} {\bibfnamefont {A.}~\bibnamefont {Ceballos}}, \bibinfo
		{author} {\bibfnamefont {C.-H.}\ \bibnamefont {Lambert}}, \bibinfo {author}
		{\bibfnamefont {A.}~\bibnamefont {Pattabi}}, \bibinfo {author} {\bibfnamefont
			{S.}~\bibnamefont {Salahuddin}}, \bibinfo {author} {\bibfnamefont
			{F.}~\bibnamefont {Hellman}},\ and\ \bibinfo {author} {\bibfnamefont
			{J.}~\bibnamefont {Bokor}},\ }\href {https://doi.org/10.1063/1.5098453}
	{\bibfield  {journal} {\bibinfo  {journal} {Appl. Phys. Lett.}\ }\textbf
		{\bibinfo {volume} {114}},\ \bibinfo {pages} {232407} (\bibinfo {year}
		{2019})}\BibitemShut {NoStop}%
	\bibitem [{\citenamefont {Hofherr}\ \emph {et~al.}(2020)\citenamefont
		{Hofherr}, \citenamefont {H{\"a}user}, \citenamefont {Dewhurst},
		\citenamefont {Tengdin}, \citenamefont {Sakshath}, \citenamefont {Nembach},
		\citenamefont {Weber}, \citenamefont {Shaw}, \citenamefont {Silva},
		\citenamefont {Kapteyn}, \citenamefont {Cinchetti}, \citenamefont {Rethfeld},
		\citenamefont {Murnane}, \citenamefont {Steil}, \citenamefont
		{Stadtm{\"u}ller}, \citenamefont {Sharma}, \citenamefont {Aeschlimann},\ and\
		\citenamefont {Mathias}}]{hofherrUltrafastOpticallyInduced2020}%
	\BibitemOpen
	\bibfield  {author} {\bibinfo {author} {\bibfnamefont {M.}~\bibnamefont
			{Hofherr}}, \bibinfo {author} {\bibfnamefont {S.}~\bibnamefont {H{\"a}user}},
		\bibinfo {author} {\bibfnamefont {J.~K.}\ \bibnamefont {Dewhurst}}, \bibinfo
		{author} {\bibfnamefont {P.}~\bibnamefont {Tengdin}}, \bibinfo {author}
		{\bibfnamefont {S.}~\bibnamefont {Sakshath}}, \bibinfo {author}
		{\bibfnamefont {H.~T.}\ \bibnamefont {Nembach}}, \bibinfo {author}
		{\bibfnamefont {S.~T.}\ \bibnamefont {Weber}}, \bibinfo {author}
		{\bibfnamefont {J.~M.}\ \bibnamefont {Shaw}}, \bibinfo {author}
		{\bibfnamefont {T.~J.}\ \bibnamefont {Silva}}, \bibinfo {author}
		{\bibfnamefont {H.~C.}\ \bibnamefont {Kapteyn}}, \bibinfo {author}
		{\bibfnamefont {M.}~\bibnamefont {Cinchetti}}, \bibinfo {author}
		{\bibfnamefont {B.}~\bibnamefont {Rethfeld}}, \bibinfo {author}
		{\bibfnamefont {M.~M.}\ \bibnamefont {Murnane}}, \bibinfo {author}
		{\bibfnamefont {D.}~\bibnamefont {Steil}}, \bibinfo {author} {\bibfnamefont
			{B.}~\bibnamefont {Stadtm{\"u}ller}}, \bibinfo {author} {\bibfnamefont
			{S.}~\bibnamefont {Sharma}}, \bibinfo {author} {\bibfnamefont
			{M.}~\bibnamefont {Aeschlimann}},\ and\ \bibinfo {author} {\bibfnamefont
			{S.}~\bibnamefont {Mathias}},\ }\href
	{https://doi.org/10.1126/sciadv.aay8717} {\bibfield  {journal} {\bibinfo
			{journal} {Sci. Adv.}\ }\textbf {\bibinfo {volume} {6}},\ \bibinfo {pages}
		{eaay8717} (\bibinfo {year} {2020})}\BibitemShut {NoStop}%
	\bibitem [{\citenamefont {Schultze}\ \emph {et~al.}(2014)\citenamefont
		{Schultze}, \citenamefont {Ramasesha}, \citenamefont {Pemmaraju},
		\citenamefont {Sato}, \citenamefont {Whitmore}, \citenamefont {Gandman},
		\citenamefont {Prell}, \citenamefont {Borja}, \citenamefont {Prendergast},
		\citenamefont {Yabana}, \citenamefont {Neumark},\ and\ \citenamefont
		{Leone}}]{Schultze2014}%
	\BibitemOpen
	\bibfield  {author} {\bibinfo {author} {\bibfnamefont {M.}~\bibnamefont
			{Schultze}}, \bibinfo {author} {\bibfnamefont {K.}~\bibnamefont {Ramasesha}},
		\bibinfo {author} {\bibfnamefont {C.~D.}\ \bibnamefont {Pemmaraju}}, \bibinfo
		{author} {\bibfnamefont {S.~A.}\ \bibnamefont {Sato}}, \bibinfo {author}
		{\bibfnamefont {D.}~\bibnamefont {Whitmore}}, \bibinfo {author}
		{\bibfnamefont {A.}~\bibnamefont {Gandman}}, \bibinfo {author} {\bibfnamefont
			{J.~S.}\ \bibnamefont {Prell}}, \bibinfo {author} {\bibfnamefont {L.~J.}\
			\bibnamefont {Borja}}, \bibinfo {author} {\bibfnamefont {D.}~\bibnamefont
			{Prendergast}}, \bibinfo {author} {\bibfnamefont {K.}~\bibnamefont {Yabana}},
		\bibinfo {author} {\bibfnamefont {D.~M.}\ \bibnamefont {Neumark}},\ and\
		\bibinfo {author} {\bibfnamefont {S.~R.}\ \bibnamefont {Leone}},\ }\href
	{https://doi.org/10.1126/science.1260311} {\bibfield  {journal} {\bibinfo
			{journal} {Science}\ }\textbf {\bibinfo {volume} {346}},\ \bibinfo {pages}
		{1348} (\bibinfo {year} {2014})}\BibitemShut {NoStop}%
	\bibitem [{\citenamefont {Z{\"u}rch}\ \emph
		{et~al.}(2017{\natexlab{a}})\citenamefont {Z{\"u}rch}, \citenamefont {Chang},
		\citenamefont {Borja}, \citenamefont {Kraus}, \citenamefont {Cushing},
		\citenamefont {Gandman}, \citenamefont {Kaplan}, \citenamefont {Oh},
		\citenamefont {Prell}, \citenamefont {Prendergast}, \citenamefont
		{Pemmaraju}, \citenamefont {Neumark},\ and\ \citenamefont
		{Leone}}]{zurch2017direct}%
	\BibitemOpen
	\bibfield  {author} {\bibinfo {author} {\bibfnamefont {M.}~\bibnamefont
			{Z{\"u}rch}}, \bibinfo {author} {\bibfnamefont {H.-T.}\ \bibnamefont
			{Chang}}, \bibinfo {author} {\bibfnamefont {L.~J.}\ \bibnamefont {Borja}},
		\bibinfo {author} {\bibfnamefont {P.~M.}\ \bibnamefont {Kraus}}, \bibinfo
		{author} {\bibfnamefont {S.~K.}\ \bibnamefont {Cushing}}, \bibinfo {author}
		{\bibfnamefont {A.}~\bibnamefont {Gandman}}, \bibinfo {author} {\bibfnamefont
			{C.~J.}\ \bibnamefont {Kaplan}}, \bibinfo {author} {\bibfnamefont {M.~H.}\
			\bibnamefont {Oh}}, \bibinfo {author} {\bibfnamefont {J.~S.}\ \bibnamefont
			{Prell}}, \bibinfo {author} {\bibfnamefont {D.}~\bibnamefont {Prendergast}},
		\bibinfo {author} {\bibfnamefont {C.~D.}\ \bibnamefont {Pemmaraju}}, \bibinfo
		{author} {\bibfnamefont {D.~M.}\ \bibnamefont {Neumark}},\ and\ \bibinfo
		{author} {\bibfnamefont {S.~R.}\ \bibnamefont {Leone}},\ }\href
	{https://doi.org/10.1038/ncomms15734} {\bibfield  {journal} {\bibinfo
			{journal} {Nat. Commun.}\ }\textbf {\bibinfo {volume} {8}},\ \bibinfo {pages}
		{15734} (\bibinfo {year} {2017}{\natexlab{a}})}\BibitemShut {NoStop}%
	\bibitem [{\citenamefont {Z{\"u}rch}\ \emph
		{et~al.}(2017{\natexlab{b}})\citenamefont {Z{\"u}rch}, \citenamefont {Chang},
		\citenamefont {Kraus}, \citenamefont {Cushing}, \citenamefont {Borja},
		\citenamefont {Gandman}, \citenamefont {Kaplan}, \citenamefont {Oh},
		\citenamefont {Prell}, \citenamefont {Prendergast}, \citenamefont
		{Pemmaraju}, \citenamefont {Neumark},\ and\ \citenamefont
		{Leone}}]{zurchUltrafastCarrierThermalization2017}%
	\BibitemOpen
	\bibfield  {author} {\bibinfo {author} {\bibfnamefont {M.}~\bibnamefont
			{Z{\"u}rch}}, \bibinfo {author} {\bibfnamefont {H.-T.}\ \bibnamefont
			{Chang}}, \bibinfo {author} {\bibfnamefont {P.~M.}\ \bibnamefont {Kraus}},
		\bibinfo {author} {\bibfnamefont {S.~K.}\ \bibnamefont {Cushing}}, \bibinfo
		{author} {\bibfnamefont {L.~J.}\ \bibnamefont {Borja}}, \bibinfo {author}
		{\bibfnamefont {A.}~\bibnamefont {Gandman}}, \bibinfo {author} {\bibfnamefont
			{C.~J.}\ \bibnamefont {Kaplan}}, \bibinfo {author} {\bibfnamefont {M.~H.}\
			\bibnamefont {Oh}}, \bibinfo {author} {\bibfnamefont {J.~S.}\ \bibnamefont
			{Prell}}, \bibinfo {author} {\bibfnamefont {D.}~\bibnamefont {Prendergast}},
		\bibinfo {author} {\bibfnamefont {C.~D.}\ \bibnamefont {Pemmaraju}}, \bibinfo
		{author} {\bibfnamefont {D.~M.}\ \bibnamefont {Neumark}},\ and\ \bibinfo
		{author} {\bibfnamefont {S.~R.}\ \bibnamefont {Leone}},\ }\href
	{https://doi.org/10.1063/1.4985056} {\bibfield  {journal} {\bibinfo
			{journal} {Struct. Dyn.}\ }\textbf {\bibinfo {volume} {4}},\ \bibinfo {pages}
		{044029} (\bibinfo {year} {2017}{\natexlab{b}})}\BibitemShut {NoStop}%
	\bibitem [{\citenamefont {Schlaepfer}\ \emph {et~al.}(2018)\citenamefont
		{Schlaepfer}, \citenamefont {Lucchini}, \citenamefont {Sato}, \citenamefont
		{Volkov}, \citenamefont {Kasmi}, \citenamefont {Hartmann}, \citenamefont
		{Rubio}, \citenamefont {Gallmann},\ and\ \citenamefont
		{Keller}}]{schlaepferAttosecondOpticalfieldenhancedCarrier2018}%
	\BibitemOpen
	\bibfield  {author} {\bibinfo {author} {\bibfnamefont {F.}~\bibnamefont
			{Schlaepfer}}, \bibinfo {author} {\bibfnamefont {M.}~\bibnamefont
			{Lucchini}}, \bibinfo {author} {\bibfnamefont {S.~A.}\ \bibnamefont {Sato}},
		\bibinfo {author} {\bibfnamefont {M.}~\bibnamefont {Volkov}}, \bibinfo
		{author} {\bibfnamefont {L.}~\bibnamefont {Kasmi}}, \bibinfo {author}
		{\bibfnamefont {N.}~\bibnamefont {Hartmann}}, \bibinfo {author}
		{\bibfnamefont {A.}~\bibnamefont {Rubio}}, \bibinfo {author} {\bibfnamefont
			{L.}~\bibnamefont {Gallmann}},\ and\ \bibinfo {author} {\bibfnamefont
			{U.}~\bibnamefont {Keller}},\ }\href
	{https://doi.org/10.1038/s41567-018-0069-0} {\bibfield  {journal} {\bibinfo
			{journal} {Nat. Phys.}\ }\textbf {\bibinfo {volume} {14}},\ \bibinfo {pages}
		{560} (\bibinfo {year} {2018})}\BibitemShut {NoStop}%
	\bibitem [{\citenamefont {Lin}\ \emph {et~al.}(2017)\citenamefont {Lin},
		\citenamefont {Verkamp}, \citenamefont {Leveillee}, \citenamefont {Ryland},
		\citenamefont {Benke}, \citenamefont {Zhang}, \citenamefont {Weninger},
		\citenamefont {Shen}, \citenamefont {Li}, \citenamefont {Fritz},
		\citenamefont {Bergmann}, \citenamefont {Wang}, \citenamefont {Schleife},\
		and\ \citenamefont {{Vura-Weis}}}]{linCarrierSpecificFemtosecondXUV2017a}%
	\BibitemOpen
	\bibfield  {author} {\bibinfo {author} {\bibfnamefont {M.~F.}\ \bibnamefont
			{Lin}}, \bibinfo {author} {\bibfnamefont {M.~A.}\ \bibnamefont {Verkamp}},
		\bibinfo {author} {\bibfnamefont {J.}~\bibnamefont {Leveillee}}, \bibinfo
		{author} {\bibfnamefont {E.~S.}\ \bibnamefont {Ryland}}, \bibinfo {author}
		{\bibfnamefont {K.}~\bibnamefont {Benke}}, \bibinfo {author} {\bibfnamefont
			{K.}~\bibnamefont {Zhang}}, \bibinfo {author} {\bibfnamefont
			{C.}~\bibnamefont {Weninger}}, \bibinfo {author} {\bibfnamefont
			{X.}~\bibnamefont {Shen}}, \bibinfo {author} {\bibfnamefont {R.}~\bibnamefont
			{Li}}, \bibinfo {author} {\bibfnamefont {D.}~\bibnamefont {Fritz}}, \bibinfo
		{author} {\bibfnamefont {U.}~\bibnamefont {Bergmann}}, \bibinfo {author}
		{\bibfnamefont {X.}~\bibnamefont {Wang}}, \bibinfo {author} {\bibfnamefont
			{A.}~\bibnamefont {Schleife}},\ and\ \bibinfo {author} {\bibfnamefont
			{J.}~\bibnamefont {{Vura-Weis}}},\ }\href
	{https://doi.org/10.1021/acs.jpcc.7b11147} {\bibfield  {journal} {\bibinfo
			{journal} {J. Phys. Chem. C}\ }\textbf {\bibinfo {volume} {121}},\ \bibinfo
		{pages} {27886} (\bibinfo {year} {2017})},\ \Eprint
	{https://arxiv.org/abs/1703.03135} {arXiv:1703.03135} \BibitemShut {NoStop}%
	\bibitem [{\citenamefont {Verkamp}\ \emph {et~al.}(2019)\citenamefont
		{Verkamp}, \citenamefont {Leveillee}, \citenamefont {Sharma}, \citenamefont
		{Schleife},\ and\ \citenamefont
		{{Vura-Weis}}}]{verkampBottleneckFreeHotHole2019}%
	\BibitemOpen
	\bibfield  {author} {\bibinfo {author} {\bibfnamefont {M.~A.}\ \bibnamefont
			{Verkamp}}, \bibinfo {author} {\bibfnamefont {J.}~\bibnamefont {Leveillee}},
		\bibinfo {author} {\bibfnamefont {A.}~\bibnamefont {Sharma}}, \bibinfo
		{author} {\bibfnamefont {A.}~\bibnamefont {Schleife}},\ and\ \bibinfo
		{author} {\bibfnamefont {J.}~\bibnamefont {{Vura-Weis}}},\ }\bibfield
	{journal} {\bibinfo  {journal} {ChemRxiv}\ }\href
	{https://doi.org/10.26434/chemrxiv.8323289.v1} {10.26434/chemrxiv.8323289.v1}
	(\bibinfo {year} {2019})\BibitemShut {NoStop}%
	\bibitem [{\citenamefont {Carneiro}\ \emph {et~al.}(2017)\citenamefont
		{Carneiro}, \citenamefont {Cushing}, \citenamefont {Liu}, \citenamefont {Su},
		\citenamefont {Yang}, \citenamefont {Alivisatos},\ and\ \citenamefont
		{Leone}}]{Carneiro2017}%
	\BibitemOpen
	\bibfield  {author} {\bibinfo {author} {\bibfnamefont {L.~M.}\ \bibnamefont
			{Carneiro}}, \bibinfo {author} {\bibfnamefont {S.~K.}\ \bibnamefont
			{Cushing}}, \bibinfo {author} {\bibfnamefont {C.}~\bibnamefont {Liu}},
		\bibinfo {author} {\bibfnamefont {Y.}~\bibnamefont {Su}}, \bibinfo {author}
		{\bibfnamefont {P.}~\bibnamefont {Yang}}, \bibinfo {author} {\bibfnamefont
			{A.~P.}\ \bibnamefont {Alivisatos}},\ and\ \bibinfo {author} {\bibfnamefont
			{S.~R.}\ \bibnamefont {Leone}},\ }\href {https://doi.org/10.1038/nmat4936}
	{\bibfield  {journal} {\bibinfo  {journal} {Nat. Mater.}\ }\textbf {\bibinfo
			{volume} {16}},\ \bibinfo {pages} {819} (\bibinfo {year} {2017})}\BibitemShut
	{NoStop}%
	\bibitem [{\citenamefont {Cushing}\ \emph {et~al.}(2018)\citenamefont
		{Cushing}, \citenamefont {Z{\"u}rch}, \citenamefont {Kraus}, \citenamefont
		{Carneiro}, \citenamefont {Lee}, \citenamefont {Chang}, \citenamefont
		{Kaplan},\ and\ \citenamefont {Leone}}]{Cushing2018}%
	\BibitemOpen
	\bibfield  {author} {\bibinfo {author} {\bibfnamefont {S.~K.}\ \bibnamefont
			{Cushing}}, \bibinfo {author} {\bibfnamefont {M.}~\bibnamefont {Z{\"u}rch}},
		\bibinfo {author} {\bibfnamefont {P.~M.}\ \bibnamefont {Kraus}}, \bibinfo
		{author} {\bibfnamefont {L.~M.}\ \bibnamefont {Carneiro}}, \bibinfo {author}
		{\bibfnamefont {A.}~\bibnamefont {Lee}}, \bibinfo {author} {\bibfnamefont
			{H.-T.}\ \bibnamefont {Chang}}, \bibinfo {author} {\bibfnamefont {C.~J.}\
			\bibnamefont {Kaplan}},\ and\ \bibinfo {author} {\bibfnamefont {S.~R.}\
			\bibnamefont {Leone}},\ }\href {https://doi.org/10.1063/1.5038015} {\bibfield
		{journal} {\bibinfo  {journal} {Struct. Dyn.}\ }\textbf {\bibinfo {volume}
			{5}},\ \bibinfo {pages} {054302} (\bibinfo {year} {2018})}\BibitemShut
	{NoStop}%
	\bibitem [{\citenamefont {Porter}\ \emph {et~al.}(2018)\citenamefont {Porter},
		\citenamefont {Cushing}, \citenamefont {Carneiro}, \citenamefont {Lee},
		\citenamefont {Ondry}, \citenamefont {Dahl}, \citenamefont {Chang},
		\citenamefont {Alivisatos},\ and\ \citenamefont {Leone}}]{Porter2018}%
	\BibitemOpen
	\bibfield  {author} {\bibinfo {author} {\bibfnamefont {I.~J.}\ \bibnamefont
			{Porter}}, \bibinfo {author} {\bibfnamefont {S.~K.}\ \bibnamefont {Cushing}},
		\bibinfo {author} {\bibfnamefont {L.~M.}\ \bibnamefont {Carneiro}}, \bibinfo
		{author} {\bibfnamefont {A.}~\bibnamefont {Lee}}, \bibinfo {author}
		{\bibfnamefont {J.~C.}\ \bibnamefont {Ondry}}, \bibinfo {author}
		{\bibfnamefont {J.~C.}\ \bibnamefont {Dahl}}, \bibinfo {author}
		{\bibfnamefont {H.-T.}\ \bibnamefont {Chang}}, \bibinfo {author}
		{\bibfnamefont {A.~P.}\ \bibnamefont {Alivisatos}},\ and\ \bibinfo {author}
		{\bibfnamefont {S.~R.}\ \bibnamefont {Leone}},\ }\href
	{https://doi.org/10.1021/acs.jpclett.8b01525} {\bibfield  {journal} {\bibinfo
			{journal} {J. Phys. Chem. Lett.}\ }\textbf {\bibinfo {volume} {9}},\
		\bibinfo {pages} {4120} (\bibinfo {year} {2018})}\BibitemShut {NoStop}%
	\bibitem [{\citenamefont {Cushing}\ \emph {et~al.}(2019)\citenamefont
		{Cushing}, \citenamefont {Lee}, \citenamefont {Porter}, \citenamefont
		{Carneiro}, \citenamefont {Chang}, \citenamefont {Z{\"u}rch},\ and\
		\citenamefont {Leone}}]{Cushing2019}%
	\BibitemOpen
	\bibfield  {author} {\bibinfo {author} {\bibfnamefont {S.~K.}\ \bibnamefont
			{Cushing}}, \bibinfo {author} {\bibfnamefont {A.}~\bibnamefont {Lee}},
		\bibinfo {author} {\bibfnamefont {I.~J.}\ \bibnamefont {Porter}}, \bibinfo
		{author} {\bibfnamefont {L.~M.}\ \bibnamefont {Carneiro}}, \bibinfo {author}
		{\bibfnamefont {H.-T.}\ \bibnamefont {Chang}}, \bibinfo {author}
		{\bibfnamefont {M.}~\bibnamefont {Z{\"u}rch}},\ and\ \bibinfo {author}
		{\bibfnamefont {S.~R.}\ \bibnamefont {Leone}},\ }\href
	{https://doi.org/10.1021/acs.jpcc.8b10887} {\bibfield  {journal} {\bibinfo
			{journal} {J. Phys. Chem. C}\ }\textbf {\bibinfo {volume} {123}},\ \bibinfo
		{pages} {3343} (\bibinfo {year} {2019})}\BibitemShut {NoStop}%
	\bibitem [{\citenamefont {Cushing}\ \emph {et~al.}(2020)\citenamefont
		{Cushing}, \citenamefont {Porter}, \citenamefont {de~Roulet}, \citenamefont
		{Lee}, \citenamefont {Marsh}, \citenamefont {Szoke}, \citenamefont {Vaida},\
		and\ \citenamefont {Leone}}]{cushingLayerresolvedUltrafastExtreme2020}%
	\BibitemOpen
	\bibfield  {author} {\bibinfo {author} {\bibfnamefont {S.~K.}\ \bibnamefont
			{Cushing}}, \bibinfo {author} {\bibfnamefont {I.~J.}\ \bibnamefont {Porter}},
		\bibinfo {author} {\bibfnamefont {B.~R.}\ \bibnamefont {de~Roulet}}, \bibinfo
		{author} {\bibfnamefont {A.}~\bibnamefont {Lee}}, \bibinfo {author}
		{\bibfnamefont {B.~M.}\ \bibnamefont {Marsh}}, \bibinfo {author}
		{\bibfnamefont {S.}~\bibnamefont {Szoke}}, \bibinfo {author} {\bibfnamefont
			{M.~E.}\ \bibnamefont {Vaida}},\ and\ \bibinfo {author} {\bibfnamefont
			{S.~R.}\ \bibnamefont {Leone}},\ }\href
	{https://doi.org/10.1126/sciadv.aay6650} {\bibfield  {journal} {\bibinfo
			{journal} {Sci. Adv.}\ }\textbf {\bibinfo {volume} {6}},\ \bibinfo {pages}
		{eaay6650} (\bibinfo {year} {2020})}\BibitemShut {NoStop}%
	\bibitem [{\citenamefont {Volkov}\ \emph {et~al.}(2019)\citenamefont {Volkov},
		\citenamefont {Sato}, \citenamefont {Schlaepfer}, \citenamefont {Kasmi},
		\citenamefont {Hartmann}, \citenamefont {Lucchini}, \citenamefont {Gallmann},
		\citenamefont {Rubio},\ and\ \citenamefont
		{Keller}}]{volkovAttosecondScreeningDynamics2019a}%
	\BibitemOpen
	\bibfield  {author} {\bibinfo {author} {\bibfnamefont {M.}~\bibnamefont
			{Volkov}}, \bibinfo {author} {\bibfnamefont {S.~A.}\ \bibnamefont {Sato}},
		\bibinfo {author} {\bibfnamefont {F.}~\bibnamefont {Schlaepfer}}, \bibinfo
		{author} {\bibfnamefont {L.}~\bibnamefont {Kasmi}}, \bibinfo {author}
		{\bibfnamefont {N.}~\bibnamefont {Hartmann}}, \bibinfo {author}
		{\bibfnamefont {M.}~\bibnamefont {Lucchini}}, \bibinfo {author}
		{\bibfnamefont {L.}~\bibnamefont {Gallmann}}, \bibinfo {author}
		{\bibfnamefont {A.}~\bibnamefont {Rubio}},\ and\ \bibinfo {author}
		{\bibfnamefont {U.}~\bibnamefont {Keller}},\ }\href
	{https://doi.org/10.1038/s41567-019-0602-9} {\bibfield  {journal} {\bibinfo
			{journal} {Nat. Phys.}\ }\textbf {\bibinfo {volume} {15}},\ \bibinfo {pages}
		{1145} (\bibinfo {year} {2019})},\ \Eprint {https://arxiv.org/abs/1811.00801}
	{arXiv:1811.00801} \BibitemShut {NoStop}%
	\bibitem [{\citenamefont {Rehr}(2003)}]{Rehr2003}%
	\BibitemOpen
	\bibfield  {author} {\bibinfo {author} {\bibfnamefont {J.~J.}\ \bibnamefont
			{Rehr}},\ }\href {https://doi.org/10.1023/A:1026277520940} {\bibfield
		{journal} {\bibinfo  {journal} {Found. Phys.}\ }\textbf {\bibinfo {volume}
			{33}},\ \bibinfo {pages} {1735} (\bibinfo {year} {2003})}\BibitemShut
	{NoStop}%
	\bibitem [{\citenamefont {Attar}\ \emph {et~al.}(2020)\citenamefont {Attar},
		\citenamefont {Chang}, \citenamefont {Britz}, \citenamefont {Zhang},
		\citenamefont {Lin}, \citenamefont {Krishnamoorthy}, \citenamefont {Linker},
		\citenamefont {Fritz}, \citenamefont {Neumark}, \citenamefont {Kalia},
		\citenamefont {Nakano}, \citenamefont {Ajayan}, \citenamefont {Vashishta},
		\citenamefont {Bergmann},\ and\ \citenamefont
		{Leone}}]{attarSimultaneousObservationCarrierSpecific2020}%
	\BibitemOpen
	\bibfield  {author} {\bibinfo {author} {\bibfnamefont {A.~R.}\ \bibnamefont
			{Attar}}, \bibinfo {author} {\bibfnamefont {H.-T.}\ \bibnamefont {Chang}},
		\bibinfo {author} {\bibfnamefont {A.}~\bibnamefont {Britz}}, \bibinfo
		{author} {\bibfnamefont {X.}~\bibnamefont {Zhang}}, \bibinfo {author}
		{\bibfnamefont {M.-F.}\ \bibnamefont {Lin}}, \bibinfo {author} {\bibfnamefont
			{A.}~\bibnamefont {Krishnamoorthy}}, \bibinfo {author} {\bibfnamefont
			{T.}~\bibnamefont {Linker}}, \bibinfo {author} {\bibfnamefont
			{D.}~\bibnamefont {Fritz}}, \bibinfo {author} {\bibfnamefont {D.~M.}\
			\bibnamefont {Neumark}}, \bibinfo {author} {\bibfnamefont {R.~K.}\
			\bibnamefont {Kalia}}, \bibinfo {author} {\bibfnamefont {A.}~\bibnamefont
			{Nakano}}, \bibinfo {author} {\bibfnamefont {P.}~\bibnamefont {Ajayan}},
		\bibinfo {author} {\bibfnamefont {P.}~\bibnamefont {Vashishta}}, \bibinfo
		{author} {\bibfnamefont {U.}~\bibnamefont {Bergmann}},\ and\ \bibinfo
		{author} {\bibfnamefont {S.~R.}\ \bibnamefont {Leone}},\ }\href
	{https://doi.org/10.1021/acsnano.0c06988} {\bibfield  {journal} {\bibinfo
			{journal} {ACS Nano}\ }\textbf {\bibinfo {volume} {14}},\ \bibinfo {pages}
		{15829} (\bibinfo {year} {2020})},\ \Eprint
	{https://arxiv.org/abs/2009.00721} {arXiv:2009.00721} \BibitemShut {NoStop}%
	\bibitem [{\citenamefont {Mahan}(2000)}]{mahanManyParticlePhysics2000}%
	\BibitemOpen
	\bibfield  {author} {\bibinfo {author} {\bibfnamefont {G.~D.}\ \bibnamefont
			{Mahan}},\ }\href {https://doi.org/10.1007/978-1-4757-5714-9} {\emph
		{\bibinfo {title} {Many-{{Particle Physics}}}}}\ (\bibinfo  {publisher}
	{{Springer US}},\ \bibinfo {address} {{Boston, MA}},\ \bibinfo {year}
	{2000})\BibitemShut {NoStop}%
	\bibitem [{\citenamefont {Mahan}(1967)}]{mahanExcitonsMetalsInfinite1967}%
	\BibitemOpen
	\bibfield  {author} {\bibinfo {author} {\bibfnamefont {G.~D.}\ \bibnamefont
			{Mahan}},\ }\href {https://doi.org/10.1103/PhysRev.163.612} {\bibfield
		{journal} {\bibinfo  {journal} {Phys. Rev.}\ }\textbf {\bibinfo {volume}
			{163}},\ \bibinfo {pages} {612} (\bibinfo {year} {1967})}\BibitemShut
	{NoStop}%
	\bibitem [{\citenamefont {Roulet}\ \emph {et~al.}(1969)\citenamefont {Roulet},
		\citenamefont {Gavoret},\ and\ \citenamefont
		{Nozi{\`e}res}}]{rouletSingularitiesXRayAbsorption1969}%
	\BibitemOpen
	\bibfield  {author} {\bibinfo {author} {\bibfnamefont {B.}~\bibnamefont
			{Roulet}}, \bibinfo {author} {\bibfnamefont {J.}~\bibnamefont {Gavoret}},\
		and\ \bibinfo {author} {\bibfnamefont {P.}~\bibnamefont {Nozi{\`e}res}},\
	}\href {https://doi.org/10.1103/PhysRev.178.1072} {\bibfield  {journal}
		{\bibinfo  {journal} {Phys. Rev.}\ }\textbf {\bibinfo {volume} {178}},\
		\bibinfo {pages} {1072} (\bibinfo {year} {1969})}\BibitemShut {NoStop}%
	\bibitem [{\citenamefont {Nozi{\`e}res}\ and\ \citenamefont
		{De~Dominicis}(1969)}]{nozieresSingularitiesXRayAbsorption1969}%
	\BibitemOpen
	\bibfield  {author} {\bibinfo {author} {\bibfnamefont {P.}~\bibnamefont
			{Nozi{\`e}res}}\ and\ \bibinfo {author} {\bibfnamefont {C.~T.}\ \bibnamefont
			{De~Dominicis}},\ }\href {https://doi.org/10.1103/PhysRev.178.1097}
	{\bibfield  {journal} {\bibinfo  {journal} {Phys. Rev.}\ }\textbf {\bibinfo
			{volume} {178}},\ \bibinfo {pages} {1097} (\bibinfo {year}
		{1969})}\BibitemShut {NoStop}%
	\bibitem [{\citenamefont {Mahan}(1975)}]{mahanCollectiveExcitationsXray1975}%
	\BibitemOpen
	\bibfield  {author} {\bibinfo {author} {\bibfnamefont {G.~D.}\ \bibnamefont
			{Mahan}},\ }\href {https://doi.org/10.1103/PhysRevB.11.4814} {\bibfield
		{journal} {\bibinfo  {journal} {Phys. Rev. B}\ }\textbf {\bibinfo {volume}
			{11}},\ \bibinfo {pages} {4814} (\bibinfo {year} {1975})}\BibitemShut
	{NoStop}%
	\bibitem [{\citenamefont {Ohtaka}\ and\ \citenamefont
		{Tanabe}(1990)}]{Ohtaka1990}%
	\BibitemOpen
	\bibfield  {author} {\bibinfo {author} {\bibfnamefont {K.}~\bibnamefont
			{Ohtaka}}\ and\ \bibinfo {author} {\bibfnamefont {Y.}~\bibnamefont
			{Tanabe}},\ }\href {https://doi.org/10.1103/RevModPhys.62.929} {\bibfield
		{journal} {\bibinfo  {journal} {Rev. Mod. Phys.}\ }\textbf {\bibinfo {volume}
			{62}},\ \bibinfo {pages} {929} (\bibinfo {year} {1990})}\BibitemShut
	{NoStop}%
	\bibitem [{\citenamefont {Hohlfeld}\ \emph {et~al.}(1997)\citenamefont
		{Hohlfeld}, \citenamefont {Matthias}, \citenamefont {Knorren},\ and\
		\citenamefont {Bennemann}}]{hohlfeldNonequilibriumMagnetizationDynamics1997}%
	\BibitemOpen
	\bibfield  {author} {\bibinfo {author} {\bibfnamefont {J.}~\bibnamefont
			{Hohlfeld}}, \bibinfo {author} {\bibfnamefont {E.}~\bibnamefont {Matthias}},
		\bibinfo {author} {\bibfnamefont {R.}~\bibnamefont {Knorren}},\ and\ \bibinfo
		{author} {\bibfnamefont {K.~H.}\ \bibnamefont {Bennemann}},\ }\href
	{https://doi.org/10.1103/PhysRevLett.78.4861} {\bibfield  {journal} {\bibinfo
			{journal} {Phys. Rev. Lett.}\ }\textbf {\bibinfo {volume} {78}},\ \bibinfo
		{pages} {4861} (\bibinfo {year} {1997})}\BibitemShut {NoStop}%
	\bibitem [{\citenamefont {Conrad}\ \emph {et~al.}(1999)\citenamefont {Conrad},
		\citenamefont {G{\"u}dde}, \citenamefont {J{\"a}hnke},\ and\ \citenamefont
		{Matthias}}]{conradUltrafastElectronMagnetization1999}%
	\BibitemOpen
	\bibfield  {author} {\bibinfo {author} {\bibfnamefont {U.}~\bibnamefont
			{Conrad}}, \bibinfo {author} {\bibfnamefont {J.}~\bibnamefont {G{\"u}dde}},
		\bibinfo {author} {\bibfnamefont {V.}~\bibnamefont {J{\"a}hnke}},\ and\
		\bibinfo {author} {\bibfnamefont {E.}~\bibnamefont {Matthias}},\ }\href
	{https://doi.org/10.1007/s003400050658} {\bibfield  {journal} {\bibinfo
			{journal} {Appl Phys B}\ }\textbf {\bibinfo {volume} {68}},\ \bibinfo {pages}
		{511} (\bibinfo {year} {1999})}\BibitemShut {NoStop}%
	\bibitem [{\citenamefont {Regensburger}\ \emph {et~al.}(2000)\citenamefont
		{Regensburger}, \citenamefont {Vollmer},\ and\ \citenamefont
		{Kirschner}}]{regensburgerTimeresolvedMagnetizationinducedSecondharmonic2000}%
	\BibitemOpen
	\bibfield  {author} {\bibinfo {author} {\bibfnamefont {H.}~\bibnamefont
			{Regensburger}}, \bibinfo {author} {\bibfnamefont {R.}~\bibnamefont
			{Vollmer}},\ and\ \bibinfo {author} {\bibfnamefont {J.}~\bibnamefont
			{Kirschner}},\ }\href {https://doi.org/10.1103/PhysRevB.61.14716} {\bibfield
		{journal} {\bibinfo  {journal} {Phys. Rev. B}\ }\textbf {\bibinfo {volume}
			{61}},\ \bibinfo {pages} {14716} (\bibinfo {year} {2000})}\BibitemShut
	{NoStop}%
	\bibitem [{\citenamefont {Melnikov}\ \emph {et~al.}(2002)\citenamefont
		{Melnikov}, \citenamefont {G{\"u}dde},\ and\ \citenamefont
		{Matthias}}]{melnikovDemagnetizationFollowingOptical2002}%
	\BibitemOpen
	\bibfield  {author} {\bibinfo {author} {\bibfnamefont {A.}~\bibnamefont
			{Melnikov}}, \bibinfo {author} {\bibfnamefont {J.}~\bibnamefont
			{G{\"u}dde}},\ and\ \bibinfo {author} {\bibfnamefont {E.}~\bibnamefont
			{Matthias}},\ }\href {https://doi.org/10.1007/s00340-002-0927-3} {\bibfield
		{journal} {\bibinfo  {journal} {Appl Phys B}\ }\textbf {\bibinfo {volume}
			{74}},\ \bibinfo {pages} {735} (\bibinfo {year} {2002})}\BibitemShut
	{NoStop}%
	\bibitem [{\citenamefont {van Kampen}\ \emph {et~al.}(2005)\citenamefont {van
			Kampen}, \citenamefont {Kohlhepp}, \citenamefont {de~Jonge}, \citenamefont
		{Koopmans},\ and\ \citenamefont {Coehoorn}}]{Kampen2005}%
	\BibitemOpen
	\bibfield  {author} {\bibinfo {author} {\bibfnamefont {M.}~\bibnamefont {van
				Kampen}}, \bibinfo {author} {\bibfnamefont {J.~T.}\ \bibnamefont {Kohlhepp}},
		\bibinfo {author} {\bibfnamefont {W.~J.~M.}\ \bibnamefont {de~Jonge}},
		\bibinfo {author} {\bibfnamefont {B.}~\bibnamefont {Koopmans}},\ and\
		\bibinfo {author} {\bibfnamefont {R.}~\bibnamefont {Coehoorn}},\ }\href
	{https://doi.org/10.1088/0953-8984/17/43/004} {\bibfield  {journal} {\bibinfo
			{journal} {J. Phys. Condens. Matter}\ }\textbf {\bibinfo {volume} {17}},\
		\bibinfo {pages} {6823} (\bibinfo {year} {2005})}\BibitemShut {NoStop}%
	\bibitem [{\citenamefont {You}\ \emph {et~al.}(2018)\citenamefont {You},
		\citenamefont {Tengdin}, \citenamefont {Chen}, \citenamefont {Shi},
		\citenamefont {Zusin}, \citenamefont {Zhang}, \citenamefont {Gentry},
		\citenamefont {Blonsky}, \citenamefont {Keller}, \citenamefont {Oppeneer},
		\citenamefont {Kapteyn}, \citenamefont {Tao},\ and\ \citenamefont
		{Murnane}}]{youRevealingNatureUltrafast2018}%
	\BibitemOpen
	\bibfield  {author} {\bibinfo {author} {\bibfnamefont {W.}~\bibnamefont
			{You}}, \bibinfo {author} {\bibfnamefont {P.}~\bibnamefont {Tengdin}},
		\bibinfo {author} {\bibfnamefont {C.}~\bibnamefont {Chen}}, \bibinfo {author}
		{\bibfnamefont {X.}~\bibnamefont {Shi}}, \bibinfo {author} {\bibfnamefont
			{D.}~\bibnamefont {Zusin}}, \bibinfo {author} {\bibfnamefont
			{Y.}~\bibnamefont {Zhang}}, \bibinfo {author} {\bibfnamefont
			{C.}~\bibnamefont {Gentry}}, \bibinfo {author} {\bibfnamefont
			{A.}~\bibnamefont {Blonsky}}, \bibinfo {author} {\bibfnamefont
			{M.}~\bibnamefont {Keller}}, \bibinfo {author} {\bibfnamefont {P.~M.}\
			\bibnamefont {Oppeneer}}, \bibinfo {author} {\bibfnamefont {H.}~\bibnamefont
			{Kapteyn}}, \bibinfo {author} {\bibfnamefont {Z.}~\bibnamefont {Tao}},\ and\
		\bibinfo {author} {\bibfnamefont {M.}~\bibnamefont {Murnane}},\ }\href
	{https://doi.org/10.1103/PhysRevLett.121.077204} {\bibfield  {journal}
		{\bibinfo  {journal} {Phys. Rev. Lett.}\ }\textbf {\bibinfo {volume} {121}},\
		\bibinfo {pages} {077204} (\bibinfo {year} {2018})}\BibitemShut {NoStop}%
	\bibitem [{\citenamefont {Tengdin}\ \emph {et~al.}(2018)\citenamefont
		{Tengdin}, \citenamefont {You}, \citenamefont {Chen}, \citenamefont {Shi},
		\citenamefont {Zusin}, \citenamefont {Zhang}, \citenamefont {Gentry},
		\citenamefont {Blonsky}, \citenamefont {Keller}, \citenamefont {Oppeneer},
		\citenamefont {Kapteyn}, \citenamefont {Tao},\ and\ \citenamefont
		{Murnane}}]{Tengdin2018}%
	\BibitemOpen
	\bibfield  {author} {\bibinfo {author} {\bibfnamefont {P.}~\bibnamefont
			{Tengdin}}, \bibinfo {author} {\bibfnamefont {W.}~\bibnamefont {You}},
		\bibinfo {author} {\bibfnamefont {C.}~\bibnamefont {Chen}}, \bibinfo {author}
		{\bibfnamefont {X.}~\bibnamefont {Shi}}, \bibinfo {author} {\bibfnamefont
			{D.}~\bibnamefont {Zusin}}, \bibinfo {author} {\bibfnamefont
			{Y.}~\bibnamefont {Zhang}}, \bibinfo {author} {\bibfnamefont
			{C.}~\bibnamefont {Gentry}}, \bibinfo {author} {\bibfnamefont
			{A.}~\bibnamefont {Blonsky}}, \bibinfo {author} {\bibfnamefont
			{M.}~\bibnamefont {Keller}}, \bibinfo {author} {\bibfnamefont {P.~M.}\
			\bibnamefont {Oppeneer}}, \bibinfo {author} {\bibfnamefont {H.~C.}\
			\bibnamefont {Kapteyn}}, \bibinfo {author} {\bibfnamefont {Z.}~\bibnamefont
			{Tao}},\ and\ \bibinfo {author} {\bibfnamefont {M.~M.}\ \bibnamefont
			{Murnane}},\ }\href {https://doi.org/10.1126/sciadv.aap9744} {\bibfield
		{journal} {\bibinfo  {journal} {Sci. Adv.}\ }\textbf {\bibinfo {volume}
			{4}},\ \bibinfo {pages} {eaap9744} (\bibinfo {year} {2018})}\BibitemShut
	{NoStop}%
	\bibitem [{\citenamefont {Maldonado}\ \emph {et~al.}(2020)\citenamefont
		{Maldonado}, \citenamefont {Chase}, \citenamefont {Reid}, \citenamefont
		{Shen}, \citenamefont {Li}, \citenamefont {Carva}, \citenamefont {Payer},
		\citenamefont {{Horn von Hoegen}}, \citenamefont {{Sokolowski-Tinten}},
		\citenamefont {Wang}, \citenamefont {Oppeneer},\ and\ \citenamefont
		{D{\"u}rr}}]{maldonadoTrackingUltrafastNonequilibrium2020}%
	\BibitemOpen
	\bibfield  {author} {\bibinfo {author} {\bibfnamefont {P.}~\bibnamefont
			{Maldonado}}, \bibinfo {author} {\bibfnamefont {T.}~\bibnamefont {Chase}},
		\bibinfo {author} {\bibfnamefont {A.~H.}\ \bibnamefont {Reid}}, \bibinfo
		{author} {\bibfnamefont {X.}~\bibnamefont {Shen}}, \bibinfo {author}
		{\bibfnamefont {R.~K.}\ \bibnamefont {Li}}, \bibinfo {author} {\bibfnamefont
			{K.}~\bibnamefont {Carva}}, \bibinfo {author} {\bibfnamefont
			{T.}~\bibnamefont {Payer}}, \bibinfo {author} {\bibfnamefont
			{M.}~\bibnamefont {{Horn von Hoegen}}}, \bibinfo {author} {\bibfnamefont
			{K.}~\bibnamefont {{Sokolowski-Tinten}}}, \bibinfo {author} {\bibfnamefont
			{X.~J.}\ \bibnamefont {Wang}}, \bibinfo {author} {\bibfnamefont {P.~M.}\
			\bibnamefont {Oppeneer}},\ and\ \bibinfo {author} {\bibfnamefont {H.~A.}\
			\bibnamefont {D{\"u}rr}},\ }\href
	{https://doi.org/10.1103/PhysRevB.101.100302} {\bibfield  {journal} {\bibinfo
			{journal} {Phys. Rev. B}\ }\textbf {\bibinfo {volume} {101}},\ \bibinfo
		{pages} {100302} (\bibinfo {year} {2020})}\BibitemShut {NoStop}%
	\bibitem [{\citenamefont {Mueller}\ and\ \citenamefont
		{Rethfeld}(2013)}]{Mueller2013}%
	\BibitemOpen
	\bibfield  {author} {\bibinfo {author} {\bibfnamefont {B.~Y.}\ \bibnamefont
			{Mueller}}\ and\ \bibinfo {author} {\bibfnamefont {B.}~\bibnamefont
			{Rethfeld}},\ }\href {https://doi.org/10.1103/PhysRevB.87.035139} {\bibfield
		{journal} {\bibinfo  {journal} {Phys. Rev. B}\ }\textbf {\bibinfo {volume}
			{87}},\ \bibinfo {pages} {035139} (\bibinfo {year} {2013})}\BibitemShut
	{NoStop}%
	\bibitem [{\citenamefont {Dietz}\ \emph {et~al.}(1980)\citenamefont {Dietz},
		\citenamefont {McRae},\ and\ \citenamefont {Weaver}}]{Dietz1980}%
	\BibitemOpen
	\bibfield  {author} {\bibinfo {author} {\bibfnamefont {R.~E.}\ \bibnamefont
			{Dietz}}, \bibinfo {author} {\bibfnamefont {E.~G.}\ \bibnamefont {McRae}},\
		and\ \bibinfo {author} {\bibfnamefont {J.~H.}\ \bibnamefont {Weaver}},\
	}\href {https://doi.org/10.1103/PhysRevB.21.2229} {\bibfield  {journal}
		{\bibinfo  {journal} {Phys. Rev. B}\ }\textbf {\bibinfo {volume} {21}},\
		\bibinfo {pages} {2229} (\bibinfo {year} {1980})}\BibitemShut {NoStop}%
	\bibitem [{\citenamefont {Carva}\ \emph {et~al.}(2009)\citenamefont {Carva},
		\citenamefont {Legut},\ and\ \citenamefont
		{Oppeneer}}]{carvaInfluenceLaserexcitedElectron2009}%
	\BibitemOpen
	\bibfield  {author} {\bibinfo {author} {\bibfnamefont {K.}~\bibnamefont
			{Carva}}, \bibinfo {author} {\bibfnamefont {D.}~\bibnamefont {Legut}},\ and\
		\bibinfo {author} {\bibfnamefont {P.~M.}\ \bibnamefont {Oppeneer}},\ }\href
	{https://doi.org/10.1209/0295-5075/86/57002} {\bibfield  {journal} {\bibinfo
			{journal} {Europhys. Lett.}\ }\textbf {\bibinfo {volume} {86}},\ \bibinfo
		{pages} {57002} (\bibinfo {year} {2009})}\BibitemShut {NoStop}%
	\bibitem [{\citenamefont {Dietz}\ \emph {et~al.}(1974)\citenamefont {Dietz},
		\citenamefont {McRae}, \citenamefont {Yafet},\ and\ \citenamefont
		{Caldwell}}]{dietzLineShapeExcitation1974}%
	\BibitemOpen
	\bibfield  {author} {\bibinfo {author} {\bibfnamefont {R.~E.}\ \bibnamefont
			{Dietz}}, \bibinfo {author} {\bibfnamefont {E.~G.}\ \bibnamefont {McRae}},
		\bibinfo {author} {\bibfnamefont {Y.}~\bibnamefont {Yafet}},\ and\ \bibinfo
		{author} {\bibfnamefont {C.~W.}\ \bibnamefont {Caldwell}},\ }\href
	{https://doi.org/10.1103/PhysRevLett.33.1372} {\bibfield  {journal} {\bibinfo
			{journal} {Phys. Rev. Lett.}\ }\textbf {\bibinfo {volume} {33}},\ \bibinfo
		{pages} {1372} (\bibinfo {year} {1974})}\BibitemShut {NoStop}%
	\bibitem [{\citenamefont {Davis}\ and\ \citenamefont
		{Feldkamp}(1976)}]{davisInterpretation3pcoreexcitationSpectra1976}%
	\BibitemOpen
	\bibfield  {author} {\bibinfo {author} {\bibfnamefont {L.~C.}\ \bibnamefont
			{Davis}}\ and\ \bibinfo {author} {\bibfnamefont {L.~A.}\ \bibnamefont
			{Feldkamp}},\ }\href {https://doi.org/10.1016/0038-1098(76)91179-0}
	{\bibfield  {journal} {\bibinfo  {journal} {Solid State Commun.}\ }\textbf
		{\bibinfo {volume} {19}},\ \bibinfo {pages} {413} (\bibinfo {year}
		{1976})}\BibitemShut {NoStop}%
	\bibitem [{\citenamefont {Valencia}\ \emph {et~al.}(2010)\citenamefont
		{Valencia}, \citenamefont {Kleibert}, \citenamefont {Gaupp}, \citenamefont
		{Rusz}, \citenamefont {Legut}, \citenamefont {Bansmann}, \citenamefont
		{Gudat},\ and\ \citenamefont
		{Oppeneer}}]{valenciaQuadraticXRayMagnetoOptical2010}%
	\BibitemOpen
	\bibfield  {author} {\bibinfo {author} {\bibfnamefont {S.}~\bibnamefont
			{Valencia}}, \bibinfo {author} {\bibfnamefont {A.}~\bibnamefont {Kleibert}},
		\bibinfo {author} {\bibfnamefont {A.}~\bibnamefont {Gaupp}}, \bibinfo
		{author} {\bibfnamefont {J.}~\bibnamefont {Rusz}}, \bibinfo {author}
		{\bibfnamefont {D.}~\bibnamefont {Legut}}, \bibinfo {author} {\bibfnamefont
			{J.}~\bibnamefont {Bansmann}}, \bibinfo {author} {\bibfnamefont
			{W.}~\bibnamefont {Gudat}},\ and\ \bibinfo {author} {\bibfnamefont {P.~M.}\
			\bibnamefont {Oppeneer}},\ }\href
	{https://doi.org/10.1103/PhysRevLett.104.187401} {\bibfield  {journal}
		{\bibinfo  {journal} {Phys. Rev. Lett.}\ }\textbf {\bibinfo {volume} {104}},\
		\bibinfo {pages} {187401} (\bibinfo {year} {2010})}\BibitemShut {NoStop}%
	\bibitem [{Note1()}]{Note1}%
	\BibitemOpen
	\bibinfo {note} {Tengdin et al. detected thermalized hot electron
		distribution with time-resolved angle-resolved photoemission 24 fs after
		photoexcitation by pulses centered at 780 nm with fluence of <6 mJ/cm$^{2}$
		\protect \citet {Tengdin2018}. The reported pulse duration in Ref. \protect
		\citet {Tengdin2018} is 28 fs. As the rate of electron scattering increases
		with carrier temperature, the carrier thermalization time is expected to be
		<30 fs long at fluences used in this study (8 -- 62
		mJ/cm$^{2}$).}\BibitemShut {Stop}%
	\bibitem [{\citenamefont {Almbladh}\ and\ \citenamefont
		{Minnhagen}(1978)}]{almbladhThermalBroadeningCore1978}%
	\BibitemOpen
	\bibfield  {author} {\bibinfo {author} {\bibfnamefont {C.-O.}\ \bibnamefont
			{Almbladh}}\ and\ \bibinfo {author} {\bibfnamefont {P.}~\bibnamefont
			{Minnhagen}},\ }\href {https://doi.org/10.1002/pssb.2220850114} {\bibfield
		{journal} {\bibinfo  {journal} {Phys. Status Solidi B}\ }\textbf {\bibinfo
			{volume} {85}},\ \bibinfo {pages} {135} (\bibinfo {year} {1978})}\BibitemShut
	{NoStop}%
	\bibitem [{\citenamefont {Ohtaka}\ and\ \citenamefont
		{Tanabe}(1983)}]{Ohtaka1983}%
	\BibitemOpen
	\bibfield  {author} {\bibinfo {author} {\bibfnamefont {K.}~\bibnamefont
			{Ohtaka}}\ and\ \bibinfo {author} {\bibfnamefont {Y.}~\bibnamefont
			{Tanabe}},\ }\href {https://doi.org/10.1103/PhysRevB.28.6833} {\bibfield
		{journal} {\bibinfo  {journal} {Phys. Rev. B}\ }\textbf {\bibinfo {volume}
			{28}},\ \bibinfo {pages} {6833} (\bibinfo {year} {1983})}\BibitemShut
	{NoStop}%
	\bibitem [{\citenamefont {Tanabe}\ and\ \citenamefont
		{Ohtaka}(1984)}]{Tanabe1984}%
	\BibitemOpen
	\bibfield  {author} {\bibinfo {author} {\bibfnamefont {Y.}~\bibnamefont
			{Tanabe}}\ and\ \bibinfo {author} {\bibfnamefont {K.}~\bibnamefont
			{Ohtaka}},\ }\href {https://doi.org/10.1103/PhysRevB.29.1653} {\bibfield
		{journal} {\bibinfo  {journal} {Phys. Rev. B}\ }\textbf {\bibinfo {volume}
			{29}},\ \bibinfo {pages} {1653} (\bibinfo {year} {1984})}\BibitemShut
	{NoStop}%
	\bibitem [{\citenamefont {Ohtaka}\ and\ \citenamefont
		{Tanabe}(1984)}]{ohtakaGoldenruleApproachSoftxrayabsorption1984}%
	\BibitemOpen
	\bibfield  {author} {\bibinfo {author} {\bibfnamefont {K.}~\bibnamefont
			{Ohtaka}}\ and\ \bibinfo {author} {\bibfnamefont {Y.}~\bibnamefont
			{Tanabe}},\ }\href {https://doi.org/10.1103/PhysRevB.30.4235} {\bibfield
		{journal} {\bibinfo  {journal} {Phys. Rev. B}\ }\textbf {\bibinfo {volume}
			{30}},\ \bibinfo {pages} {4235} (\bibinfo {year} {1984})}\BibitemShut
	{NoStop}%
	\bibitem [{\citenamefont {Adamjan}\ \emph {et~al.}(1995)\citenamefont
		{Adamjan}, \citenamefont {Ortner}, \citenamefont {Salistra},\ and\
		\citenamefont {Tkachenko}}]{adamjanXrayabsorptionProblemMetals1995}%
	\BibitemOpen
	\bibfield  {author} {\bibinfo {author} {\bibfnamefont {V.~M.}\ \bibnamefont
			{Adamjan}}, \bibinfo {author} {\bibfnamefont {J.}~\bibnamefont {Ortner}},
		\bibinfo {author} {\bibfnamefont {A.~G.}\ \bibnamefont {Salistra}},\ and\
		\bibinfo {author} {\bibfnamefont {I.~M.}\ \bibnamefont {Tkachenko}},\ }\href
	{https://doi.org/10.1103/PhysRevB.52.13827} {\bibfield  {journal} {\bibinfo
			{journal} {Phys. Rev. B}\ }\textbf {\bibinfo {volume} {52}},\ \bibinfo
		{pages} {13827} (\bibinfo {year} {1995})}\BibitemShut {NoStop}%
	\bibitem [{\citenamefont
		{Ortner}(1996)}]{ortnerXrayabsorptionProblemMetals1996}%
	\BibitemOpen
	\bibfield  {author} {\bibinfo {author} {\bibfnamefont {J.}~\bibnamefont
			{Ortner}},\ }\href {https://doi.org/10.1103/PhysRevB.54.4401} {\bibfield
		{journal} {\bibinfo  {journal} {Phys. Rev. B}\ }\textbf {\bibinfo {volume}
			{54}},\ \bibinfo {pages} {4401} (\bibinfo {year} {1996})}\BibitemShut
	{NoStop}%
	\bibitem [{\citenamefont {Ortner}(1997)}]{Ortner1997}%
	\BibitemOpen
	\bibfield  {author} {\bibinfo {author} {\bibfnamefont {J.}~\bibnamefont
			{Ortner}},\ }\href {https://doi.org/10.1016/S0921-4526(97)00336-0} {\bibfield
		{journal} {\bibinfo  {journal} {Phys. B Condens. Matter}\ }\textbf {\bibinfo
			{volume} {239}},\ \bibinfo {pages} {328} (\bibinfo {year}
		{1997})}\BibitemShut {NoStop}%
	\bibitem [{\citenamefont {Johansson}\ and\ \citenamefont
		{M{\aa}rtensson}(1980)}]{johanssonCorelevelBindingenergyShifts1980}%
	\BibitemOpen
	\bibfield  {author} {\bibinfo {author} {\bibfnamefont {B.}~\bibnamefont
			{Johansson}}\ and\ \bibinfo {author} {\bibfnamefont {N.}~\bibnamefont
			{M{\aa}rtensson}},\ }\href {https://doi.org/10.1103/PhysRevB.21.4427}
	{\bibfield  {journal} {\bibinfo  {journal} {Phys. Rev. B}\ }\textbf {\bibinfo
			{volume} {21}},\ \bibinfo {pages} {4427} (\bibinfo {year}
		{1980})}\BibitemShut {NoStop}%
	\bibitem [{\citenamefont {Olson}\ and\ \citenamefont
		{Lynch}(1980{\natexlab{a}})}]{olsonThermomodulatedExafsSoft1980}%
	\BibitemOpen
	\bibfield  {author} {\bibinfo {author} {\bibfnamefont {C.~G.}\ \bibnamefont
			{Olson}}\ and\ \bibinfo {author} {\bibfnamefont {D.~W.}\ \bibnamefont
			{Lynch}},\ }\href {https://doi.org/10.1016/0038-1098(80)90377-4} {\bibfield
		{journal} {\bibinfo  {journal} {Solid State Commun.}\ }\textbf {\bibinfo
			{volume} {36}},\ \bibinfo {pages} {513} (\bibinfo {year}
		{1980}{\natexlab{a}})}\BibitemShut {NoStop}%
	\bibitem [{\citenamefont {Olson}\ and\ \citenamefont
		{Lynch}(1980{\natexlab{b}})}]{olsonTemperatureDependenceM21980}%
	\BibitemOpen
	\bibfield  {author} {\bibinfo {author} {\bibfnamefont {C.~G.}\ \bibnamefont
			{Olson}}\ and\ \bibinfo {author} {\bibfnamefont {D.~W.}\ \bibnamefont
			{Lynch}},\ }\href {https://doi.org/10.1016/0038-1098(80)91203-X} {\bibfield
		{journal} {\bibinfo  {journal} {Solid State Commun.}\ }\textbf {\bibinfo
			{volume} {33}},\ \bibinfo {pages} {849} (\bibinfo {year}
		{1980}{\natexlab{b}})}\BibitemShut {NoStop}%
	\bibitem [{Note2()}]{Note2}%
	\BibitemOpen
	\bibinfo {note} {Note that the results only indicate that the TA signal is
		not sensitive to the particular phonon heating dynamics in the experiment. It
		does not imply that core-level absorption spectroscopy is insensitive to
		phonon dynamics overall and depending on the system measured, lattice
		dynamics can contribute to core-level TA signals (e.g. Ref. \protect \citet
		{rothenbachMicroscopicNonequilibriumEnergy2019}).}\BibitemShut {Stop}%
	\bibitem [{\citenamefont {Waldecker}\ \emph {et~al.}(2016)\citenamefont
		{Waldecker}, \citenamefont {Bertoni}, \citenamefont {Ernstorfer},\ and\
		\citenamefont {Vorberger}}]{Waldecker2016}%
	\BibitemOpen
	\bibfield  {author} {\bibinfo {author} {\bibfnamefont {L.}~\bibnamefont
			{Waldecker}}, \bibinfo {author} {\bibfnamefont {R.}~\bibnamefont {Bertoni}},
		\bibinfo {author} {\bibfnamefont {R.}~\bibnamefont {Ernstorfer}},\ and\
		\bibinfo {author} {\bibfnamefont {J.}~\bibnamefont {Vorberger}},\ }\href
	{https://doi.org/10.1103/PhysRevX.6.021003} {\bibfield  {journal} {\bibinfo
			{journal} {Phys. Rev. X}\ }\textbf {\bibinfo {volume} {6}},\ \bibinfo {pages}
		{021003} (\bibinfo {year} {2016})}\BibitemShut {NoStop}%
	\bibitem [{Note3()}]{Note3}%
	\BibitemOpen
	\bibinfo {note} {Note that, however, because the core-level absorption
		spectrum at the Ni M$_{2,3}$ edge cannot be directly mapped on to the CB DOS,
		it is impossible to directly quantify the deviation of the carrier
		distribution from a hot Fermi-Dirac function with the core-level TA
		spectra.}\BibitemShut {Stop}%
	\bibitem [{\citenamefont {Pines}\ and\ \citenamefont
		{Nozi{\`e}res}(1989)}]{pinesTheoryQuantumLiquids1989}%
	\BibitemOpen
	\bibfield  {author} {\bibinfo {author} {\bibfnamefont {D.}~\bibnamefont
			{Pines}}\ and\ \bibinfo {author} {\bibfnamefont {P.}~\bibnamefont
			{Nozi{\`e}res}},\ }\href@noop {} {\emph {\bibinfo {title} {The Theory of
				Quantum Liquids}}},\ Advanced Book Classics Series\ (\bibinfo  {publisher}
	{{Addison-Wesley Pub. Co., Advanced Book Program}},\ \bibinfo {address}
	{{Redwood City, Calif}},\ \bibinfo {year} {1989})\BibitemShut {NoStop}%
	\bibitem [{\citenamefont {Rothenbach}\ \emph {et~al.}(2019)\citenamefont
		{Rothenbach}, \citenamefont {Gruner}, \citenamefont {Ollefs}, \citenamefont
		{{Schmitz-Antoniak}}, \citenamefont {Salamon}, \citenamefont {Zhou},
		\citenamefont {Li}, \citenamefont {Mo}, \citenamefont {Park}, \citenamefont
		{Shen}, \citenamefont {Weathersby}, \citenamefont {Yang}, \citenamefont
		{Wang}, \citenamefont {Pentcheva}, \citenamefont {Wende}, \citenamefont
		{Bovensiepen}, \citenamefont {{Sokolowski-Tinten}},\ and\ \citenamefont
		{Eschenlohr}}]{rothenbachMicroscopicNonequilibriumEnergy2019}%
	\BibitemOpen
	\bibfield  {author} {\bibinfo {author} {\bibfnamefont {N.}~\bibnamefont
			{Rothenbach}}, \bibinfo {author} {\bibfnamefont {M.~E.}\ \bibnamefont
			{Gruner}}, \bibinfo {author} {\bibfnamefont {K.}~\bibnamefont {Ollefs}},
		\bibinfo {author} {\bibfnamefont {C.}~\bibnamefont {{Schmitz-Antoniak}}},
		\bibinfo {author} {\bibfnamefont {S.}~\bibnamefont {Salamon}}, \bibinfo
		{author} {\bibfnamefont {P.}~\bibnamefont {Zhou}}, \bibinfo {author}
		{\bibfnamefont {R.}~\bibnamefont {Li}}, \bibinfo {author} {\bibfnamefont
			{M.}~\bibnamefont {Mo}}, \bibinfo {author} {\bibfnamefont {S.}~\bibnamefont
			{Park}}, \bibinfo {author} {\bibfnamefont {X.}~\bibnamefont {Shen}}, \bibinfo
		{author} {\bibfnamefont {S.}~\bibnamefont {Weathersby}}, \bibinfo {author}
		{\bibfnamefont {J.}~\bibnamefont {Yang}}, \bibinfo {author} {\bibfnamefont
			{X.~J.}\ \bibnamefont {Wang}}, \bibinfo {author} {\bibfnamefont
			{R.}~\bibnamefont {Pentcheva}}, \bibinfo {author} {\bibfnamefont
			{H.}~\bibnamefont {Wende}}, \bibinfo {author} {\bibfnamefont
			{U.}~\bibnamefont {Bovensiepen}}, \bibinfo {author} {\bibfnamefont
			{K.}~\bibnamefont {{Sokolowski-Tinten}}},\ and\ \bibinfo {author}
		{\bibfnamefont {A.}~\bibnamefont {Eschenlohr}},\ }\href
	{https://doi.org/10.1103/PhysRevB.100.174301} {\bibfield  {journal} {\bibinfo
			{journal} {Phys. Rev. B}\ }\textbf {\bibinfo {volume} {100}},\ \bibinfo
		{pages} {174301} (\bibinfo {year} {2019})}\BibitemShut {NoStop}%
	\bibitem [{\citenamefont {Lin}\ and\ \citenamefont
		{Zhigilei}(2007)}]{linTemperatureDependencesElectronphonon2007}%
	\BibitemOpen
	\bibfield  {author} {\bibinfo {author} {\bibfnamefont {Z.}~\bibnamefont
			{Lin}}\ and\ \bibinfo {author} {\bibfnamefont {L.~V.}\ \bibnamefont
			{Zhigilei}},\ }\href {https://doi.org/10.1016/j.apsusc.2007.01.032}
	{\bibfield  {journal} {\bibinfo  {journal} {Appl. Surf. Sci.}\ }\textbf
		{\bibinfo {volume} {253}},\ \bibinfo {pages} {6295} (\bibinfo {year}
		{2007})}\BibitemShut {NoStop}%
	\bibitem [{\citenamefont {Lin}\ \emph {et~al.}(2008)\citenamefont {Lin},
		\citenamefont {Zhigilei},\ and\ \citenamefont
		{Celli}}]{linElectronphononCouplingElectron2008}%
	\BibitemOpen
	\bibfield  {author} {\bibinfo {author} {\bibfnamefont {Z.}~\bibnamefont
			{Lin}}, \bibinfo {author} {\bibfnamefont {L.~V.}\ \bibnamefont {Zhigilei}},\
		and\ \bibinfo {author} {\bibfnamefont {V.}~\bibnamefont {Celli}},\ }\href
	{https://doi.org/10.1103/PhysRevB.77.075133} {\bibfield  {journal} {\bibinfo
			{journal} {Phys. Rev. B}\ }\textbf {\bibinfo {volume} {77}},\ \bibinfo
		{pages} {075133} (\bibinfo {year} {2008})}\BibitemShut {NoStop}%
	\bibitem [{\citenamefont {B{\'e}villon}\ \emph {et~al.}(2014)\citenamefont
		{B{\'e}villon}, \citenamefont {Colombier}, \citenamefont {Recoules},\ and\
		\citenamefont {Stoian}}]{bevillonFreeelectronPropertiesMetals2014}%
	\BibitemOpen
	\bibfield  {author} {\bibinfo {author} {\bibfnamefont {E.}~\bibnamefont
			{B{\'e}villon}}, \bibinfo {author} {\bibfnamefont {J.~P.}\ \bibnamefont
			{Colombier}}, \bibinfo {author} {\bibfnamefont {V.}~\bibnamefont
			{Recoules}},\ and\ \bibinfo {author} {\bibfnamefont {R.}~\bibnamefont
			{Stoian}},\ }\href {https://doi.org/10.1103/PhysRevB.89.115117} {\bibfield
		{journal} {\bibinfo  {journal} {Phys. Rev. B}\ }\textbf {\bibinfo {volume}
			{89}},\ \bibinfo {pages} {115117} (\bibinfo {year} {2014})}\BibitemShut
	{NoStop}%
	\bibitem [{\citenamefont {Kasap}\ and\ \citenamefont
		{Capper}(2017)}]{kasapSpringerHandbookElectronic2017}%
	\BibitemOpen
	\bibinfo {editor} {\bibfnamefont {S.}~\bibnamefont {Kasap}}\ and\ \bibinfo
	{editor} {\bibfnamefont {P.}~\bibnamefont {Capper}},\ eds.,\ \href@noop {}
	{\emph {\bibinfo {title} {Springer Handbook of Electronic and Photonic
				Materials}}},\ \bibinfo {edition} {2nd}\ ed.,\ Springer {{Handbooks}}\
	(\bibinfo  {publisher} {{Springer}},\ \bibinfo {address} {{Cham,
			Switzerland}},\ \bibinfo {year} {2017})\BibitemShut {NoStop}%
	\bibitem [{\citenamefont {Timmers}\ \emph {et~al.}(2017)\citenamefont
		{Timmers}, \citenamefont {Kobayashi}, \citenamefont {Chang}, \citenamefont
		{Reduzzi}, \citenamefont {Neumark},\ and\ \citenamefont
		{Leone}}]{Timmers2017}%
	\BibitemOpen
	\bibfield  {author} {\bibinfo {author} {\bibfnamefont {H.}~\bibnamefont
			{Timmers}}, \bibinfo {author} {\bibfnamefont {Y.}~\bibnamefont {Kobayashi}},
		\bibinfo {author} {\bibfnamefont {K.~F.}\ \bibnamefont {Chang}}, \bibinfo
		{author} {\bibfnamefont {M.}~\bibnamefont {Reduzzi}}, \bibinfo {author}
		{\bibfnamefont {D.~M.}\ \bibnamefont {Neumark}},\ and\ \bibinfo {author}
		{\bibfnamefont {S.~R.}\ \bibnamefont {Leone}},\ }\href
	{https://doi.org/10.1364/OL.42.000811} {\bibfield  {journal} {\bibinfo
			{journal} {Opt. Lett.}\ }\textbf {\bibinfo {volume} {42}},\ \bibinfo {pages}
		{811} (\bibinfo {year} {2017})}\BibitemShut {NoStop}%
	\bibitem [{\citenamefont {Silva}\ \emph {et~al.}(2014)\citenamefont {Silva},
		\citenamefont {Miranda}, \citenamefont {Alonso}, \citenamefont
		{Rauschenberger}, \citenamefont {Pervak},\ and\ \citenamefont
		{Crespo}}]{Silva2014}%
	\BibitemOpen
	\bibfield  {author} {\bibinfo {author} {\bibfnamefont {F.}~\bibnamefont
			{Silva}}, \bibinfo {author} {\bibfnamefont {M.}~\bibnamefont {Miranda}},
		\bibinfo {author} {\bibfnamefont {B.}~\bibnamefont {Alonso}}, \bibinfo
		{author} {\bibfnamefont {J.}~\bibnamefont {Rauschenberger}}, \bibinfo
		{author} {\bibfnamefont {V.}~\bibnamefont {Pervak}},\ and\ \bibinfo {author}
		{\bibfnamefont {H.}~\bibnamefont {Crespo}},\ }\href
	{https://doi.org/10.1364/OE.22.010181} {\bibfield  {journal} {\bibinfo
			{journal} {Opt. Express}\ }\textbf {\bibinfo {volume} {22}},\ \bibinfo
		{pages} {10181} (\bibinfo {year} {2014})}\BibitemShut {NoStop}%
	\bibitem [{\citenamefont {Kaldun}\ \emph {et~al.}(2016)\citenamefont {Kaldun},
		\citenamefont {Bl{\"a}ttermann}, \citenamefont {Stoo{\ss}}, \citenamefont
		{Donsa}, \citenamefont {Wei}, \citenamefont {Pazourek}, \citenamefont
		{Nagele}, \citenamefont {Ott}, \citenamefont {Lin}, \citenamefont
		{Burgd{\"o}rfer},\ and\ \citenamefont {Pfeifer}}]{Kaldun2016}%
	\BibitemOpen
	\bibfield  {author} {\bibinfo {author} {\bibfnamefont {A.}~\bibnamefont
			{Kaldun}}, \bibinfo {author} {\bibfnamefont {A.}~\bibnamefont
			{Bl{\"a}ttermann}}, \bibinfo {author} {\bibfnamefont {V.}~\bibnamefont
			{Stoo{\ss}}}, \bibinfo {author} {\bibfnamefont {S.}~\bibnamefont {Donsa}},
		\bibinfo {author} {\bibfnamefont {H.}~\bibnamefont {Wei}}, \bibinfo {author}
		{\bibfnamefont {R.}~\bibnamefont {Pazourek}}, \bibinfo {author}
		{\bibfnamefont {S.}~\bibnamefont {Nagele}}, \bibinfo {author} {\bibfnamefont
			{C.}~\bibnamefont {Ott}}, \bibinfo {author} {\bibfnamefont {C.~D.}\
			\bibnamefont {Lin}}, \bibinfo {author} {\bibfnamefont {J.}~\bibnamefont
			{Burgd{\"o}rfer}},\ and\ \bibinfo {author} {\bibfnamefont {T.}~\bibnamefont
			{Pfeifer}},\ }\href {https://doi.org/10.1126/science.aah6972} {\bibfield
		{journal} {\bibinfo  {journal} {Science}\ }\textbf {\bibinfo {volume}
			{354}},\ \bibinfo {pages} {738} (\bibinfo {year} {2016})}\BibitemShut
	{NoStop}%
	\bibitem [{\citenamefont {Guggenmos}\ \emph {et~al.}(2013)\citenamefont
		{Guggenmos}, \citenamefont {Rauhut}, \citenamefont {Hofstetter},
		\citenamefont {Hertrich}, \citenamefont {Nickel}, \citenamefont {Schmidt},
		\citenamefont {Gullikson}, \citenamefont {Seibald}, \citenamefont {Schnick},\
		and\ \citenamefont {Kleineberg}}]{guggenmosAperiodicCrScMultilayer2013}%
	\BibitemOpen
	\bibfield  {author} {\bibinfo {author} {\bibfnamefont {A.}~\bibnamefont
			{Guggenmos}}, \bibinfo {author} {\bibfnamefont {R.}~\bibnamefont {Rauhut}},
		\bibinfo {author} {\bibfnamefont {M.}~\bibnamefont {Hofstetter}}, \bibinfo
		{author} {\bibfnamefont {S.}~\bibnamefont {Hertrich}}, \bibinfo {author}
		{\bibfnamefont {B.}~\bibnamefont {Nickel}}, \bibinfo {author} {\bibfnamefont
			{J.}~\bibnamefont {Schmidt}}, \bibinfo {author} {\bibfnamefont {E.~M.}\
			\bibnamefont {Gullikson}}, \bibinfo {author} {\bibfnamefont {M.}~\bibnamefont
			{Seibald}}, \bibinfo {author} {\bibfnamefont {W.}~\bibnamefont {Schnick}},\
		and\ \bibinfo {author} {\bibfnamefont {U.}~\bibnamefont {Kleineberg}},\
	}\href {https://doi.org/10.1364/OE.21.021728} {\bibfield  {journal} {\bibinfo
			{journal} {Opt. Express}\ }\textbf {\bibinfo {volume} {21}},\ \bibinfo
		{pages} {21728} (\bibinfo {year} {2013})}\BibitemShut {NoStop}%
	\bibitem [{\citenamefont {Burkhard}\ \emph {et~al.}(2010)\citenamefont
		{Burkhard}, \citenamefont {Hoke},\ and\ \citenamefont
		{McGehee}}]{burkhardAccountingInterferenceScattering2010}%
	\BibitemOpen
	\bibfield  {author} {\bibinfo {author} {\bibfnamefont {G.~F.}\ \bibnamefont
			{Burkhard}}, \bibinfo {author} {\bibfnamefont {E.~T.}\ \bibnamefont {Hoke}},\
		and\ \bibinfo {author} {\bibfnamefont {M.~D.}\ \bibnamefont {McGehee}},\
	}\href {https://doi.org/10.1002/adma.201000883} {\bibfield  {journal}
		{\bibinfo  {journal} {Adv. Mater.}\ }\textbf {\bibinfo {volume} {22}},\
		\bibinfo {pages} {3293} (\bibinfo {year} {2010})}\BibitemShut {NoStop}%
	\bibitem [{\citenamefont {Johnson}\ and\ \citenamefont
		{Christy}(1974)}]{johnsonOpticalConstantsTransition1974}%
	\BibitemOpen
	\bibfield  {author} {\bibinfo {author} {\bibfnamefont {P.}~\bibnamefont
			{Johnson}}\ and\ \bibinfo {author} {\bibfnamefont {R.}~\bibnamefont
			{Christy}},\ }\href {https://doi.org/10.1103/PhysRevB.9.5056} {\bibfield
		{journal} {\bibinfo  {journal} {Phys. Rev. B}\ }\textbf {\bibinfo {volume}
			{9}},\ \bibinfo {pages} {5056} (\bibinfo {year} {1974})}\BibitemShut
	{NoStop}%
	\bibitem [{\citenamefont {Vogt}(2015)}]{vogtDevelopmentPhysicalModels2015}%
	\BibitemOpen
	\bibfield  {author} {\bibinfo {author} {\bibfnamefont {M.~R.}\ \bibnamefont
			{Vogt}},\ }\emph {\bibinfo {title} {Development of Physical Models for the
			Simulation of Optical Properties of Solar Cell Modules}},\ \href@noop {}
	{Ph.D. thesis},\ \bibinfo  {school} {Dissertation, Gottfried Wilhelm Leibniz
		Universit\"at Hannover} (\bibinfo {year} {2015})\BibitemShut {NoStop}%
	\bibitem [{\citenamefont {Lax}(1952)}]{laxFranckCondonPrinciple1952}%
	\BibitemOpen
	\bibfield  {author} {\bibinfo {author} {\bibfnamefont {M.}~\bibnamefont
			{Lax}},\ }\href {https://doi.org/10.1063/1.1700283} {\bibfield  {journal}
		{\bibinfo  {journal} {J. Chem. Phys.}\ }\textbf {\bibinfo {volume} {20}},\
		\bibinfo {pages} {1752} (\bibinfo {year} {1952})}\BibitemShut {NoStop}%
	\bibitem [{\citenamefont {Kubo}(1962)}]{kuboGeneralizedCumulantExpansion1962}%
	\BibitemOpen
	\bibfield  {author} {\bibinfo {author} {\bibfnamefont {R.}~\bibnamefont
			{Kubo}},\ }\href {https://doi.org/10.1143/JPSJ.17.1100} {\bibfield  {journal}
		{\bibinfo  {journal} {J. Phys. Soc. Jpn.}\ }\textbf {\bibinfo {volume}
			{17}},\ \bibinfo {pages} {1100} (\bibinfo {year} {1962})}\BibitemShut
	{NoStop}%
	\bibitem [{\citenamefont {de~Groot}\ and\ \citenamefont
		{Kotani}(2008)}]{grootCoreLevelSpectroscopy2008}%
	\BibitemOpen
	\bibfield  {author} {\bibinfo {author} {\bibfnamefont {F.}~\bibnamefont
			{de~Groot}}\ and\ \bibinfo {author} {\bibfnamefont {A.}~\bibnamefont
			{Kotani}},\ }\href@noop {} {\emph {\bibinfo {title} {Core Level Spectroscopy
				of Solids}}},\ \bibinfo {series} {Advances in Condensed Matter Science}\ No.\
	\bibinfo {number} {v. 6}\ (\bibinfo  {publisher} {{CRC Press}},\ \bibinfo
	{address} {{Boca Raton}},\ \bibinfo {year} {2008})\BibitemShut {NoStop}%
	\bibitem [{\citenamefont {Abramowitz}\ and\ \citenamefont
		{Stegun}(1970)}]{abramowitzHandbookMathematicalFunctions1970}%
	\BibitemOpen
	\bibinfo {editor} {\bibfnamefont {M.}~\bibnamefont {Abramowitz}}\ and\
	\bibinfo {editor} {\bibfnamefont {I.~A.}\ \bibnamefont {Stegun}},\ eds.,\
	\href@noop {} {\emph {\bibinfo {title} {Handbook of Mathematical Functions:
				With Formulas, Graphs, and Mathematical Tables}}},\ Dover Books on Advanced
	Mathematics\ (\bibinfo  {publisher} {{Dover Publications}},\ \bibinfo
	{address} {{New York}},\ \bibinfo {year} {1970})\BibitemShut {NoStop}%
	\bibitem [{\citenamefont {Perdew}\ \emph {et~al.}(1996)\citenamefont {Perdew},
		\citenamefont {Burke},\ and\ \citenamefont
		{Ernzerhof}}]{perdewGeneralizedGradientApproximation1996}%
	\BibitemOpen
	\bibfield  {author} {\bibinfo {author} {\bibfnamefont {J.~P.}\ \bibnamefont
			{Perdew}}, \bibinfo {author} {\bibfnamefont {K.}~\bibnamefont {Burke}},\ and\
		\bibinfo {author} {\bibfnamefont {M.}~\bibnamefont {Ernzerhof}},\ }\href
	{https://doi.org/10.1103/PhysRevLett.77.3865} {\bibfield  {journal} {\bibinfo
			{journal} {Phys. Rev. Lett.}\ }\textbf {\bibinfo {volume} {77}},\ \bibinfo
		{pages} {3865} (\bibinfo {year} {1996})}\BibitemShut {NoStop}%
	\bibitem [{\citenamefont {Giannozzi}\ \emph {et~al.}(2009)\citenamefont
		{Giannozzi}, \citenamefont {Baroni}, \citenamefont {Bonini}, \citenamefont
		{Calandra}, \citenamefont {Car}, \citenamefont {Cavazzoni}, \citenamefont
		{Ceresoli}, \citenamefont {Chiarotti}, \citenamefont {Cococcioni},
		\citenamefont {Dabo}, \citenamefont {Dal~Corso}, \citenamefont {{de
				Gironcoli}}, \citenamefont {Fabris}, \citenamefont {Fratesi}, \citenamefont
		{Gebauer}, \citenamefont {Gerstmann}, \citenamefont {Gougoussis},
		\citenamefont {Kokalj}, \citenamefont {Lazzeri}, \citenamefont
		{{Martin-Samos}}, \citenamefont {Marzari}, \citenamefont {Mauri},
		\citenamefont {Mazzarello}, \citenamefont {Paolini}, \citenamefont
		{Pasquarello}, \citenamefont {Paulatto}, \citenamefont {Sbraccia},
		\citenamefont {Scandolo}, \citenamefont {Sclauzero}, \citenamefont
		{Seitsonen}, \citenamefont {Smogunov}, \citenamefont {Umari},\ and\
		\citenamefont {Wentzcovitch}}]{giannozziQUANTUMESPRESSOModular2009}%
	\BibitemOpen
	\bibfield  {author} {\bibinfo {author} {\bibfnamefont {P.}~\bibnamefont
			{Giannozzi}}, \bibinfo {author} {\bibfnamefont {S.}~\bibnamefont {Baroni}},
		\bibinfo {author} {\bibfnamefont {N.}~\bibnamefont {Bonini}}, \bibinfo
		{author} {\bibfnamefont {M.}~\bibnamefont {Calandra}}, \bibinfo {author}
		{\bibfnamefont {R.}~\bibnamefont {Car}}, \bibinfo {author} {\bibfnamefont
			{C.}~\bibnamefont {Cavazzoni}}, \bibinfo {author} {\bibfnamefont
			{D.}~\bibnamefont {Ceresoli}}, \bibinfo {author} {\bibfnamefont {G.~L.}\
			\bibnamefont {Chiarotti}}, \bibinfo {author} {\bibfnamefont {M.}~\bibnamefont
			{Cococcioni}}, \bibinfo {author} {\bibfnamefont {I.}~\bibnamefont {Dabo}},
		\bibinfo {author} {\bibfnamefont {A.}~\bibnamefont {Dal~Corso}}, \bibinfo
		{author} {\bibfnamefont {S.}~\bibnamefont {{de Gironcoli}}}, \bibinfo
		{author} {\bibfnamefont {S.}~\bibnamefont {Fabris}}, \bibinfo {author}
		{\bibfnamefont {G.}~\bibnamefont {Fratesi}}, \bibinfo {author} {\bibfnamefont
			{R.}~\bibnamefont {Gebauer}}, \bibinfo {author} {\bibfnamefont
			{U.}~\bibnamefont {Gerstmann}}, \bibinfo {author} {\bibfnamefont
			{C.}~\bibnamefont {Gougoussis}}, \bibinfo {author} {\bibfnamefont
			{A.}~\bibnamefont {Kokalj}}, \bibinfo {author} {\bibfnamefont
			{M.}~\bibnamefont {Lazzeri}}, \bibinfo {author} {\bibfnamefont
			{L.}~\bibnamefont {{Martin-Samos}}}, \bibinfo {author} {\bibfnamefont
			{N.}~\bibnamefont {Marzari}}, \bibinfo {author} {\bibfnamefont
			{F.}~\bibnamefont {Mauri}}, \bibinfo {author} {\bibfnamefont
			{R.}~\bibnamefont {Mazzarello}}, \bibinfo {author} {\bibfnamefont
			{S.}~\bibnamefont {Paolini}}, \bibinfo {author} {\bibfnamefont
			{A.}~\bibnamefont {Pasquarello}}, \bibinfo {author} {\bibfnamefont
			{L.}~\bibnamefont {Paulatto}}, \bibinfo {author} {\bibfnamefont
			{C.}~\bibnamefont {Sbraccia}}, \bibinfo {author} {\bibfnamefont
			{S.}~\bibnamefont {Scandolo}}, \bibinfo {author} {\bibfnamefont
			{G.}~\bibnamefont {Sclauzero}}, \bibinfo {author} {\bibfnamefont {A.~P.}\
			\bibnamefont {Seitsonen}}, \bibinfo {author} {\bibfnamefont {A.}~\bibnamefont
			{Smogunov}}, \bibinfo {author} {\bibfnamefont {P.}~\bibnamefont {Umari}},\
		and\ \bibinfo {author} {\bibfnamefont {R.~M.}\ \bibnamefont {Wentzcovitch}},\
	}\href {https://doi.org/10.1088/0953-8984/21/39/395502} {\bibfield  {journal}
		{\bibinfo  {journal} {J. Phys. Condens. Matter}\ }\textbf {\bibinfo {volume}
			{21}},\ \bibinfo {pages} {395502} (\bibinfo {year} {2009})}\BibitemShut
	{NoStop}%
	\bibitem [{\citenamefont {Giannozzi}\ \emph {et~al.}(2017)\citenamefont
		{Giannozzi}, \citenamefont {Andreussi}, \citenamefont {Brumme}, \citenamefont
		{Bunau}, \citenamefont {Buongiorno~Nardelli}, \citenamefont {Calandra},
		\citenamefont {Car}, \citenamefont {Cavazzoni}, \citenamefont {Ceresoli},
		\citenamefont {Cococcioni}, \citenamefont {Colonna}, \citenamefont
		{Carnimeo}, \citenamefont {Dal~Corso}, \citenamefont {{de Gironcoli}},
		\citenamefont {Delugas}, \citenamefont {DiStasio}, \citenamefont {Ferretti},
		\citenamefont {Floris}, \citenamefont {Fratesi}, \citenamefont {Fugallo},
		\citenamefont {Gebauer}, \citenamefont {Gerstmann}, \citenamefont {Giustino},
		\citenamefont {Gorni}, \citenamefont {Jia}, \citenamefont {Kawamura},
		\citenamefont {Ko}, \citenamefont {Kokalj}, \citenamefont {K{\"u}{\c
				c}{\"u}kbenli}, \citenamefont {Lazzeri}, \citenamefont {Marsili},
		\citenamefont {Marzari}, \citenamefont {Mauri}, \citenamefont {Nguyen},
		\citenamefont {Nguyen}, \citenamefont {{Otero-de-la-Roza}}, \citenamefont
		{Paulatto}, \citenamefont {Ponc{\'e}}, \citenamefont {Rocca}, \citenamefont
		{Sabatini}, \citenamefont {Santra}, \citenamefont {Schlipf}, \citenamefont
		{Seitsonen}, \citenamefont {Smogunov}, \citenamefont {Timrov}, \citenamefont
		{Thonhauser}, \citenamefont {Umari}, \citenamefont {Vast}, \citenamefont
		{Wu},\ and\ \citenamefont
		{Baroni}}]{giannozziAdvancedCapabilitiesMaterials2017}%
	\BibitemOpen
	\bibfield  {author} {\bibinfo {author} {\bibfnamefont {P.}~\bibnamefont
			{Giannozzi}}, \bibinfo {author} {\bibfnamefont {O.}~\bibnamefont
			{Andreussi}}, \bibinfo {author} {\bibfnamefont {T.}~\bibnamefont {Brumme}},
		\bibinfo {author} {\bibfnamefont {O.}~\bibnamefont {Bunau}}, \bibinfo
		{author} {\bibfnamefont {M.}~\bibnamefont {Buongiorno~Nardelli}}, \bibinfo
		{author} {\bibfnamefont {M.}~\bibnamefont {Calandra}}, \bibinfo {author}
		{\bibfnamefont {R.}~\bibnamefont {Car}}, \bibinfo {author} {\bibfnamefont
			{C.}~\bibnamefont {Cavazzoni}}, \bibinfo {author} {\bibfnamefont
			{D.}~\bibnamefont {Ceresoli}}, \bibinfo {author} {\bibfnamefont
			{M.}~\bibnamefont {Cococcioni}}, \bibinfo {author} {\bibfnamefont
			{N.}~\bibnamefont {Colonna}}, \bibinfo {author} {\bibfnamefont
			{I.}~\bibnamefont {Carnimeo}}, \bibinfo {author} {\bibfnamefont
			{A.}~\bibnamefont {Dal~Corso}}, \bibinfo {author} {\bibfnamefont
			{S.}~\bibnamefont {{de Gironcoli}}}, \bibinfo {author} {\bibfnamefont
			{P.}~\bibnamefont {Delugas}}, \bibinfo {author} {\bibfnamefont {R.~A.}\
			\bibnamefont {DiStasio}}, \bibinfo {author} {\bibfnamefont {A.}~\bibnamefont
			{Ferretti}}, \bibinfo {author} {\bibfnamefont {A.}~\bibnamefont {Floris}},
		\bibinfo {author} {\bibfnamefont {G.}~\bibnamefont {Fratesi}}, \bibinfo
		{author} {\bibfnamefont {G.}~\bibnamefont {Fugallo}}, \bibinfo {author}
		{\bibfnamefont {R.}~\bibnamefont {Gebauer}}, \bibinfo {author} {\bibfnamefont
			{U.}~\bibnamefont {Gerstmann}}, \bibinfo {author} {\bibfnamefont
			{F.}~\bibnamefont {Giustino}}, \bibinfo {author} {\bibfnamefont
			{T.}~\bibnamefont {Gorni}}, \bibinfo {author} {\bibfnamefont
			{J.}~\bibnamefont {Jia}}, \bibinfo {author} {\bibfnamefont {M.}~\bibnamefont
			{Kawamura}}, \bibinfo {author} {\bibfnamefont {H.-Y.}\ \bibnamefont {Ko}},
		\bibinfo {author} {\bibfnamefont {A.}~\bibnamefont {Kokalj}}, \bibinfo
		{author} {\bibfnamefont {E.}~\bibnamefont {K{\"u}{\c c}{\"u}kbenli}},
		\bibinfo {author} {\bibfnamefont {M.}~\bibnamefont {Lazzeri}}, \bibinfo
		{author} {\bibfnamefont {M.}~\bibnamefont {Marsili}}, \bibinfo {author}
		{\bibfnamefont {N.}~\bibnamefont {Marzari}}, \bibinfo {author} {\bibfnamefont
			{F.}~\bibnamefont {Mauri}}, \bibinfo {author} {\bibfnamefont {N.~L.}\
			\bibnamefont {Nguyen}}, \bibinfo {author} {\bibfnamefont {H.-V.}\
			\bibnamefont {Nguyen}}, \bibinfo {author} {\bibfnamefont {A.}~\bibnamefont
			{{Otero-de-la-Roza}}}, \bibinfo {author} {\bibfnamefont {L.}~\bibnamefont
			{Paulatto}}, \bibinfo {author} {\bibfnamefont {S.}~\bibnamefont {Ponc{\'e}}},
		\bibinfo {author} {\bibfnamefont {D.}~\bibnamefont {Rocca}}, \bibinfo
		{author} {\bibfnamefont {R.}~\bibnamefont {Sabatini}}, \bibinfo {author}
		{\bibfnamefont {B.}~\bibnamefont {Santra}}, \bibinfo {author} {\bibfnamefont
			{M.}~\bibnamefont {Schlipf}}, \bibinfo {author} {\bibfnamefont {A.~P.}\
			\bibnamefont {Seitsonen}}, \bibinfo {author} {\bibfnamefont {A.}~\bibnamefont
			{Smogunov}}, \bibinfo {author} {\bibfnamefont {I.}~\bibnamefont {Timrov}},
		\bibinfo {author} {\bibfnamefont {T.}~\bibnamefont {Thonhauser}}, \bibinfo
		{author} {\bibfnamefont {P.}~\bibnamefont {Umari}}, \bibinfo {author}
		{\bibfnamefont {N.}~\bibnamefont {Vast}}, \bibinfo {author} {\bibfnamefont
			{X.}~\bibnamefont {Wu}},\ and\ \bibinfo {author} {\bibfnamefont
			{S.}~\bibnamefont {Baroni}},\ }\href
	{https://doi.org/10.1088/1361-648X/aa8f79} {\bibfield  {journal} {\bibinfo
			{journal} {J. Phys. Condens. Matter}\ }\textbf {\bibinfo {volume} {29}},\
		\bibinfo {pages} {465901} (\bibinfo {year} {2017})}\BibitemShut {NoStop}%
	\bibitem [{\citenamefont {Monkhorst}\ and\ \citenamefont
		{Pack}(1976)}]{monkhorstSpecialPointsBrillouinzone1976}%
	\BibitemOpen
	\bibfield  {author} {\bibinfo {author} {\bibfnamefont {H.~J.}\ \bibnamefont
			{Monkhorst}}\ and\ \bibinfo {author} {\bibfnamefont {J.~D.}\ \bibnamefont
			{Pack}},\ }\href {https://doi.org/10.1103/PhysRevB.13.5188} {\bibfield
		{journal} {\bibinfo  {journal} {Phys. Rev. B}\ }\textbf {\bibinfo {volume}
			{13}},\ \bibinfo {pages} {5188} (\bibinfo {year} {1976})}\BibitemShut
	{NoStop}%
\end{thebibliography}
%

\end{document}